\documentclass[draft,10pt]{article}
\usepackage{amsmath}
\usepackage{amssymb}
\usepackage{amsthm,amsxtra}
\usepackage{epsf}
\newlength{\mytopmargin}
\newlength{\myleftmargin}
\newtheorem{lemma}{Lemma}
\newtheorem{prop}[lemma]{Proposition}
\newcommand{\zz}{\mathbb Z}
\newcommand{\qq}{\mathbb Q}

\newcommand{\uU}{\overset{\frown}{U}}
\newcommand{\uuU}{\overset{\overset{\frown}{\frown}}{U}}
\newcommand{\dU}{\overset{\smile}{U}}

\newcommand{\PV}{${\rm P}_{\rm V}\;$}
\newcommand{\PII}{${\rm P}_{\rm II}\;$}
\newcommand{\PIII}{${\rm P}_{\rm III}\;$}
\newcommand{\III}{{\rm III'}}
\newcommand{\PIIIa}{${\rm P}_{\rm III^{\prime}}\;$}
\newcommand{\PIV}{${\rm P}_{\rm IV}\;$}
\newcommand{\PVI}{${\rm P}_{\rm VI}\;$}
\newcommand{\dPV}{${\rm dP}_{\rm V}\;$}
\newcommand{\uK}{\overset{\frown}{K}}
\newcommand{\Uq}{\overset{\frown}{q}}
\newcommand{\Up}{\overset{\frown}{p}}
\newcommand{\Dq}{\overset{\smile}{q}}
\newcommand{\Dp}{\overset{\smile}{p}}
\newcommand{\Df}{\overset{\smile}{f}}
\newcommand{\dK}{\overset{\smile}{K}}
\newcommand{\uuK}{\overset{\overset{\frown}{\frown}}{K}}
\newcommand{\DZ}{\overset{\smile}{Z}}
\newcommand{\UZ}{\overset{\frown}{Z}}
\newcommand{\threehalf}{
        {\lower0.00ex\hbox{\raise.6ex\hbox{\the\scriptfont0 3}
                           \kern-.5em\slash\kern-.1em\lower.45ex
                                     \hbox{\the\scriptfont0 2}}}}
\newcommand{\half}{
        {\lower0.00ex\hbox{\raise.6ex\hbox{\the\scriptfont0 1}
                           \kern-.5em\slash\kern-.1em\lower.45ex
                                     \hbox{\the\scriptfont0 2}}}}
\newcommand{\quarter}{
        {\lower0.00ex\hbox{\raise.6ex\hbox{\the\scriptfont0 1}
                           \kern-.5em\slash\kern-.1em\lower.45ex
                                     \hbox{\the\scriptfont0 4}}}}
\newcommand{\threequarter}{
        {\lower0.00ex\hbox{\raise.6ex\hbox{\the\scriptfont0 3}
                           \kern-.5em\slash\kern-.1em\lower.45ex
                                     \hbox{\the\scriptfont0 4}}}}
\newcommand{\eighth}{
        {\lower0.00ex\hbox{\raise.6ex\hbox{\the\scriptfont0 1}
                           \kern-.5em\slash\kern-.1em\lower.45ex
                                     \hbox{\the\scriptfont0 8}}}}
\newcommand{\sixteenth}{
        {\lower0.00ex\hbox{\raise.6ex\hbox{\the\scriptfont0 1}
                           \kern-.5em\slash\kern-.1em\lower.45ex
                                     \hbox{\the\scriptfont0 16}}}}
\newcommand{\thirtytwo}{
        {\lower0.00ex\hbox{\raise.6ex\hbox{\the\scriptfont0 1}
                           \kern-.5em\slash\kern-.1em\lower.45ex
                                     \hbox{\the\scriptfont0 32}}}}

\begin{document}
\vspace{4cm}
\noindent
{\bf Application of the $\tau$-function theory of Painlev\'e equations to
random matrices: \\
\PVI, the JUE, CyUE, cJUE and scaled limits}

\vspace{5mm}
\noindent
P.J.~Forrester and N.S.~Witte${}^\dagger$

\noindent
Department of Mathematics and Statistics
${}^\dagger$(and School of Physics),
University of Melbourne, \\
Victoria 3010, Australia ;
email: p.forrester@ms.unimelb.edu.au; n.witte@ms.unimelb.edu.au

\small
\begin{quote}
Okamoto has obtained a sequence of $\tau$-functions for the \PVI system
expressed as a double Wronskian determinant based on a solution of the
Gauss hypergeometric equation. Starting with integral solutions
of the Gauss hypergeometric equation, we show that the determinant can be
re-expressed as multi-dimensional integrals, and these in turn can be
identified with averages over the eigenvalue probability density function
for the Jacobi unitary ensemble (JUE), and the Cauchy  unitary ensemble
(CyUE) (the latter being equivalent to the circular Jacobi unitary
ensemble (cJUE)). Hence these averages, which depend on four continuous
parameters and the discrete parameter $N$, can be characterised as the solution
of the second order second degree equation satisfied by the Hamiltonian in
the \PVI theory. We show that the Hamiltonian also satisfies an equation
related to the discrete \PV equation, 
thus providing an alternative characterisation in terms of a
difference equation. In the case of the cJUE, the spectrum singularity
scaled limit is considered, and the evaluation of a certain
four parameter average is given in terms of the general \PV
transcendent in $\sigma$ form. Applications are given to the evaluation of
the spacing distribution for the 
circular unitary ensemble (CUE) and its scaled counterpart, giving
formulas more succinct than those known previously; to expressions for the
hard edge gap probability in the scaled Laguerre orthogonal ensemble
(LOE) (parameter $a$ a non-negative integer) and Laguerre symplectic
ensemble (LSE) (parameter $a$ an even non-negative integer) as finite
dimensional combinatorial integrals over the symplectic and orthogonal
groups respectively; to the evaluation of the cumulative distribution
function for the last passage time in certain models of directed
percolation; to the $\tau$-function evaluation of the largest
eigenvalue in the finite LOE and LSE with parameter $a=0$; and to the
characterisation of the diagonal-diagonal spin-spin correlation in the
two-dimensional Ising model. 
\end{quote}

\tableofcontents

\section{Introduction and summary}
\subsection{Setting and objectives}
This paper is the last in a series devoted to a systematic account of
the application of the Okamoto $\tau$-function theory of
Painlev\'e equations to the characterisation of certain averages in random
matrix theory. The $\tau$-function theory applies directly to random matrix
ensembles defined by a probability density function (PDF) of the form
\begin{equation}\label{a1}
{1 \over C} \prod_{j=1}^N w(x_j) \prod_{1 \le j < k \le N} (x_k - x_j)^2,
\end{equation}
where the weight function $w(x)$ is one of the classical forms
\begin{equation}\label{a2}
w(x) = \left \{ \begin{array}{ll} e^{-x^2}, & {\rm Gaussian} \\
x^a e^{-x} \: \: (x >0), & {\rm Laguerre} \\
x^a(1-x)^b \: \: (0<x<1), & {\rm Jacobi} \\
(1+x^2)^{-\eta}, & {\rm Cauchy} \end{array} \right.
\end{equation}
The symbol $C$, which in (\ref{a1}) denotes the normalisation, will be
used throughout to denote {\it some} constant (i.e.~quantity independent
of the primary variables of the equation). 
The PDFs (\ref{a1}) can be realised as the joint eigenvalue distribution
of Hermitian matrices with
independent complex Gaussian entries for the first three weights
of (\ref{a2}), and by a stereographic projection of
random unitary matrices for the Cauchy weight (with $\eta = N$)
(see (\ref{pw2}) below). 
The underlying matrix distributions giving rise to (\ref{a1}) are invariant
with respect to similarity transformations involving unitary matrices,
and for this reason are termed matrix ensembles with a unitary symmetry
or simply
unitary ensembles
(not to be confused with unitary matrices). The name of the weight
function is then prefixed to the term unitary ensemble. 
In this work our
interest is in the Jacobi unitary ensemble (JUE) and the Cauchy unitary
ensemble (CyUE). We will see that the Cauchy unitary ensemble is
equivalent to the circular Jacobi unitary ensemble (cJUE) in which the
eigenvalues are on the unit circle in the complex plane.
The averages of interest are
\begin{equation}\label{a3}
\tilde{E}_N(s;\mu) := \Big \langle \prod_{l=1}^N \chi_{(-\infty, s]}^{(l)}
(s - x_l)^\mu \Big \rangle, \qquad
 \chi_{(-\infty, s]}^{(l)} = \left \{ \begin{array}{ll}
1, & x_l \in (-\infty, s] \\
0, & {\rm otherwise} \end{array} \right.
\end{equation}
and
\begin{equation}\label{a4}
F_N(s;\mu) := \Big \langle \prod_{l=1}^N
(s - x_l)^\mu \Big \rangle
\end{equation}
(in the latter, for $\mu \notin \zz$ a suitable branch must be specified).

For $N=1$ (\ref{a3}) and (\ref{a4}) read
\begin{equation}\label{a5}
\tilde{E}_1(s;\mu) = \int_{-\infty}^s (s-x)^\mu w(x) \, dx, \qquad
F_1(s;\mu) = \int_{-\infty}^\infty (s-x)^\mu w(x) \, dx.
\end{equation}
A fundamental fact is that as a function of $s$, and after
multiplication by a suitable elementary function of $s$, these functions
satisfy a classical second order linear differential equation
(Hermite-Weber equation in the Gaussian case \cite{FW_2001a},
confluent hypergeometric equation in the Laguerre case \cite{FW_2002a},
and, as will be shown below, Gauss hypergeometric equation in the Jacobi
and Cauchy cases). In the Okamoto $\tau$-function theory of \PIV
\cite{Ok_1986}, \PV \cite{Ok_1987b} and \PVI \cite{Ok_1987a}, these same linear
differential equations respectively characterise the first member of an
infinite sequence of $\tau$-functions (the zeroth member, as with
the random matrix averages (\ref{a3}) and (\ref{a4}), is unity).
The general $N$th member of the $\tau$ function sequence is
characterised by the fact that its logarithmic derivative satisfies a
second order second degree differential equation. Furthermore  the
$N$th member can be written explicitly as
a Wronskian determinant and we have shown in the Gaussian case in
\cite{FW_2001a}, and in the Laguerre case in \cite{FW_2002a}, that the
Wronskian determinant is just a rewrite of the average
(\ref{a3}) (or (\ref{a4}) as appropriate). This way we have been able
to characterise (\ref{a3}) and (\ref{a4}) in terms of the solution of
second order second degree differential equations from the Painlev\'e
theory. 
B\"acklund transformations of quantities associated with the
Hamiltonian formalism of the particular Painlev\'e system have also
allowed us to characterise these quantities in terms of difference
equations related to discrete Painlev\'e equations.
In this work we will show that the same strategy allows
(\ref{a3}) and (\ref{a4}) in the Jacobi and Cauchy cases to be
characterised similarly. 
Scaled limits of these averages can then be characterised as solutions
of limiting forms of the differential equations.
Applications are  given to the exact evaluation
of eigenvalue spacing distribution functions,
probabilities of last passage times in directed percolation models, and
the diagonal spin-spin correlation of the two-dimensional Ising
model. 

\subsection{Definition of averages in the various ensembles}
In the Jacobi case the explicit form of (\ref{a3}) and (\ref{a4}) is
\begin{align}
 \tilde{E}_N^{\rm J}(s;a,b,\mu) 
  & = {1 \over C}
        \int_0^s dx_1 \, x_1^a (1-x_1)^b (s - x_1)^\mu \cdots 
        \int_0^s dx_N \, x_N^a (1-x_N)^b (s - x_N)^\mu 
        \prod_{1 \le j < k \le N} (x_k - x_j)^2  
  \nonumber \\
  & = s^{N(a+\mu+N)} {1 \over C}
        \int_0^1 dx_1 \, x_1^a (1-s x_1)^b (1 - x_1)^\mu \cdots
        \int_0^1 dx_N \, x_N^a (1-s x_N)^b (1 -  x_N)^\mu 
  \nonumber \\ 
  & \qquad\times \prod_{1 \le j < k \le N} (x_k - x_j)^2
\label{E1} \\
  {F}_N^{\rm J}(s;a,b,\mu) 
  & = {1 \over C} 
         \int_0^1 dx_1 \, x_1^a (1-x_1)^b (s - x_1)^\mu \cdots 
         \int_0^1 dx_N \, x_N^a (1-x_N)^b (s - x_N)^\mu 
         \prod_{1 \le j < k \le N} (x_k - x_j)^2,
\label{F1}
\end{align}
while the explicit form of (\ref{a3}) and (\ref{a4}) in the Cauchy case is
\begin{align}\label{E2}
  \tilde{E}_N^{\rm Cy}(s;\eta,\mu) 
  & = {1 \over C} 
       \int_{-\infty}^s dx_1 \, {(s - x_1)^\mu \over (1+x_1^2)^\eta} \cdots
       \int_{-\infty}^s dx_N \, {(s - x_N)^\mu \over (1+x_N^2)^\eta} \,
       \prod_{1 \le j < k \le N} (x_k - x_j)^2 \\
  {F}_N^{\rm Cy}(s;\eta,\mu) 
  & = {1 \over C} 
       \int_{-\infty}^\infty dx_1 \, {(s - x_1)^\mu \over (1+x_1^2)^\eta} \cdots
       \int_{-\infty}^\infty dx_N \, {(s - x_N)^\mu \over (1+x_N^2)^\eta} \,
       \prod_{1 \le j < k \le N} (x_k - x_j)^2. 
\label{F2}
\end{align}
A single quantity which combines both (\ref{E1}) and (\ref{F1}) is
\begin{multline}\label{1.17a}
  \tilde{E}_N^{\rm J}(s;a,b,\mu;\xi) := {1 \over C} 
        \Big( \int_0^1 - \xi \int_s^1 \Big ) dx_1 \, x_1^a(1-x_1)^b(s-x_1)^\mu \cdots
        \Big ( \int_0^1 - \xi \int_s^1 \Big ) dx_N \, x_N^a(1-x_N)^b(s-x_N)^\mu \\
   \qquad\times \prod_{1 \le j < k \le N} (x_k - x_j)^2.
\end{multline}
Similarly, a single quantity which combines both (\ref{E2}) and (\ref{F2}) is
\begin{multline}\label{1.14a}
  \tilde{E}_N^{\rm Cy}(s;\eta,\mu;\xi) \\
   := {1 \over C} 
      \Big( \int_{-\infty}^\infty - \xi \int_s^\infty \Big)dx_1 \, 
          {(s - x_1)^\mu \over (1 + x_1^2)^\eta} \cdots
      \Big( \int_{-\infty}^\infty - \xi \int_s^\infty \Big)dx_N \, 
          {(s - x_N)^\mu \over (1 + x_N^2)^\eta}
      \prod_{1 \le j < k \le N} (x_k - x_j)^2.
\end{multline}
As will be revised in Section \ref{s5.1}, the power series expansions in
$\xi$ of (\ref{1.17a}) and (\ref{1.14a}) are quantities of relevance to
conditional spacing distributions in the corresponding random matrix
ensembles. 

The integrand in (\ref{1.14a}) requires $\eta$ to be real for itself to
be real. However, if we write
\begin{equation*}
(1 + x^2)^{-\eta} \mapsto (1 + i x)^{-{\eta}}
(1 - ix)^{-\bar{\eta}},
\end{equation*}
where $\bar{\eta}$ denotes the complex conjugate of $\eta$, then the
integrand remains real for $\eta$ complex, giving a meaningful
generalisation of the original Cauchy ensemble \cite{BO-2001a}. Doing this,
and setting $\eta = \eta_1 + i \eta_2$ we can generalise
(\ref{1.14a}) to read
\begin{multline}\label{1.15a}
  \tilde{E}_N^{\rm Cy}(s;(\eta_1,\eta_2),\mu;\xi) := {1 \over C} 
  \Big( \int_{-\infty}^\infty - \xi \int_s^\infty \Big)dx_1 \,
      {(s - x_1)^\mu \over (1 + i x_1)^{\eta_1 + i \eta_2}
       (1 - i x_1)^{\eta_1 - i \eta_2}} \cdots \\ 
  \times
  \Big( \int_{-\infty}^\infty - \xi \int_s^\infty \Big)dx_N \, 
      {(s - x_N)^\mu \over (1 + i x_N)^{\eta_1 + i \eta_2}
       (1 - i x_N)^{\eta_1 - i \eta_2}}
       \prod_{1 \le j < k \le N} (x_k - x_j)^2.
\end{multline}

It was remarked below (\ref{a4}) that for $(s-x_l)^\mu$ to be well
defined for $\mu \notin \zz$ a definite branch must be specified. For
$s$ real a natural choice is that $(s - x_l)^\mu$ is real for
$x_l < s$ and $(s-x_l)^\mu = e^{-\pi i \mu} |x_l - s|^\mu$ for
$x_l > s$. In particular this shows
\begin{multline}\label{1.20a}
  \tilde{E}_N^{\rm J}(s;a,b,\mu;\xi) := {1 \over C}
    \Big( \int_0^1 - \xi^* \int_s^1 \Big) dx_1 \, x_1^a(1-x_1)^b|s-x_1|^\mu \cdots
    \Big( \int_0^1 - \xi^* \int_s^1 \Big) dx_N \, x_N^a(1-x_N)^b|s-x_N|^\mu \\
  \times \prod_{1 \le j < k \le N} (x_k - x_j)^2,
\end{multline}
where $\xi^* := 1 - (1 - \xi) 
e^{-\pi i \mu}$
and similarly for (\ref{1.14a}) and (\ref{1.15a}).

In the opening paragraph it was noted that the Cauchy ensemble results
from a stereographic projection of the eigenvalue PDF for random unitary
matrices. To be more explicit, following \cite{WF_2000}, consider the ensemble
of random unitary matrices specified by the eigenvalue PDF
\begin{equation}\label{pw}
{1 \over C} \prod_{l=1}^N |1+z_l|^{2\omega_1} \prod_{1 \le j < k \le N}
|z_k - z_j|^2, \qquad z_l = e^{i \theta_l}, \quad
 \theta_l \in [-\pi, \pi].
\end{equation}
In \cite{WF_2000} this was referred to as the circular Jacobi ensemble with
unitary symmetry, and denoted cJUE. When $\omega_1=0$ (\ref{pw}) is realised 
by random unitary matrices with the uniform (Haar) measure and is
then referred to as the circular unitary ensemble (CUE),
while (\ref{pw}) with $\omega_1=1$ gives the eigenvalue PDF of
$(N+1) \times (N+1)$ CUE matrices with all angles $\theta$
measured from any one eigenvalue (taken to be $\theta = \pi$).
Making the change of variable
\begin{equation}\label{pw1}
e^{i \theta} = {1 + ix \over 1 - ix} \qquad x = \tan{\theta \over 2}
\end{equation}
(note that $\theta = \pm\pi$ corresponds to $x \to \pm \infty$) shows
\begin{equation}\label{pw2}
\prod_{l=1}^N |1+z_l|^{2\omega_1} \prod_{1 \le j < k \le N} |z_k - z_j |^2
d \theta_1 \cdots d \theta_N
= 2^{N(N+2\omega_1)} \prod_{l=1}^N {1 \over (1 + x_l^2)^{N+\omega_1}}
\prod_{1 \le j < k \le N} |x_j - x_k|^2 dx_1 \cdots dx_N,
\end{equation}
thus specifying the relation with the Cauchy unitary ensemble.
A generalisation of the cJUE eigenvalue PDF (\ref{pw}) is the PDF
\begin{equation}\label{pw1a}
{1 \over C} \prod_{l=1}^N e^{\omega_2 \theta_l}
|1+z_l|^{2\omega_1} \prod_{1 \le j < k \le N}
|z_k - z_j|^2, \qquad z_l = e^{i \theta_l}, \quad
 \theta_l \in [-\pi, \pi].
\end{equation}
Under the change of variable (\ref{pw1}) this transforms into the PDF 
\begin{equation*}
{1 \over C} \prod_{l=1}^N {1 \over (1+ix_l)^{\omega_1+i\omega_2+N}
(1 - ix_l)^{\omega_1-i\omega_2+N}} \prod_{1 \le j < k \le N} |x_k - x_j|^2
\end{equation*}
which corresponds to the generalised Cauchy probability density in
(\ref{1.15a}). Consequently, changing variables
according to (\ref{pw1}) in 
\begin{equation}\label{pw4}
\tilde{E}_N^{\rm cJ}(\phi;(\omega_1,\omega_2),\mu;\xi) :=
\Big \langle \prod_{l=1}^N(1 - \xi \chi_{(\pi - \phi,\pi)}^{(l)})
e^{\omega_2 \theta_l}
| e^{i(\pi - \phi)} - e^{i \theta_l}|^\mu \Big \rangle_{\rm cJUE}.
\end{equation}
and making use of (\ref{pw2}) and the analogue of (\ref{1.20a}) for
$\tilde{E}_N^{\rm cJ}$
shows
\begin{equation}\label{1.28a}
\tilde{E}_N^{\rm cJ}(\phi;(\omega_1,\omega_2),\mu;\xi^*) \propto 
      {1 \over (1 + s^2)^{N\mu/2}}
\tilde{E}_N^{\rm Cy}(s;(N+\omega_1+\mu/2,\omega_2),\mu;\xi) \Big|_{s=\cot \phi/2}.
\end{equation}

The second Jacobi ensemble integral in (\ref{E1}) shares the feature of the
Cauchy ensemble integral (\ref{1.14a}) of being related to an
average in the CUE. This is done using the integral identity \cite{Fo-95}
\begin{multline}\label{1.19a}
  \int_0^1 dt_1 \, t_1^{\epsilon - 1} \cdots
  \int_0^1 dt_N \, t_N^{\epsilon - 1} \, f(t_1,\dots, t_N) \\
  = \Big( {\pi \over \sin \pi \epsilon} \Big)^N
    \int_{-1/2}^{1/2} dx_1 \, e^{2 \pi i \epsilon x_1} \cdots
    \int_{-1/2}^{1/2}  dx_N \, e^{2 \pi i \epsilon x_N} \,
    f(-e^{2\pi i x_1}, \dots, - e^{2\pi i x_N})
\end{multline}
valid for $f$ a Laurent polynomial, and making use of Carlson's theorem.
One finds
\begin{equation}\label{ju}
\Big \langle \prod_{l=1}^N (1 - t x_l)^\mu \Big \rangle_{\rm JUE} =
{M_N(0,0) \over M_N(a',b')} \Big \langle
\prod_{l=1}^N z_l^{(a'-b')/2} |1 + z_l|^{a'+b'} (1 + tz_l)^\mu
 \Big \rangle_{\rm CUE}
\end{equation}
where $a'=N+a+b$, $b'=-(N+a)$ and
\begin{equation}\label{ju1}
M_N(a',b') := \int_{-1/2}^{1/2} dx_1 \cdots \int_{-1/2}^{1/2} dx_N
\, \prod_{l=1}^N z_l^{(a'-b')/2} |1 + z_l|^{a'+b'}
\prod_{1 \le j < k \le N} |z_j - z_k|^2, \quad z_l := e^{2\pi i x_l}
\end{equation}
(note that the left hand side of (\ref{ju}) corresponds to the second
integral in (\ref{E1}) after interchanging $\mu$ and $b$). The
normalisation (\ref{ju1}) results from the Jacobi ensemble normalisation
\begin{equation}\label{nj}
J_N(a,b) := \int_0^1 dx_1 \, x_1^a(1-x_1)^b \cdots
\int_0^1 dx_N \, x_N^a(1-x_N)^b \prod_{1 \le j < k \le N} (x_k - x_j)^2.
\end{equation}
Both (\ref{ju1}) and (\ref{nj}) have gamma function evaluations,
\begin{align}
  M_N(a,b) 
  & = \prod_{j=0}^{N-1} {\Gamma(a+b+1+j) \Gamma(2+j) \over
                         \Gamma(a+1+j) \Gamma(b+1+j)}
  \label{mna} \\
  J_N(a,b)
  & = \prod_{j=0}^{N-1} {\Gamma(a+1+j) \Gamma(b+1+j)
                         \Gamma(2+j) \over \Gamma(a+b+1+N+j)},
  \label{jna}
\end{align}
with the latter following from the Selberg integral
\cite{Se-44}, and the former by applying (\ref{1.19a}) to the evaluation
(\ref{jna}). 

The Jacobi average (\ref{1.17a}) and the circular Jacobi average
(\ref{pw4}) (which according to (\ref{1.28a}) is equivalent to the
Cauchy average (\ref{1.14a})) admit various scaled $N \to \infty$
limits. There are three distinct possibilities in that the limiting
averages can correspond either to the soft edge, hard edge or a spectrum
singularity in the bulk. The averages at the soft and hard edge have
been studied in our previous papers \cite{FW_2001a,FW_2002a}, giving rise to
\PII and \PIII transcendents respectively, and will not be discussed here.
Instead, attention will be focussed on the scaling to a spectrum
singularity in the bulk. This results by replacing
$X \mapsto X/N$ in (\ref{pw4}), making a suitable choice of the constant
$C$, and taking the $N \to \infty$ limit. The problem with $C$ can
be avoided by taking the logarithmic derivative with respect to $X$, leading
us to consider the scaled quantity
\begin{equation}\label{vpu}
u(X;(\omega_1,\omega_2),\mu;\xi) :=
\lim_{N \to \infty} X {d \over dX} \log \Big \langle
\prod_{l=1}^N(1 - \xi^* \chi^{(l)}_{(\pi - X/N, \pi)}) e^{\omega_2 \theta_l}
|1 + z_l|^{2\omega_1} | e^{i(\pi - X/N )} - z_l |^{2 \mu}
\Big \rangle_{\rm CUE} 
\end{equation} 
(${d \over dX}$ is multiplied by $X$ for later convenience).

\subsection{Summary of the characterisation of the averages as
Painlev\'e $\sigma$-functions}
Previous studies \cite{TW_1994,HS-99,WFC-00a,BD-2001} have characterised 
$\tilde{E}_N^{\rm J}(s;a,b,\mu=0;\xi)$ 
in terms of the solution of nonlinear equations related to
\PVI. The nonlinear equations obtained have been third order \cite{TW_1994,HS-99}
and second order second degree \cite{HS-99,WFC-00a,BD-2001}. By the work of Cosgrove
and Scoufis \cite{CS-93}, these equations are equivalent, and are in fact
examples of the so called Jimbo-Miwa-Okamoto $\sigma$-form of the
\PVI differential equation,
\begin{equation}\label{4}
h' \Big ( t(1-t) h'' \Big )^2 +
\Big(h' [2h - (2t - 1) h'] + b_1 b_2 b_3 b_4 \Big)^2 =
\prod_{k=1}^4( h' + b_k^2).
\end{equation}
This fact is most explicit in the work of Borodin and 
Deift \cite{BD-2001} who have shown that
\begin{equation}\label{xx1}
\sigma(t) := -t(t-1) {d \over dt} \log \tilde{E}_N^{\rm J}(1-t;b,a,0;\xi) +
b_1 b_2 t + {1 \over 2}( - b_1 b_2 + b_3 b_4)
\end{equation}
with
\begin{equation}\label{xx1a}
b_1 = b_2 = N + {a + b \over 2}, \quad
b_3 = {a + b \over 2}, \quad b_4 = {a - b \over 2}
\end{equation}
satisfies (\ref{4}). The quantity (\ref{1.14a}) in the case
$\mu = 0$ has been similarly characterised. In particular, we know
from \cite{WF_2000} that
\begin{equation}\label{yy1}
\sigma(s) := (1+s^2) {d \over ds} \log \tilde{E}_N^{\rm Cy}(s;(a+N,0),0;\xi)
\end{equation}
satisfies the equation
\begin{equation}\label{yy2}
  (1+s^2)^2(\sigma'')^2 + 4(1+s^2)(\sigma')^3 - 8s \sigma (\sigma')^2
  + 4 \sigma^2(\sigma' - a^2)
  + 8 a^2 s \sigma\sigma' + 4\big[N(N+2a) - a^2 s^2 \big](\sigma')^2
   = 0.
\end{equation}
To relate (\ref{yy2}) to (\ref{4}), change
variables $t \mapsto (is+1)/2$, $h(t) \mapsto {i \over 2} h(s)$ in the latter
so it reads
\begin{equation}\label{iv}
  h' \Big( (1+s^2) h'' \Big)^2 + 4\Big( h'(h-sh') - i b_1 b_2 b_3 b_4 \Big)^2 
  + 4\prod_{k=1}^4(h'+b_k^2) = 0.
\end{equation}
With
\begin{equation}\label{iva}
h = \sigma - a^2 s, \qquad \mathbf{b} = (-a,0,N+a,a),
\end{equation}
(\ref{iv}) reduces to (\ref{yy2}).

In this paper we generalise the \PVI $\sigma$-function characterisations
(\ref{xx1}) and (\ref{yy1}). Consider first the Jacobi case. In
Proposition \ref{p9} we show that
\begin{equation}\label{1.40}
\hat{U}_N^{\rm J}(t;a,b,\mu;\xi) := e_2'[\hat{\mathbf{b}}] t -
{1 \over 2} e_2[\hat{\mathbf{b}}] + t(t-1) {d \over dt}
\log \tilde{E}_N^{\rm J}(t;a,b,\mu;\xi)
\end{equation}
with
\begin{equation}\label{1.34e}
\hat{\mathbf{b}}= \Big ({1 \over 2}(a+b) + N, {1 \over 2}(b-a),
- {1 \over 2}(a+b), - {1 \over 2}(a+b) - N - \mu \Big )
\end{equation}
satisfies (\ref{4}) (the quantities $e_2'[\hat{\mathbf{b}}]$ and
$e_2[\hat{\mathbf{b}}]$ are defined in Proposition \ref{p1}).
This characterisation can be made unique in the cases $\xi =0,$ 1, when
we have the boundary conditions
\begin{align}
  \hat{U}_N^{\rm J}(t;a,b,\mu;0)
  & \mathop{\sim}\limits_{|t| \to \infty} (e_2'[\hat{\mathbf{b}}] + N\mu)t + O(1)
  \label{r.1} \\
  \hat{U}_N^{\rm J}(t;a,b,\mu;1)
  & \mathop{\sim}\limits_{t \to 0}
    - {1 \over 2} e_2[\hat{\mathbf{b}}] - N(a+\mu+N) + 
       \Big( e_2'[\hat{\mathbf{b}}] + N(N+a+\mu) + bN{a+N \over a+\mu+2N} \Big)t
  \label{r.2}
\end{align}
((\ref{bc.4}) and (\ref{bc.3})). Also, when $\mu = 0$ with $\xi$
general we have
\begin{align}
  \hat{U}_N^{\rm J}(t;a,b,0;\xi) 
       -e_2'[\hat{\mathbf{b}}] t + {1 \over 2} e_2[\hat{\mathbf{b}}] \,
  & \mathop{\sim}\limits_{t \to 1^-} 
    \, \xi (t-1) \rho^{\rm J}(t) \nonumber \\
  \rho^{\rm J}(t) \,
  & \mathop{\sim}\limits_{t \to 1^-}
    \, {(1-t)^b } {\Gamma(a+b+N+1) \Gamma(b+N+1) \over
                   \Gamma(N) \Gamma(a+N) \Gamma(b+1) \Gamma(b+2)} 
  \label{1.34r}
\end{align}
where $ \rho^{\rm J}(t)$ denotes the eigenvalue density in the JUE.
Regarding the Cauchy case, in Proposition \ref{p12} we show
\begin{equation}\label{1.42}
  U_N^{\rm Cy}(t;(\eta,0),\mu;\xi) =
  (t^2 + 1) {d \over dt} \log \Big(
  (t^2 + 1)^{e_2'[\mathbf{b}]/2} \tilde{E}_N^{\rm Cy}(t;(\eta,0),\mu;\xi) \Big),
\end{equation}
with
\begin{equation}\label{1.45a}
\mathbf{b} = (N-\eta, 0,\eta, -\mu+\eta-N)
\end{equation}
satisfies (\ref{iv}). 
Proposition \ref{p12} in fact contains the characterisation of the more
general Cauchy average (\ref{1.15a}). It follows from this and (\ref{1.28a})
that
\begin{multline}\label{3.35d'}
  \tilde{E}_N^{\rm cJ}(\phi;(\omega_1,\omega_2),\mu;\xi^*) 
   = {M_N(\omega_1-i\omega_2+\mu/2,\omega_1+i\omega_2+\mu/2) \over 
      M_N(\omega_1,\omega_1)} \\
  \qquad\times
  \exp \Big\{  -{1 \over 2} \int_0^\phi \Big( 
               U_N^{\rm Cy}(\cot {\theta \over 2}; (\omega,\omega_2),\mu;\xi)
               + i(e_2'[\mathbf{b}] - e_2[\mathbf{b}])
               - (e_2'[\mathbf{b}] + N\mu)\cot{\theta \over 2} \Big)
               \Big|_{\omega = N + \omega_1 + \mu/2} d\theta 
       \Big\}
\end{multline}
(eq.~(\ref{3.35d})) where $U_N^{\rm Cy}(t;(\eta_1,\eta_2),\mu;\xi)$
satisfies (\ref{iv}) with
\begin{equation*}
\mathbf{b} = (N - \eta_1, i \eta_2, \eta_1, - \mu + \eta_1 - N).
\end{equation*}

The scaled limit $\phi \mapsto X/N$, $N \to \infty$ of the logarithmic
derivative of (\ref{3.35d'}) is essentially (\ref{vpu}). This gives rise
to the differential equation \cite{Ok_1987b}
\begin{equation}\label{vp}
  (th_V'')^2 - (h_V - th_V' + 2 (h_V')^2)^2 + 4\prod_{k=1}^4(h_V' + v_k) = 0
\end{equation}
(c.f.~(\ref{4})) where
\begin{equation}\label{1.38a}
  v_1 + v_2 + v_3 + v_4 = 0
\end{equation}
satisfied by a particular auxiliary Hamiltonian
$h_V$ in the $\tau$-function of \PV. We show 
(Proposition \ref{pho}) that the scaled average (\ref{vpu}) is such that
\begin{equation*}
  h(t) = u(it;(\omega_1,\omega_2),\mu;\xi) + {i\omega_2 \over 2} t 
         + 2\omega_1 \mu + \omega_2^2/2
\end{equation*}
satisfies (\ref{vp}) with
\begin{equation}
\tilde{v}_1= \mu,\: \: \tilde{v}_2 = - \mu, \: \:
\tilde{v}_3 = \omega, \: \: \tilde{v}_4 = - \bar{\omega}, \: \:
\tilde{v}_j := v_j + {i\omega_2 \over 2}, \: \: \omega:= \omega_1 + i\omega_2.
\end{equation}  

\section{Overview of the Okamoto $\tau$-function theory of \PVI}
\setcounter{equation}{0}
\subsection{The Jimbo-Miwa-Okamoto $\sigma$-form of \PVI}
The sixth Painlev\'e equation \PVI reads
\begin{equation}\label{p6}
  q'' = {1 \over 2} \Big( {1 \over q} + {1 \over q-1} + {1 \over q-t} \Big)(q')^2
         - \Big( {1 \over t} + {1 \over t-1} + {1 \over q-t} \Big)q'
      + {q(q-1)(q-t) \over t^2 (t-1)^2} 
        \Big( \alpha + {\beta t \over q^2} + {\gamma (t-1) \over (q-1)^2}
              + {\delta t (t-1) \over (q-t)^2} \Big).
\end{equation}
It has been known since the work of Malmquist in the early 1920's \cite{Ma_1922}
that (\ref{p6}) can be obtained by eliminating $p$ from a Hamiltonian system
\begin{equation}\label{1.1}
   q' = {\partial H \over \partial p}, \qquad
   p' = - {\partial H \over \partial q}.
\end{equation}
In the notation of \cite{KMNOY_2001} the required Hamiltonian can be written
\begin{equation}\label{2}
  t(t-1) H = q(q-1)(q-t) p^2 
           - [ \alpha_4 (q-1) (q-t) + \alpha_3 q(q-t) + (\alpha_0 - 1) q (q-1) ] p
           + \alpha_2(\alpha_1 + \alpha_2) (q-t),
\end{equation}
where the parameters $\alpha_0, \dots, \alpha_4$ in (\ref{2}) are
inter-related by
\begin{equation}\label{2.1}
\alpha_0 + \alpha_1 + 2 \alpha_2 + \alpha_3 + \alpha_4 = 1
\end{equation}
and are related to the parameters $\alpha, \dots, \delta$ in (\ref{p6}) by
\begin{equation}
\alpha  = {1 \over 2} \alpha_1^2, \quad
\beta  = - {1 \over 2} \alpha_4^2, \quad
\gamma  = {1 \over 2} \alpha_3^2,  \qquad
\delta = {1 \over 2} (1 - \alpha_0^2).
\end{equation}

One sees that the Hamiltonian can be written as an explicit rational function
of the \PVI transcendent and its derivative. This follows from the fact that
with the substitution (\ref{2}), the first of the Hamilton equations is
linear in $p$, so $p$ can be written as a rational function of $q$, $q'$
and $t$. The sought form of $H$ then follows by substituting this
expression for $p$ in (\ref{2}).

After the addition of a certain linear function in $t$, the Hamiltonian
(\ref{2}) has the crucial feature for random matrix applications of
satisfying the second order, second degree differential equation
(\ref{4}).

\begin{prop}\label{p1}
\cite{JM_1981b,Ok_1987a}
Rewrite the parameters $\alpha_0,\dots,\alpha_4$ of (\ref{2}) in favour of 
the parameters
\begin{equation}\label{3.1}
b_1 = {1 \over 2}(\alpha_3 + \alpha_4), \quad
b_2 = {1 \over 2}(\alpha_4 - \alpha_3), \quad
b_3 = {1 \over 2}(\alpha_0 + \alpha_1 - 1), \quad
b_4 = {1 \over 2}(\alpha_0 - \alpha_1 - 1),
\end{equation}
and introduce the auxiliary Hamiltonian $h$ by
\begin{align}\label{3}
  h & = t(t-1) H + e_2'[\mathbf b] t - {1 \over 2} e_2[\mathbf b] \nonumber \\
    & = t(t-1)H + (b_1 b_3 + b_1 b_4 + b_3 b_4) t
                - {1 \over 2} \sum_{1 \le j < k \le 4} b_j b_k ,
\end{align}
where $e_j'[\mathbf b]$ denotes the $j$th degree elementary symmetric
function in $b_1,b_3$ and $b_4$ while $e_j[\mathbf b]$ denotes the
$j$th degree elementary symmetric function in $b_1, \dots, b_4$.
The auxiliary Hamiltonian satisfies the 
Jimbo-Miwa-Okamoto $\sigma$-form of \PVI, (\ref{4}).
\end{prop}

\noindent
Proof. \quad Following \cite{Ok_1987a}, we note from (\ref{2}), (\ref{1.1})
and (\ref{3.1}) that
\begin{equation}\label{4.1}
h' = - q(q-1)p^2 + \{b_1(2q-1) - b_2\}p - b_1^2,
\end{equation}
or equivalently
\begin{equation}\label{4.2}
q(q-1)(h' + b_1^2) = - \Big ( q(q-1)p \Big )^2 +
\Big ( b_1 (2q-1) - b_2 \Big ) q(q-1) p.
\end{equation}
We see from (\ref{4.1}) and (\ref{4.2}) that a differential equation for
$h$ will result if we can express $q$ and $q(q-1)p$ in terms of $h$ and
its derivatives. For this purpose we note from (\ref{3}) and (\ref{4.1})
that
\begin{equation}\label{5}
h - th' = q( - h' + e_2'[\mathbf b] ) - (b_3 + b_4) q(q-1) p -
{1 \over 2} e_2[\mathbf b], 
\end{equation}
while differentiation of this formula and use of the Hamilton equations shows
\begin{equation}\label{6}
t(t-1) h'' = 2q(e_1'[\mathbf b] h' - e_3'[\mathbf b] ) -
2q(q-1) p (h' - b_3 b_4) - e_1[\mathbf b] h' +  e_3[\mathbf b].
\end{equation}
The equations (\ref{5}) and (\ref{6}) are linear in $q$ and $q(q-1)p$.
Solving for these quantities and substituting in (\ref{4.2}) gives
(\ref{4}). \hfill $\square$ 

The $\tau$-function is defined in terms of the Hamiltonian by
\begin{equation}\label{2.12a}
H = {d \over d t} \log \tau(t).
\end{equation}
In terms of $h$, it follows from (\ref{3}) that
\begin{equation}\label{th}
h = t(t-1) {d \over dt} \log \Big (  (t-1)^{e_2'[\mathbf{b}] -
{1 \over 2} e_2[\mathbf{b}]}t^{{1 \over 2} e_2[\mathbf{b}]}
\tau(t) \Big ).
\end{equation}

\subsection{B\"acklund transformations and Toda lattice equation}
B\"acklund transformations 
\begin{equation*}
T(\mathbf b;q,p,t,H) = (\bar{\mathbf b}; \bar{q}, \bar{p}, \bar{t},
\bar{H})
\end{equation*}
are birational canonical transformations of the symplectic form. Thus the
Hamilton equations are satisfied in the variables 
$(\bar{\mathbf b}; \bar{q}, \bar{p}, \bar{t},
\bar{H})$. Because there are particular $T$ possessing the property
\begin{equation}\label{5d}
T H = H \Big |_{\mathbf b \mapsto T  \mathbf b},
\end{equation}
B\"acklund transformations allow an infinite family of solutions of the
\PVI system to be generated from one seed solution.

Okamoto \cite{Ok_1987a} identified the affine Weyl group 
$W_a(D_4^{(1)})$ as being
realised by a set of $t$-invariant B\"acklund transformations of the \PVI
system (if B\"acklund transformations altering $t$ are permitted,
a realisation of the affine $F_4$ reflection group is obtained
\cite{Ok_1987a}). The group $W_a(D_4^{(1)})$ is generated by the operators
$s_0,\dots,s_4$ obeying the algebraic relations 
\begin{equation}\label{5e}
(s_i s_j)^{m_{ij}} = 1, \qquad 0 \le i,j \le 4
\end{equation}
where
\begin{equation}\label{5f}
[ m_{ij} ] = \left [ \begin{array}{ccccc} 1 & 2 & 3 & 2 & 2 \\
2 & 1 & 3 & 2 & 2 \\
3 & 3 & 1 & 3 & 3 \\
2 & 2 & 3 & 1 & 2 \\
2 & 2 & 3 & 2 & 1 \end{array} \right ].
\end{equation}
The entries $m_{ij}$ are related to the Dynkin diagram for the affine
root system $D_4^{(1)}$,

\begin{figure}[h]
\epsfxsize=2cm
\centerline{\epsfbox{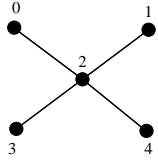}}
\caption{\label{Fig1} $D_4^{(1)}$ Dynkin diagram.}\label{PVI.fig}
\end{figure}

\noindent
Thus for $i \ne j$, ($0 \le i < j \le 4$),
$m_{ij} = 2$ if the nodes $i$ and $j$ are not connected,
while $m_{ij} = 3$ if the nodes are connected. An equivalent way to specify
the algebra (\ref{5e}) is via the relations
\begin{equation}\label{5e'}
s_i^2 = 1 \: \: (0 \le i,j \le 4) , \quad 
s_i s_j = s_j s_i \: \: (i,j \ne 2), \quad
(s_i s_2)^3 = (s_2 s_i)^3 = 1 \: \: (i \ne 2).
\end{equation}

The operators $s_0,\dots,s_4$ are associated with affine vectors
$\alpha_0, \dots, \alpha_4$ 
$(\alpha_0 + \alpha_1 +2 \alpha_2 + \alpha_3 + \alpha_4 = 1$, 
with the $\alpha$'s regarded as coordinates)
in a four dimensional vector space such that $s_i$ corresponds to a
reflection in the subspace perpendicular to $\alpha_i$, and thus
$s_i \alpha_i = - \alpha_i$. The action of $s_i$ on the other
affine vectors is given by
\begin{equation}\label{sA}
s_i(\alpha_j) = \alpha_j - \alpha_i a_{ij}
\end{equation}
where $A = (a_{ij})$ is the Cartan matrix
\begin{equation}\label{A}
A = \left [ \begin{array}{ccccc}
               2 & 0 & -1 & 0 & 0 \\
               0 & 2 & -1 & 0 & 0 \\
               -1 & -1 & 2 & -1 & -1 \\
               0 & 0 & -1 & 2 & 0 \\
               0 & 0 & -1 & 0 & 2
            \end{array} \right ]
\end{equation}
(the off diagonal elements in (\ref{A}) are obtained from those in
(\ref{5f}) by replacing all 2's in the latter by 0's, and
all 3's by $-1$'s). As first identified by Okamoto
\cite{Ok_1987a}, the operators $s_i$ are $t$-invariant
B\"acklund transformations for the \PVI system with action on the
parameters specified by (\ref{sA}). 

It remains to specify the action of the $s_i$ on $p$ and $q$. For this
purpose the most systematic way to proceed is to make use of
recent work of Noumi and Yamada \cite{NY-2000}
(see also their subsequent work \cite{NY_2002}), 
who give a symmetric formulation
of the B\"acklund transformations for Painlev\'e type systems. In
the general formalism of \cite{NY-2000}, the action (\ref{sA}) has as its
counterpart the action
\begin{equation}\label{BA}
s_i(f_j) = f_j + {\alpha_i \over f_i} u_{ij}
\end{equation}
where $U = [u_{ij}] $ is the orientation matrix defined by
\begin{equation}\label{U}
U  = \left [ \begin{array}{ccccc}
0 & 0 & 1 & 0 & 0 \\
0 & 0 & 1 & 0 & 0 \\
-1 & -1 & 0 & -1 & -1 \\
0 & 0 & 1 & 0 & 0 \\
0 & 0 & 1 & 0& 0
\end{array} \right ].
\end{equation}
The key points are that (\ref{5e'}) is realised by (\ref{BA}) with the
$f_j$ specified in terms of $p$, $q$, $t$ by
\begin{equation}\label{BA1}
f_0 = q-t, \: \: f_1 = q - \infty, \: \: f_2 = -p, \: \:
f_3 = q-1, \: \: f_4 = q ,
\end{equation}
(the quantities subtracted from $q$ are the location of the fixed singularities
of \PVI), and that the transformations (\ref{sA}) and (\ref{BA})
together are then B\"acklund transformations for \PVI. Following
\cite{KMNOY_2001}, the action of the $s_i$ on the $\alpha_j$, $p$ and $q$
is summarised in Table \ref{t1}.

\begin{table}
\begin{center}
\begin{tabular}{|c||c|c|c|c|c|c|c|}\hline
 & $\alpha_0$ & $\alpha_1$ & $\alpha_2$ & $\alpha_3$ & $\alpha_4$ & $p$ & $q$
 \\ \hline
$s_0$ 	& $-\alpha_0$ & $\alpha_1 $ & $\alpha_2 + \alpha_0$ & $\alpha_3$
	& $\alpha_4$ & $\displaystyle p - {\alpha_0 \over q - t}$ & $q$ \\

$s_1$ 	& $\alpha_0$ &  $-\alpha_1 $ & $\alpha_2 + \alpha_1$ 
	& $\alpha_3$ & $\alpha_4$ & $p$ & $q$ \\

$s_2$ 	& $\alpha_0+\alpha_2 $ &  $\alpha_1 + \alpha_2$ & $-\alpha_2$ 
	& $\alpha_3+\alpha_2$ & $\alpha_4+\alpha_2$ & $p$ 
	& $\displaystyle q + {\alpha_2 \over p}$ \\

$s_3$ 	& $\alpha_0$ & $\alpha_1$ & $\alpha_2 + \alpha_3$ & $-\alpha_3$
	& $\alpha_4$ & $\displaystyle p - {\alpha_3 \over q - 1}$ & $q$ \\

$s_4$ 	& $\alpha_0$ & $\alpha_1$ & $\alpha_2 + \alpha_4$ & $\alpha_3$
	& $-\alpha_4$ & $\displaystyle p - {\alpha_4 \over q }$ & $q$ \\

$r_1$  	& $\alpha_1$ & $\alpha_0$ & $\alpha_2$ & $\alpha_4$
	& $\alpha_3$ & $\displaystyle -{p(q-t)^2 + \alpha_2(q-t) \over t(t-1)}$ 
	& $\displaystyle \Big ( {q-1 \over q-t} \Big ) t$ \\

$r_3$  	& $\alpha_3$ & $\alpha_4$ & $\alpha_2$ & $\alpha_0$
	& $\alpha_1$ & $\displaystyle -{q \over t}[qp + \alpha_2] $ 
	& $\displaystyle {t \over q} $ \\


\hline
\end{tabular}
\end{center}
\caption{\label{t1}  B\"acklund transformations  for the \PVI
Hamiltonian (\ref{2}).}
\end{table} 

The symmetric formalism of \cite{NY-2000,NY-98a} naturally extends the
B\"acklund transformations from a realisation of $W_a(D_4^{(1)})$
to a realisation of $W_a(D_4^{(1)} \rtimes \Omega)$, where $\Omega$
denotes particular diagram automorphisms of $D_4^{(1)}$. 
The latter are operators $r_1$ and $r_3$ (another natural diagram automorphism is
$ r_4 \boldsymbol{\alpha} = (\alpha_4,\alpha_3, \alpha_2, \alpha_1, \alpha_0) $ 
and is related to $r_1$ and $r_3$ by $ r_4=r_1 r_3 $)
with the actions
\begin{equation*}
  r_1 \boldsymbol{\alpha} = (\alpha_1,\alpha_0, \alpha_2, \alpha_4, \alpha_3),
  \qquad
  r_3 \boldsymbol{\alpha} = (\alpha_3,\alpha_4, \alpha_2, \alpha_0, \alpha_1)
\end{equation*}
of interchanging outer pairs of vertices in the diagram of 
Figure \ref{PVI.fig}. These operators obey the algebraic relations
\begin{equation*}
  \begin{array}{cccc}
   r_1^2 = r_3^2 = 1, & r_1 s_2 = s_2 r_1, & r_3 s_2 = s_2 r_3
 & r_i r_j = r_j r_i, \, i\neq j, \\
   s_1 = r_1 s_0 r_1, & s_4 = r_1 s_3 r_1, & s_3 = r_3 s_0 r_3,
 & s_4 = r_3 s_1 r_3.
   \end{array}
\end{equation*}
Their action on the $p$ and $q$ is given in Table \ref{t1}.
Making use of Table \ref{t1} the action of the fundamental operators
on the Hamiltonian 
\begin{equation}\label{Ht}
t(t-1)H =: K
\end{equation}
as specified by (\ref{2})
is found to be
\begin{equation}\label{rb}
\begin{split}
  s_0 K & = K - \alpha_0 {t(t-1) \over q - t}
              + \alpha_0(\alpha_3-1)t + \alpha_0(\alpha_4-1)(t-1) \nonumber \\
  s_1 K & = K \nonumber \\
  s_2 K & = K + \alpha_2(1+\alpha_1-\alpha_0)t
              - \alpha_2(\alpha_1+\alpha_2+\alpha_3) \nonumber \\
  s_3 K & = K - \alpha_3(1-\alpha_0)t \nonumber \\
  s_4 K & = K - \alpha_4(1-\alpha_0)(t-1) \nonumber \\
  r_1 K & = K - q(q-1)p - \alpha_2 q + \alpha_2(\alpha_1 - \alpha_0)t
              + \alpha_2(\alpha_0+\alpha_2+\alpha_4) \nonumber \\
  r_3 K & = K + (1-t) qp + \alpha_2(\alpha_0+\alpha_2+\alpha_4)(1-t)
\end{split}
\end{equation}
(the first five of these equations can be found in \cite{Wa-98a}).

Consider the composite operator
\begin{equation}\label{T}
  T_3 := r_1 s_0 s_1 s_2 s_3 s_4 s_2,
\end{equation}
which from Table \ref{t1} has the action on the $\alpha$ parameters
\begin{equation}
   T_3\boldsymbol{\alpha} 
     = (\alpha_0+1,\alpha_1+1, \alpha_2-1, \alpha_3, \alpha_4)
\end{equation}
or equivalently, using (\ref{3.1}), the action on the $b$ parameters
\begin{equation}\label{2.28}
   T_3\mathbf{b} = (b_1, b_2, b_3+1, b_4)
\end{equation}
(in \cite{Ok_1987a} this was referred to as the parallel transformation
$\ell_3$). With $K$ specified by (\ref{Ht}) we see from
(\ref{rb}) and Table \ref{t1} that
\begin{equation}\label{2.35}
   T_3 K =  K \Big|_{\boldsymbol{\alpha} \mapsto T_3 \boldsymbol{\alpha}}
         = K - q(q-1)p - (\alpha_1 + \alpha_2)(q-t),
\end{equation}
which was derived
in \cite{Ok_1987a} in a different way. There are only three other fundamental
shift operators that share the property (\ref{2.35}). They have the actions
\begin{align*}
   T_{10}K 
  & := r_1 s_1 s_2 s_3 s_4 s_2 s_ 1 K 
    = K \Big|_{\boldsymbol{\alpha} \mapsto 
               (\alpha_0+1,\alpha_1-1, \alpha_2, \alpha_3, \alpha_4)} 
    = K - q(q-1)p - \alpha_2(q-t),
  \cr 
   T_{30}K 
  & := r_3 s_3 s_2 s_1 s_4 s_2 s_3 K 
    = K \Big|_{\boldsymbol{\alpha} \mapsto 
               (\alpha_0+1,\alpha_1, \alpha_2, \alpha_3-1, \alpha_4)} 
    = K -(t-1)qp,
  \cr 
   T_{40}K 
  & := r_4 s_4 s_2 s_1 s_3 s_2 s_4 K 
    = K \Big|_{\boldsymbol{\alpha} \mapsto 
               (\alpha_0+1,\alpha_1, \alpha_2, \alpha_3, \alpha_4-1)} 
    = K -t(q-1)p,
\end{align*}
although we will not develop the theory of these cases in this work.

The result (\ref{2.35}) motivates introducing the sequence of Hamiltonians
\begin{equation*}
T_3^n H = H \Big |_{\boldsymbol{\alpha} \mapsto
(\alpha_0+n,\alpha_1+n, \alpha_2-n, \alpha_3, \alpha_4)},
\end{equation*}
and the corresponding sequence of $\tau$-functions specified by
\begin{equation}\label{2.35a}
T_3^n H = {d \over dt} \log \tau_3[n], \qquad
\tau_3[n] = \tau_3[n](t) = \tau_3(t;b_1,b_2,b_3+n,b_4).
\end{equation}
Okamoto \cite{Ok_1987a} proved that $\tau_3[n]$ satisfies the Toda lattice
equation. We will give the derivation of this result using the strategy
of Kajiwara et al.~\cite{KMNOY_2001}.

\begin{prop}
The $\tau$-function sequence (\ref{2.35a}) satisfies the Toda lattice equation
\begin{equation}\label{fi}
\delta^2 \log \bar{\tau}_3[n] =
{ \bar{\tau}_3[n-1]  \bar{\tau}_3[n+1] \over  \bar{\tau}_3^2[n] }, \qquad
\delta = t(t-1) {d \over dt}
\end{equation}
where
\begin{equation}\label{fip}
\bar{\tau}_3[n] = \Big (t(t-1) \Big )^{(n+b_1+b_3)(n+b_3+b_4)/2}
{\tau}_3[n].
\end{equation}
\end{prop}

\noindent
Proof. \quad From the definitions
\begin{align}\label{Te}
  \delta \log {\tau_3[n-1] \tau_3[n+1] \over \tau_3^2[n]}
  & = \Big( T_3 K[n] - K \Big) - \Big(  K[n] - T_3^{-1} K \Big )
  \nonumber \\
  & =  -q(q-1)p - (\alpha_1 + \alpha_2)(q - t)
  \nonumber \\
  & \phantom{=}\qquad
       + T_3^{-1} \Big( q(q-1)p + (\alpha_1+\alpha_2)(q - t) \Big)
         \Big|_{\boldsymbol{\alpha} \mapsto 
                (\alpha_0+n, \alpha_1+n, \alpha_2-n,\alpha_3,\alpha_4)},
\end{align}
where the second equality follows from (\ref{2.35}). Now $T_3$ is specified
by (\ref{T}), and each of the elementary operators is an involution so
\begin{equation}\label{T1}
T_3^{-1} = s_2 s_4 s_3 s_2 s_1 s_0 r_1 = s_2 s_3 s_4 s_2 s_0 s_1 r_1
\end{equation}
where the second equality follows from the commutation relations
$s_3 s_4 = s_4 s_3$, $s_0 s_1 = s_1 s_0$ implied by (\ref{5e'}). Using
(\ref{T1}) and Table \ref{t1} we compute that
\begin{multline}
  - T_3^{-1} \Big( q(q-1)p + (\alpha_1+\alpha_2)(q - t) \Big) \\
   = q(q-1)p + \alpha_2(q-1) + (1-\alpha_0)q + {\alpha_2(1-\alpha_0) \over p}
     + (\alpha_1 +\alpha_2)t -(\alpha_1 +\alpha_2 +\alpha_3)
   \\ \quad
   + ( \alpha_2 +\alpha_3 +\alpha_4)(1 -\alpha_0) {1 \over p}
   { q(q-1) p +\alpha_2 q +\alpha_2 (q-1) +\alpha_2^2/p \over
     q(q-1) p -\alpha_3 q -\alpha_4 (q-1) -\alpha_2(\alpha_2 +\alpha_3 +\alpha_4)/p }.
\label{Tef}
\end{multline}
Substituting this in (\ref{Te}), we can verify that the resulting
expression can be written as
\begin{equation}\label{Teg}
\delta \log \Big [ q(q-1) p^2 - [ \alpha_3 q + \alpha_4 (q-1) ] p -
\alpha_2 ( \alpha_2 + \alpha_3 + \alpha_4) \Big ].
\end{equation}
To do this we first compute the derivative in terms of $p$ and $q$ using
the Hamilton equations, and then check that the expression so obtained is
indeed equal to the rational function in $p$ and $q$ obtained by substituting
(\ref{Tef}) in (\ref{Te}) using computer algebra. But according to
(\ref{2}), (\ref{Teg}) is equal to
\begin{equation}\label{Teg1}
\delta \log\Big[ {d \over dt}K + \alpha_2(1-\alpha_0) \Big].
\end{equation}
Substituting (\ref{2.12a}) in (\ref{Teg1}) and equating the resulting
expression with the left hand side of (\ref{Te}), we see that
\begin{equation*}
  A {\tau_3[n-1] \tau_3[n+1] \over \tau_3^2[n]} =
  {d \over dt}\delta \log\tau_3[n] + (\alpha_2 -n)(1-\alpha_0 -n),
\end{equation*}
where $A \ne 0$ is arbitrary. Recalling (\ref{3.1}) to replace
the $\alpha$'s by the $b$'s and choosing $A=1$ gives (\ref{fi}).
\hfill $\square$

\subsection{Classical solutions}
An identity of Sylvester (see \cite{Mu_1960}) gives that if
\begin{equation}\label{fi3}
\bar{\tau}_3[0] = 1,
\end{equation}
then the general solution of (\ref{fi}) is given by
\begin{equation}\label{fi5}
\bar{\tau}_3[n] = \det \Big [ \delta^{j+k} \bar{\tau}_3[1] \Big ]_{j,k=0,1,
\dots, n-1}.
\end{equation}
As noted by Okamoto \cite{Ok_1987a} and Watanabe \cite{Wa-98a}, 
the solution (\ref{fi3}) is permitted by restricting the parameters so that
\begin{equation}\label{fi6}
b_1 + b_3 = 0 \qquad {\rm or \, equivalently} \qquad
\alpha_2 = 0
\end{equation}
(this corresponds to a chamber wall in the affine $D^{(1)}_4$ root system).
Thus with $\alpha_2 = 0$ we see that (\ref{2}) permits the solution
\begin{equation}\label{fi7}
p=0, \qquad H = 0
\end{equation}
and thus $\tau_3[0] = \bar{\tau}_3[0] = 1$. Furthermore, with this initial
condition ${\tau}_3[1]$ is given by a solution of the
Gauss hypergeometric equation. 

\begin{prop} \label{pr3} \cite{Ok_1987a} Let the parameters 
$\boldsymbol{\alpha}$ in (\ref{2}) be
initially restricted by (\ref{fi6}), then apply the operator $T_3$ so that
\begin{equation}\label{2.46a}
T_3 H[0] = H[1] = {d \over dt} \log \tau_3[1](t).
\end{equation}
The function $\tau_3[1](t)$ satisfies the Gauss hypergeometric equation
\begin{equation}\label{ghp}
t(1-t)  \tau_3''[1](t) + \Big ( c - (a+b+1) t \Big )
\tau_3'[1](t) - ab \tau_3[1](t) = 0
\end{equation}
where
\begin{equation}\label{abc}
a=-\alpha_1 = b_1 + b_4, \quad
b = \alpha_0 = 1 + b_3 + b_4, \quad 
c = \alpha_0 + \alpha_4 = 1 + b_2 + b_4.
\end{equation}
\end{prop}

\noindent
Proof. \quad It follows from (\ref{2.35}), (\ref{Ht}) and (\ref{fi6}),
(\ref{fi7}) that
\begin{equation}\label{fu2}
T_3 K[0] = - \alpha_1 (q - t)
\end{equation}
where $q$ is the solution of the Hamilton equation
\begin{equation}\label{fu3}
  \delta(q-t) 
  = {\partial K \over \partial p}\Big|_{\alpha_2=0 \atop p=0} - t(t-1)
  = \alpha_1 (q-t)^2
     + \left[(\alpha_1+\alpha_4)t + (\alpha_1+\alpha_3)(t-1)\right](q-t)
     - \alpha_0 t(t-1).
\end{equation}
Rewriting the left hand side of (\ref{fu2}) in terms of
$\tau_3'[1](t)/\tau_3[1](t)$ 
according to (\ref{2.46a}), then substituting the resulting
expression in (\ref{fu3}) and simplifying gives (\ref{ghp}).
\hfill $\square$

Our interest is in particular integral solutions of (\ref{ghp}), which
being a second order linear equation has in general two linearly
independent solutions. Consider first the solution
analytic at the origin --- the Gauss hypergeometric function
${}_2 F_1$ --- written as its Euler integral representation
\begin{equation}\label{1m}
{}_2 F_1(a,b;c;t) = {\Gamma(c) \over \Gamma(b) \Gamma(c-b)}
\int_0^1 u^{b-1} (1-u)^{c-b-1} (1 - ut)^{-a} \, du.
\end{equation}
We then have
\begin{equation}\label{2m}
\tau_3[1](t) = \tau_3(t;b_1,b_2,-b_1+1,b_4) =
{}_2 F_1(b_1 + b_4, 1 - b_1 + b_4, 1+b_2 + b_4; t).
\end{equation}
Integral solutions of (\ref{ghp}) which in general are not analytic
at the origin are given by
\begin{equation}\label{4m}
  f(a,b,c;t) \propto \int_p^q u^{a-c}(1-u)^{c-b-1}(t-u)^{-a} \, du
\end{equation}
where $p$ and $q$ are any of $0,1,t,\nu\, \infty$ ($\nu=1$)
such that the integrand vanishes \cite{Ho-71}. 
Forming from this a linear combination with $(p,q) = (0,1)$ and
$(p,q) = (t,1)$ we deduce from this that
$\tau_3[1](t) = f^{\rm J}(a,b,c;t)$ satisfies (\ref{ghp}), where
\begin{equation}\label{cc}
  f^{\rm J}(a,b,c;t) \propto
  \Big(\int_0^1 - \xi \int_t^1 \Big)u^{a-c}(1-u)^{c-b-1}(t-u)^{-a}\, du.
\end{equation}
Another case of interest is the linear combination of
(\ref{4m}) with $(p,q) = (-\infty, \infty)$ and 
$(p,q) = (t,\infty)$. After suitably deforming these contours in the
complex plane we deduce that
$\tau_3[1](t) = f^{\rm Cy}(a,b,c;t)$
satisfies (\ref{ghp}), where
\begin{equation}\label{5m}
  f^{\rm Cy}(a,b,c; {1 + it \over 2}) \propto 
  \Big( \int_{-\infty}^\infty - \xi \int_t^\infty \Big)
      (1+iu)^{a-c} (1-iu)^{c-b-1} (t-u)^{-a} \, du.
\end{equation}

Let $F(a,b,c;t)$ be any particular solution of (\ref{ghp}). Then recalling
(\ref{fi6}), (\ref{abc}) and (\ref{fip}) we see from (\ref{fi5})
that
\begin{equation*}
\bar{\tau}_3[n] = \det[ \delta^{j+k} t^{b/2} (t-1)^{b/2} F(a,b,c;t)
                      ]_{j,k=0,\dots,n-1} .
\end{equation*}
A useful check on further working is to note that if $\{ \bar{\tau}_3[n](t)
\}$ satisfies the Toda lattice equation (\ref{fi}) with
$\bar{\tau}_3[0] = 1$, then $\{ t^{\gamma n / 2}
(t - 1)^{-\gamma n /2} \bar{\tau}[n](t) \}$, $\gamma$ arbitrary, is also
a solution which is given by the determinant formula (\ref{fi5}). Thus
\begin{equation}\label{if}
  \bar{\tau}_3[n] = t^{-\gamma n /2} (t - 1)^{\gamma n /2}
  \det [ \delta^{j+k} t^{A} (t-1)^{B} F(a,b,c;t) ]_{j,k=0,\dots,n-1}
\end{equation}
where
\begin{equation}\label{if1}
  A + B = b, \qquad A - B = \gamma
\end{equation}
and the right hand side must be independent of $\gamma$. 
Our task is to substitute the particular solutions (\ref{1m}), (\ref{cc})
and (\ref{5m}) in the $ n\times n $ determinant (\ref{if}) and show that these 
can be reduced to three distinct $ n$-dimensional multiple integrals.

Firstly we consider (\ref{if}) by taking the solution (\ref{1m}).
\begin{prop}\label{p4}
Let $F(a,b,c;t)$ be given by (\ref{1m}), and let $\delta$ be given
as in (\ref{fi}). Then assuming the constraint (\ref{if1}),
\begin{equation}\label{oD}
\bar{\tau}_3[n] 
\propto t^{bn/2} (t-1)^{bn/2+n(n-1)/2}
\det[ {}_2F_1(a-j,b+k;c;t) ]_{j,k=0,\dots,n-1}.
\end{equation}
\end{prop}

\noindent
Proof. \quad A fundamental differential-difference relation for the
Gauss hypergeometric function is
\begin{equation*}
t {d \over dt} {}_2F_1(a,b;c;t) = 
b \Big ( {}_2F_1(a,b+1;c;t) - {}_2F_1(a,b;c;t) \Big ).
\end{equation*}
It follows from this that
\begin{multline*}
  \delta \Big( t^A(t-1)^B {}_2F_1(a,b;c;t) \Big) \\
   = \Big( b-A + t(A+B-b) \Big) t^A (t-1)^B {}_2F_1(a,b;c;t)
      + b t^A (t-1)^{B+1} {}_2F_1(a,b+1;c;t).
\end{multline*}
In the special case of the constraint (\ref{if1}), this reads
\begin{equation}\label{D2}
\delta \Big ( t^A(t-1)^B {}_2F_1(a,b;c;t) \Big ) =
B t^A (t-1)^B {}_2F_1(a,b;c;t) + b t^A (t-1)^{B+1}
{}_2F_1(a,b+1;c;t).
\end{equation}
Another fundamental differential-difference relation for the
Gauss hypergeometric function is
\begin{equation*}
t(1-t) {d \over dt} {}_2F_1(a,b;c;t) = (c-a)  {}_2F_1(a-1,b;c;t) + 
(a-c+bt) {}_2F_1(a,b;c;t) .
\end{equation*} 
This implies
\begin{multline*}
  \delta \Big( t^A(t-1)^B {}_2F_1(a,b;c;t) \Big) \\
   = \Big( c-a-A + t(A+B-b) \Big) t^A (t-1)^B {}_2F_1(a,b;c;t)
      + (a-c) t^A (t-1)^{B} {}_2F_1(a-1,b;c;t),
\end{multline*}
which in the case of the constraint (\ref{if1}) reads
\begin{equation}\label{D4}
\delta \Big ( t^A(t-1)^B {}_2F_1(a,b;c;t) \Big ) =
(c-a-A) t^A (t-1)^B {}_2F_1(a,b;c;t) + (a-c) t^A (t-1)^{B}
 {}_2F_1(a-1,b;c;t).
\end{equation}

The identity (\ref{D2}) can be used to eliminate the operation
$\delta^k$ from the right hand side of (\ref{if}). Thus substituting
(\ref{D2}) in column $k$, and subtracting $B$ times column $k-1$
$(k=n-1,n-2,\dots,1$ in that order) shows
\begin{multline*}
  \det [ \delta^{j+k} t^A (t-1)^B {}_2F_1(a,b;c;t) ]_{j,k=0,\dots,n-1} \\
   = b^{n-1} \det \left[ \delta^j t^A (t-1)^B {}_2F_1(a,b;c;t) \;\;
                   \delta^{j+k-1} t^A (t-1)^{B+1} {}_2F_1(a,b+1;c;t)
                  \right]_{j=0,\dots,n-1 \atop k=1,\dots,n-1}.
\end{multline*}
Repeating this procedure on column $k$ ($k=n-1,n-2,\dots,k'$), for each of
$k' = 2,3,\dots,n-2$ in that order shows
\begin{equation*}
  \det \left[ \delta^{j+k} t^A (t-1)^B {}_2F_1(a,b;c;t) \right]_{j,k=0,\dots,n-1}
 = \prod_{l=1}^{n-1} (b)_l
  \det \left[ \delta^j t^A (t-1)^{B+k} {}_2F_1(a,b+k;c;t) \right]_{j,k=0,\dots,n-1}.
\end{equation*}
To eliminate $\delta^j$ from this expression we substitute (\ref{D4})
in row $j$ and subtract $(c-a-A)$ times row $j-1$
$(j=n-1,n-2,\dots, 1$ in that order). This shows
\begin{multline*}
  \det\left[ \delta^{j} t^A (t-1)^{B+k} {}_2F_1(a,b+k;c;t) 
      \right]_{j,k=0,\dots,n-1} \\
   = (a-c)^{n-1} \prod_{l=1}^{n-1} (b)_l \det \left[
           \begin{array}{c} 
             t^A (t-1)^{B+k} {}_2F_1(a,b+k;c;t) \\
             \delta^{j-1} t^A (t-1)^{B+k} {}_2F_1(a-1,b+k;c;t)
           \end{array} \right]_{j=1,\dots, n-1 \atop k=0,\dots,n-1}.
\end{multline*}
Repeating this procedure on row $j$, $j=n-1,n-2,\dots,j'$ for each of
$j'=3,4,\dots,n-2$ in that order gives
\begin{multline*}
  \det \left[ \delta^{j+k} t^A (t-1)^B {}_2F_1(a,b;c;t) \right]_{j,k=0,\dots,n-1} \\
   = (-1)^{n(n-1)/2} \prod_{j=1}^{n-1} (b)_j (c-a)_j
     \det \left[  t^A (t-1)^{B+k} {}_2F_1(a-j,b+k;c;t) \right]_{j,k=0,\dots,n-1}.
\end{multline*}
Removing the factor $t^A (t-1)^{B+k}$ from each column and using (\ref{if1})
gives (\ref{oD}). 
\hfill $\square$

\begin{prop}
It follows from (\ref{oD}) that
\begin{multline}\label{EIa}
  {\tau}_3[n](t) := \tau_3(t;b_1,b_2,b_3+n,b_4) \Big|_{b_1 + b_3 = 0} \\
   \propto \int_0^1 du_1 \cdots \int_0^1 du_n \,
   \prod_{i=1}^n u_i^{b_3 + b_4} (1 - u_i)^{b_2 - b_3 - n}(1 - tu_i)^{b_3 - b_4}
   \prod_{1 \le j < k \le n} (u_k - u_j)^2.
\end{multline}
\end{prop}

\noindent
Proof. \quad 
From (\ref{fip}) we have
\begin{equation*}
\tau_3[n](t) = (t(t-1))^{-n(n-1+b)/2} \bar{\tau}_3[n](t)
\end{equation*}
where use has been made of the equation $b_1+b_3=0$ from (\ref{fi6})
and $b=1-b_1+b_4$ from (\ref{abc}). Substituting (\ref{oD}) then shows
\begin{equation}\label{ob}
{\tau}_3[n](t) \propto t^{-n(n-1)/2}
\det [ {}_2F_1(a-j,b+k;c;t)
]_{j,k=0,\dots,n-1}.
\end{equation}
Substituting the integral representation (\ref{1m})
and recalling (\ref{abc}) shows
\begin{align}
  {\tau}_3[n](t)
   & \propto t^{-n(n-1)/2} \det \Big[
     \int_0^1 du \, u^{b_3 +b_4 +k}(1-u)^{b_2 -b_3 -k-1}(1-tu)^{j-b_1 -b_4}
                                \Big]_{j,k=0,\dots,n-1}
   \nonumber \\
   & = t^{-n(n-1)/2}
     \int_0^1 du_1 \cdots \int_0^1 du_n \, \prod_{i=1}^n u_i^{b_3+b_4}
      (1-u_i)^{b_2-b_3-n}(1-tu_i)^{-b_1 -b_4} 
   \nonumber \\
   & \phantom{=}\times \det \Big[ u_{j+1}^k(1-u_{j+1})^{n-1-k}(1-tu_{j+1})^j
                   \Big ]_{j,k=0,\dots,n-1}. \label{EIo}
\end{align}
The integrand can be symmetrised without changing the value of the integral
provided we divide by $n!$. The function of the $u_i$'s outside the
determinant is symmetric in the $u_i$'s, so symmetrising the integrand is
equivalent to symmetrising the determinant. We have
\begin{align}
  & {\rm Sym} \, \det \Big[ 
       u_{j+1}^k (1 - u_{j+1})^{n-1-k}(1-tu_{j+1})^j \Big]_{j,k=0,\dots,n-1}
  \nonumber \\ \qquad
  & = {\rm Sym} \, \prod_{j=0}^{n-1} (1 - t u_{j+1})^j
       \det \Big[ u_{j+1}^k (1 - u_{j+1})^{n-1-k} \Big]_{j,k=0,\dots,n-1}
  \nonumber \\
  & = \Big( {\rm Asym} \, \prod_{j=0}^{n-1} (1-t u_{j+1})^j \Big)
      \det \Big[ u_{j+1}^k (1 - u_{j+1})^{n-1-k} \Big]_{j,k=0,\dots,n-1}
  \nonumber \\
  & = \det \Big[ (1-t u_{j+1})^k \Big]_{j,k=0,\dots,n-1}
      \det \Big[ u_{j+1}^k (1-u_{j+1})^{n-1-k} \Big]_{j,k=0,\dots,n-1}
  \nonumber \\
  & = \prod_{j=0}^{n-1}(1-u_{j+1})^{n-1}
      \det \Big[ (1-t u_{j+1})^k \Big ]_{j,k=0,\dots,n-1}
      \det \Big[ \Big( {u_{j+1} \over 1-u_{j+1}} \Big)^k \Big]_{j,k=0,\dots,n-1}
  \label{EI1}
\end{align}
But the Vandermonde determinant identity gives that for any $a_1, a_2,\dots, a_{n}$,
\begin{equation*}
\det [ a_{j+1}^k ]_{j,k=0,\dots, n - 1} = \prod_{1 \le j < k \le n}
(a_k - a_j).
\end{equation*}
This allows the determinants in (\ref{EI1}) to be written as products.
Substituting the results for the determinant in (\ref{EIo}) gives
(\ref{EIa}). \hfill $\square$

Next we will show that by choosing $ F $ in (\ref{if}) to equal the solution 
(\ref{cc}), a generalised multiple integral representation for 
$ \bar{\tau}_3[n] $ can be obtained.
\begin{prop}
Let $F(a,b,c;t)$ in (\ref{if}) be given by
\begin{equation}\label{3.03a}
F(a,b,c;t) = \Big ( \int_0^1 - \xi \int_t^1 \Big )
u^{a-c} (1 - u)^{c-b-1} (t - u)^{-a} \, du.
\end{equation}
Then
\begin{equation}\label{3.14a}
\bar{\tau}_3[n](t) \propto t^{bn/2}(t-1)^{bn/2+n(n-1)/2}
\det[ F(a-j,b+k,c;t) ]_{j,k=0,\dots,n-1}.
\end{equation}
\end{prop}

\noindent
Proof. \quad The proof of Proposition \ref{p4} shows that the sufficient
conditions for reducing (\ref{if}) to (\ref{3.14a}) is that $ F $ satisfies 
the differential-difference relations
\begin{align}
  t{d \over dt} F(a,b,c;t) & = C_0 F(a,b+1,c;t) - b F(a,b,c;t)
  \label{3.3b} \\
  t(1-t) {d \over dt} F(a,b,c;t) & = C_1 F(a-1,b,c;t) + (C_2+bt) F(a,b,c;t),
  \label{3.3c}
\end{align}
independent of the explicit form of $C_0, C_1, C_2$. To derive the form
(\ref{3.3b})
we first change variables $u \mapsto tu$ in (\ref{3.03a}) so it reads
\begin{equation*}
F(a,b,c;t) = t^{-c+1} \Big ( \int_0^{1/t} - \xi \int_1^{1/t}
\Big ) u^{a-c} (1 - tu)^{c-b-1} (1 - u)^{-a} \, du.
\end{equation*}
Differentiating this shows
\begin{multline*}
  t{d \over dt} F(a,b,c;t) = (-c+1) F(a,b,c;t) \\
  \qquad + t^{-c+1} (c-b-1) \Big( \int_0^{1/t} - \xi \int_1^{1/t} \Big) 
           u^{a-c} \Big( {-tu \over 1-tu} \Big)(1-tu)^{c-b-1}(1-u)^{-a} \, du.
\end{multline*}
Writing
\begin{equation}\label{3.u}
- {tu \over 1 - tu} = 1 - {1 \over 1 - tu}
\end{equation}
we see from this that
\begin{equation}\label{3.3d}
t {d \over dt} F(a,b,c;t) = - b F(a,b,c;t) - (c-b-1) F(a,b+1,c;t),
\end{equation}
thus establishing (\ref{3.3b}).

To establish the structure (\ref{3.3c}), we first multiply both sides of
(\ref{3.3d}) by $(1-t)$ to get
\begin{equation*}
  t(1-t){d \over dt} F(a,b,c;t) 
  = btF(a,b,c;t) - bF(a,b,c;t) - (c-b-1)F(a,b+1,c;t) + (c-b-1)tF(a,b+1,c;t).
\end{equation*}
But
\begin{equation}\label{3.3e}
t F(a,b+1,c;t) = t^{-c+1}
\Big ( \int_0^{1/t} - \xi \int_1^{1/t}
\Big ) u^{a-c-1} \Big ( {tu \over 1 - tu} \Big )
(1 - tu)^{c-b-1} (1 - u)^{-a} \, du.
\end{equation}
Using (\ref{3.u}) and the equally simple manipulation
\begin{equation*}
(1 - u)^{-a+1} = (1 - u)^{-a} - u (1 - u)^{-a}
\end{equation*}
shows that the right hand side of (\ref{3.3e}) is equal to
\begin{equation*}
- \Big ( F(a,b,c;t) + F(a-1,b,c;t) \Big ) +
\Big ( F(a,b+1,c;t) + F(a-1,b+1,c;t) \Big )
\end{equation*}
and thus
\begin{equation*}
  t(1-t){d \over dt} F(a,b,c;t) = bt F(a,b,c;t) - (c-1) F(a,b,c;t)
  - (c-b-1) \Big( F(a-1,b,c;t) - F(a-1,b+1,c;t) \Big).
\end{equation*}
But
\begin{align*}
  & -(c - b -1) \Big( F(a-1,b,c;t) - F(a-1,b+1,c;t) \Big) \\
  \qquad
  & = (c-b-1) t^{-c+1} \Big( \int_0^{1/t} - \xi \int_{1}^{1/t} \Big)
              u^{a-c-1} (1-tu)^{c-b-1}(1-u)^{-a+1}{tu \over 1-tu} \, du \\
  & = -t^{-c+1} \Big( \int_0^{1/t} - \xi \int_1^{1/t} \Big)
       u^{a-c} {d \over du}(1-tu)^{c-b-1}(1-u)^{-a+1} \\
  & = t^{-c+1} (a-c) \Big( \int_0^{1/t} - \xi \int_1^{1/t} \Big)
       u^{a-c-1}(1-tu)^{c-b-1}(1-u)^{-a+1} \, du \nonumber \\
  & \quad + t^{-c+1} (a-1) \Big( \int_0^{1/t} - \xi \int_1^{1/t} \Big)
            u^{a-c}(1-tu)^{c-b-1}(1-u)^{-a} \, du \\
  & = (a-c) F(a-1,b,c;t) + (a-1) F(a,b,c;t)
\end{align*}
so we have
\begin{equation*}
t (1-t) {d \over dt} F(a,b,c;t) = bt F(a,b,c;t) +
(a-c) F(a,b,c;t) + (a-c) F(a-1,b,c;t),
\end{equation*}
in agreement with (\ref{3.3c}). \hfill $\square$
 
Following the steps from which (\ref{EIa}) was deduced from (\ref{oD})
allows us to deduce from (\ref{3.14a}) the following multiple integral
representation.

\begin{prop}\label{p8a}
We can rewrite (\ref{3.14a}) and so deduce
\begin{multline}\label{3.20a}
  {\tau}_3[n](t) = \tau_3(t;b_1,b_2,b_3+n,b_4) \Big|_{b_1 + b_3 = 0}
   \propto \Big( \int_0^1 - \xi \int_t^1 \Big) du_1 \cdots
           \Big( \int_0^1 - \xi \int_t^1 \Big) du_n \\
  \times
  \prod_{i=1}^n u_i^{-b_2 - (b_3 + n)}(1-u_i)^{b_2 - (b_3 + n)}
                (t-u_i)^{-(b_1 + b_4)}
  \prod_{1 \le j < k \le n} (u_k - u_j)^2.
\end{multline}
\end{prop}

Finally we show that by choosing $ F $ in (\ref{if}) to be given by the
linear combination (\ref{5m}), $ \tau_3[n] $ can be written as an 
$n$-dimensional integral having a form different from the previous cases.
\begin{prop}
Let $F(a,b,c;t)$ in (\ref{if})  be given by
\begin{equation}\label{3.03}
F(a,b,c;t) = \Big ( \int_{-\infty}^\infty - \xi \int_t^\infty \Big )
u^{a-c} (1 - u)^{c-b-1} (t - u)^{-a} \, du,
\end{equation}
where it is required the integrand be integrable in the
neighbourhood of $u=t, \pm \infty$ but not necessarily $u=0,1$ (the path
of integration can be deformed around these points). Then
\begin{equation}\label{3.3a}
\bar{\tau}_3[n](t) \propto t^{bn/2}(t-1)^{bn/2 + n(n-1)/2}
\det[ F(a-j,b+k,c;t) ]_{j,k=0,\dots,n-1}.
\end{equation}
\end{prop}

Following the steps from which (\ref{EIa}) was deduced from (\ref{oD})
allows us to deduce from (\ref{3.3a}) the following multiple integral
representation. 

\begin{prop}\label{p8}
We can rewrite (\ref{3.3a}) and so deduce 
\begin{multline}\label{tca}
  {\tau}_3[n](t) = \tau_3(t;b_1,b_2,b_3+n,b_4) \Big|_{b_1 + b_3 = 0}
  \propto \Big( \int_{-\infty}^\infty - \xi \int_t^\infty \Big) du_1 \cdots
          \Big( \int_{-\infty}^\infty - \xi \int_t^\infty \Big) du_n \\ 
  \times \prod_{i=1}^n u_i^{-b_2 - (b_3 + n)}(1-u_i)^{b_2 - (b_3 + n)}
                       (t-u_i)^{-(b_1 + b_4)}
         \prod_{1 \le j < k \le n} (u_k - u_j)^2.
\end{multline}
\end{prop}

The only necessary detail of the contours (intervals) of integration in
(\ref{tca}) is that the integrand vanishes at the endpoints. Deforming
the contours in this manner we see from (\ref{tca}) that
\begin{multline}\label{tca1}
  \tau_3({1 + it \over 2};b_1,b_2,b_3+n,b_4) \Big|_{b_1 + b_3 = 0}
  \propto \Big( \int_{-\infty}^\infty - \xi \int_t^\infty \Big) du_1 \cdots
          \Big( \int_{-\infty}^\infty - \xi \int_t^\infty \Big) du_n \\ 
  \times \prod_{i=1}^n (1+iu_i)^{-b_2 - (b_3 + n)}(1-iu_i)^{b_2 - (b_3 + n)}
                       (t-u_i)^{b_3 - b_4}
         \prod_{1 \le j < k \le n} (u_k - u_j)^2.
\end{multline}

\subsection{Schlesinger Transformations}
In this part we develop difference equations arising from the
sequence generated
by the action of $ T_3 $, which are also known as Schlesinger transformations
because the formal monodromy exponents are shifted by integers. In doing so
we will demonstrate that this recurrence in the canonical variables can be
expressed in a form which is precisely that of the discrete fifth Painlev\'e
equation \dPV. We thus establish directly that the discrete \dPV is the
contiguity relation of the continuous \PVI equation in contrast to other
treatments \cite{Sa_2001,ROG-2000,GORS-98,MS-95}. The $ T_3, T_3^{-1} $ operators
are equivalent to the $ R_{(9)}, R_{(10)} $ operators respectively in 
\cite{MS-95}.

\begin{prop}\label{dPv_prop}
The sequence $ \{q[n],p[n]\}^{\infty}_{n=0} $ generated by the shift operator
$ T_3 $ with parameters 
$ \boldsymbol{\alpha} = (\alpha_0+n,\alpha_1+n,\alpha_2-n,\alpha_3,\alpha_4) $
defines an auxiliary sequence, equivalent to the sequence
$ \{g[n],f[n]\}^{\infty}_{n=0} $ satisfying the discrete fifth Painlev\'e
equation \dPV
\begin{align}
   g[n+1]g[n] & = {t \over t-1}
   { (f[n]+1-\alpha_2 )(f[n]+1-\alpha_2 -\alpha_4 ) \over
      f[n](f[n]+\alpha_3 ) }
  \label{dPv_1}\\
  f[n]+f[n-1] & =
  -\alpha_3 + {\alpha_1  \over g[n]-1} + { \alpha_0 t \over t(g[n]-1)-g[n] } ,
  \label{dPv_2}
\end{align}
where
\begin{equation}
   g := {q \over q-1} ,\quad
   f := q(q-1)p+(1-\alpha_2 -\alpha_4 )(q-1)-\alpha_3 q
       -\alpha_0 {q(q-1) \over q-t} \ .
\label{fg_defn}
\end{equation}
\end{prop}

\noindent
Proof. \quad To begin with we construct the forward and backward shifts in
the canonical variables under the action of $ T_3 $ using Table \ref{t1}
($ q := q[n], p := p[n] $)
\begin{align}
   q[n+1] := \Uq & = {t \over q-t}
    \left[ (q-1)(q-t)p+(\alpha_1 +\alpha_2 )(q-t)-\alpha_0(t-1) \right]
   \nonumber\\
   & \qquad \times
    \left[ q(q-1)(q-t)p+(\alpha_1 +\alpha_2 )q(q-t)-\alpha_0(t-1)q+\alpha_4(q-t)
    \right] {1 \over X_u}
   \label{q_up}\\
   p[n+1] := \Up & = -{(q-t) \over t(t-1)}
   {\left[ (q-t)p+\alpha_1 +\alpha_2  \right] X_u \over
    \left[ q(q-t)p+(\alpha_1 +\alpha_2 )(q-t)-\alpha_0 t \right]
    \left[ (q-1)(q-t)p+(\alpha_1 +\alpha_2 )(q-t)-\alpha_0(t-1) \right]}
   \label{p_up}\\
   q[n-1] := \Dq & = t
   {\left[ (q-1)p+\alpha_2  \right]
    \left[ q(q-1)p+\alpha_2 q+\alpha_4  \right] \over X_d}
   \label{q_dn}\\
   p[n-1] := \Dp & = {X_d \over t(t-1)} \left\{
   -{\left[ (q-t)p+\alpha_2  \right] \over
     \left[ qp+\alpha_2  \right] \left[ (q-1)p+\alpha_2  \right]}
   +{\alpha_0 -1 \over
     q(q-1)p^2-(\alpha_4 (q-1)+\alpha_3 q)p
              -\alpha_2 (\alpha_2 +\alpha_3 +\alpha_4 )} \right\} ,
   \label{p_dn}
\end{align}
where 
\begin{align}
   X_u & =
    q(q-1)(q-t)^2p^2
   -[\alpha_4 (q-1)(q-t)+\alpha_3 q(q-t)+(\alpha_0 -1-\alpha_1 )q(q-1)](q-t)p
   \cr
   & \quad
   +(\alpha_1 +\alpha_2 )^2(q-t)^2
   +(\alpha_1 +\alpha_2 )[(\alpha_1 +\alpha_2 +\alpha_4 )t
                          +(\alpha_1 +\alpha_2 +\alpha_3 )(t-1)](q-t)
   \cr
   & \qquad
   -\alpha_0(\alpha_1 +1)t(t-1)
   \label{Xu}\\
   X_d & =
    q(q-1)(q-t)p^2
   -[\alpha_4 (q-1)(q-t)+\alpha_3 q(q-t)+(\alpha_0 -1+\alpha_1 )q(q-1)]p
   \cr
   & \quad
   +\alpha_2 ^2(q-t)
   +\alpha_2 (\alpha_2 +\alpha_4 )t +\alpha_2 (\alpha_2 +\alpha_3 )(t-1),
   \label{Xd}
\end{align}
(note that the factors $ X_d $, $ X_u $ are distinguished by being quadratic 
in $ p $). From these variables certain products can be constructed that have 
simple factorisable forms
\begin{align}
   \Uq\Up & =
   -{\left[ (q-t)p+\alpha_1 +\alpha_2  \right] \over t-1}
    { \left[ q(q-1)(q-t)p+(\alpha_1 +\alpha_2)q(q-t)-\alpha_0(t-1)q+\alpha_4(q-t)
      \right] \over \left[ q(q-t)p+(\alpha_1 +\alpha_2)(q-t)-\alpha_0 t \right]}
   \label{qp_up}\\
   (\Uq-1)\Up & =
   -{\left[ (q-t)p+\alpha_1 +\alpha_2  \right] \over t}
    { \left[ q(q-1)(q-t)p+(\alpha_1 +\alpha_2)(q-1)(q-t)
                         -\alpha_0 t(q-1)-\alpha_3(q-t)
      \right] \over
      \left[ (q-1)(q-t)p+(\alpha_1 +\alpha_2)(q-t)-\alpha_0(t-1) \right] }
   \label{qmop_up}\\
   (\Uq-t)\Up & =
  -{\left[ (q-t)p+\alpha_1 +\alpha_2  \right] \over
    \left[ q(q-t)p+(\alpha_1 +\alpha_2 )(q-t)-\alpha_0 t \right]
    \left[ (q-1)(q-t)p+(\alpha_1 +\alpha_2 )(q-t)-\alpha_0(t-1) \right]}
   \cr
   & \quad \times \Bigg\{
    q(q-1)(q-t)^2p^2
   -[\alpha_4 (q-1)(q-t)+\alpha_3 q(q-t)+2\alpha_0 q(q-1)](q-t)p
   \cr
   & \quad
   +\alpha_0[\alpha_4 (q-1)(q-t)+\alpha_3 q(q-t)+\alpha_0 q(q-1)]
   -(1-\alpha_2)(\alpha_0 +\alpha_1+\alpha_2)(q-t)^2
   \Bigg\}
   \label{qmtp_up}\\
   {\Dq-1 \over \Dq-t} & = {1 \over t}
   { q^2(q-1)p^2+[2\alpha_2 q(q-1)-\alpha_3 q]p
                +\alpha_2^2q-\alpha_2(\alpha_2+\alpha_3) \over
     q(q-1)p^2-[\alpha_4(q-1)+\alpha_3 q]p-\alpha_2(\alpha_2+\alpha_3+\alpha_4)}.
   \label{qm1oqmt_dn}
\end{align}
From the ratio of (\ref{qp_up},\ref{qmop_up}) one notices that it can be simply
expressed in terms of the single quantity $ f $ defined by (\ref{fg_defn}) which 
is the first of the coupled recurrences for \dPV (\ref{dPv_1}). To find the 
other member of the pair we evaluate $ \Df := f[n-1] $ using the above formulas
and find
\begin{equation}
   \Df = -(q-1)(qp+\alpha_2 ) ,
\label{f_down}
\end{equation}
so that when this is added to $ f $ we arrive at (\ref{dPv_2}). \hfill $\square$

Although of theoretical interest, Proposition \ref{dPv_prop} does not immediately 
lead to a difference equation for the Hamiltonian itself. However such an 
equation can be derived by using the workings from the derivation of Proposition 
\ref{dPv_prop}.

\begin{prop}
The sequence of Hamiltonians $ \{K[n]\}^{\infty}_{n=0} $ generated by the shift
operator $ T_3 $ with parameters 
$ \boldsymbol{\alpha} = (\alpha_0+n,\alpha_1+n,\alpha_2-n,\alpha_3,\alpha_4) $
satisfies the third order difference equation
\begin{multline}
  \left[ (1-\alpha_0)\uK+\alpha_0 K \right]
  \left[ (1+\alpha_1)K-\alpha_1 \uK +(\alpha_1+\alpha_2)(1+\alpha_1-\alpha_0)t
                -(\alpha_1+\alpha_2)(\alpha_1+\alpha_2+\alpha_3) \right]
  \\ \times
  \left[ \uK-\dK+1-\alpha_0-\alpha_4-(1+\alpha_1-\alpha_0+2\alpha_2)t \right]
  \left[ \uuK-K-\alpha_0-\alpha_4+(2\alpha_0+\alpha_3+\alpha_4)t \right]
  \\ + t(t-1) \Big[
    (1-\alpha_0-\alpha_1)K-\alpha_1(1-\alpha_0)(\uK-\dK)
    +(1-\alpha_0)(\alpha_1+\alpha_2)(1-\alpha_0+\alpha_1)t
    -(1-\alpha_0)(\alpha_1+\alpha_2)(\alpha_1+\alpha_2+\alpha_3) \Big]
  \\ \times \Big[
       (1+\alpha_0+\alpha_1)\uK-\alpha_0(1+\alpha_1)(\uuK-K)
    +\alpha_0(\alpha_1+\alpha_2)(1-\alpha_0+\alpha_1)t
    -\alpha_0(\alpha_1+\alpha_2)(\alpha_1+\alpha_2+\alpha_3) \Big] = 0
\label{dPV_Ham}
\end{multline}
where
$ K := t(t-1)H[n],\dK := t(t-1)H[n-1],\uK := t(t-1)H[n+1],\uuK := t(t-1)H[n+2] $.
\end{prop}

\noindent
Proof. \quad We focus our attention on the quantity
\begin{equation}
   Z := -q(q-1)p-(\alpha_1 +\alpha_2)(q-t) ,
\label{Z_defn}
\end{equation}
which from (\ref{2.35}) is $ Z = K[n+1]-K[n] $. First we consider
$ \DZ := Z[n-1] $.
From the definition $ Z $ is related to $ f $ so employing (\ref{f_down}) and
(\ref{qm1oqmt_dn}) for the downshifted variables we have
\begin{multline}
  \DZ = (q-1)(qp+\alpha_2)+(1-\alpha_0)q-\alpha_3+(\alpha_1+\alpha_2)(t-1)
  \\ \quad
        + (1-\alpha_0)
  { (2\alpha_2+\alpha_3+\alpha_4)[q(q-1)p+\alpha_2 q]-\alpha_2(\alpha_2+\alpha_3)
    \over
    q(q-1)p^2-[\alpha_4(q-1)+\alpha_3 q]p-\alpha_2(\alpha_2+\alpha_3+\alpha_4) }
 .
\label{Z_down}
\end{multline}
We transform this expression by replacing the canonical variables $ q, q-1, p $
where possible by $ K, Z $ and retaining $ q-t $. The resulting equation is then
one which is a linear equation for $ q-t $ in terms of $ K, Z, \DZ $,
\begin{multline}
  q-t = \left[ K+(1-\alpha_0)Z \right]
  \left[ Z+\DZ+\alpha_2+\alpha_3-(\alpha_1+\alpha_2)(t-1)-(1+\alpha_2-\alpha_0)t
  \right]
  \\ \Big\slash
  \Big[ (1-\alpha_0-\alpha_1)K-\alpha_1(1-\alpha_0)(Z+\DZ)
  \\ \qquad
       +(1-\alpha_0)[-\alpha_1(\alpha_2+\alpha_3)
         +(\alpha_1^2+\alpha_1\alpha_2+\alpha_2^2+\alpha_2\alpha_3)(t-1)
         +(\alpha_1+\alpha_1\alpha_2-
\alpha_0\alpha_1+\alpha_2^2+\alpha_2\alpha_4)t] \Big] .
\label{qmt_1}
\end{multline}
We use a similar strategy and evaluate the upshifted $ \UZ := Z[n+1] $ using
(\ref{qmop_up},\ref{qp_up},\ref{qmtp_up},\ref{p_up}).
We find after considerable simplification
\begin{multline}
  \UZ = q(q-1)p+(\alpha_1+\alpha_2)(q-t)-(2\alpha_0+\alpha_3+\alpha_4)t
               -(1+\alpha_0+\alpha_1){t(t-1) \over q-t}
  \\ + (1+\alpha_1){t(t-1) \over (q-t)X_u} \Bigg\{
     (1+\alpha_0+\alpha_1)q(q-1)(q-t)p+(\alpha_1+\alpha_2)(1+\alpha_0+\alpha_1)q^2
  \\ \qquad
     +[-(1+\alpha_0+\alpha_1)(\alpha_0+\alpha_1+\alpha_2)t
       +\alpha_0(\alpha_0+\alpha_4)-(\alpha_1+\alpha_2)
                                    (\alpha_1+\alpha_2+\alpha_3)]q
  \\ \qquad\quad
     +(\alpha_0+\alpha_1+\alpha_2)(1-\alpha_2-\alpha_4)t \Bigg\},
\label{Z_up}
\end{multline}
and after the variable replacements described above one has a linear equation for
\begin{multline}
   {t(t-1) \over q-t} =
   -\Big[ K-\alpha_1 Z
          +(\alpha_1+\alpha_2)[(1-\alpha_0+\alpha_1)t-\alpha_1-\alpha_2-\alpha_3]
    \Big]
     \left[ Z+\UZ-\alpha_0-\alpha_4+(2\alpha_0+\alpha_3+\alpha_4)t \right]
   \\ \Big\slash \Bigg\{
    (1+\alpha_0+\alpha_1)K+(1+\alpha_1-\alpha_0\alpha_1)Z
    -\alpha_0(1+\alpha_1)\UZ
   \\ \qquad
    +\alpha_0(\alpha_1+\alpha_2)(1-\alpha_0+\alpha_1)t
    -\alpha_0(\alpha_1+\alpha_2)(\alpha_1+\alpha_2+\alpha_3) \Bigg\}
\label{qmt_2}
\end{multline}
By eliminating $ q-t $ between (\ref{qmt_1}) and (\ref{qmt_2}) and further
simplification one arrives at the stated result (\ref{dPV_Ham}). \hfill $\square$

We remark that the right-hand side of (\ref{Z_defn}) can be written in terms
of $ f $ and $ g $, and then coupled with the \dPV equations 
(\ref{dPv_1},\ref{dPv_2}) to provide an alternative scheme to calculate $ K[n] $. A
similar recurrence appears in the recent work of Borodin \cite{Bo_2001}.

\section{Application to the finite JUE and CyUE}
\setcounter{equation}{0}
We are now in a position to identify the multiple integral representations for the
$ \tau$-functions of the classical solutions to the \PVI system, (\ref{EIa}), 
(\ref{3.20a}) and (\ref{tca1}), with the spectral averages defined by (\ref{E1}), 
(\ref{1.17a}) and (\ref{1.15a}) respectively.

\subsection{The JUE}
The $ \tau$-function solution (\ref{2m}) is relevant to the average (\ref{E1}), 
for with $N=1$ we have
\begin{align}
  \tilde{E}_1^J(t;a,b,\mu)
  & \propto t^{a+\mu+1} {}_2 F_1(-b,a+1;\mu+a+2;t)
  \nonumber \\
  & = t^{a+\mu+1} \tau_3(t;-(a+b)/2, \mu + 1+(a+b)/2, 1 + (a+b)/2, (a-b)/2).
\end{align}
The multiple integral representation (\ref{EIa}) of ${\tau}_3[n](t)$
is of the type occuring in the definition (\ref{E1}) of $\tilde{E}_N(t;a,b;\mu)$. 
This allows the latter to be identified with a $\tau$-function for the
\PVI system, and its logarithmic derivative identified with an
auxiliary Hamiltonian (\ref{3}) and so characterised as a solution
of the Jimbo-Miwa-Okamoto $\sigma$-form of the \PVI equation
(\ref{4}).

\begin{prop}\label{pr6}
Let $\tau_3[n](t) = \tau_3(t;b_1,b_2,b_3+n,b_4) \Big |_{b_1+b_3=0}$ 
refer to the $\tau$-function sequence (\ref{2.35a}) with
$\tau_3[0] = 1$ and $\tau_3[1](t)$ given by (\ref{2m}). Then we have
\begin{equation}\label{Et}
   \tilde{E}_N^{\rm J}(t;a,b,\mu) = C t^{(b_1+b_3)(b_2+b_4)}\tau_3(t;\mathbf{b}),
   \quad \mathbf{b} = ( -(a+b)/2, \mu+N+(a+b)/2, N+(a+b)/2,(a-b)/2 )
\end{equation}
and consequently
\begin{equation}\label{Et1}
  t(t-1){d \over dt} \log\Big(
        (t-1)^{e_2'[\mathbf{b}] - {1 \over 2}e_2[\mathbf{b}]}
         t^{{1 \over 2}e_2[\mathbf{b}]}
         t^{-(b_1+b_3)(b_2+b_4)}
        \tilde{E}_N^{\rm J}(t;a,b,\mu) \Big) = U_N^{\rm J}(t;a,b,\mu)
\end{equation}
where $U_N^{\rm J}(t;a,b,\mu)$ satisfies the Jimbo-Miwa--Okamoto $\sigma$-form 
of the \PVI equation (\ref{4}) with $\mathbf{b}$ specified as in (\ref{Et}), 
and $e_2'[\mathbf{b}]$, $e_2[\mathbf{b}]$ defined as in Proposition \ref{p1}.
The latter is to be solved subject to the boundary condition
\begin{equation}\label{bc.1}
\begin{split}
  U_N^{\rm J}(t;a,b,\mu) 
  & \mathop{\sim}\limits_{t \to 0}
    - {1 \over 2}e_2[\mathbf{b}] 
    + \Big( e_2'[\mathbf{b}] + bN{a+N \over a+\mu+2N} \Big)t
  \cr
  & \mathop{\sim}\limits_{t \to 0}
    -{1 \over 2}N^2-{1 \over 2}(\mu+a-b)N+{1 \over 4}b(a+b)-{1 \over 4}\mu(a-b)
    +\left( -bN-{1 \over 4}(a+b)^2+bN {a+N \over a+\mu+2N} \right)t
\end{split}
\end{equation}
with $U_N^{\rm J}$ given by a power series in $t$ about $t=0$.
\end{prop}

\noindent
Proof. \quad The first equation follows immediately upon comparing
(\ref{EIa}) with (\ref{E1}), and rewriting the exponent in the first factor in
(\ref{E1}) in terms of the $b$'s. The equation (\ref{Et1}) then follows from
(\ref{th}). For the boundary condition, we see from the second
integral in (\ref{E1}) and the definition of the $b$'s that
\begin{equation}\label{bc.2}
  t^{-N(a+\mu+N)} \tilde{E}_N(t;a,b,\mu) 
  = t^{-(b_1+b_3)(b_2+b_4)} \tilde{E}_N^{\rm J}(t;a,b,\mu)
  \mathop{\sim}\limits_{t \to 0}
  {J_N(a,\mu) \over C}\Big( 1 - b{J_N(a,\mu)[ \sum_{j=1}^N x_j] \over
                                  J_N(a,\mu)}t \Big)
\end{equation}
where $J_N(a,\mu)[ \sum_{j=1}^N x_j]$ is the integral (\ref{nj}) with an
extra factor of $\sum_{j=1}^N x_j$ in the integrand. The ratio of integrals
in (\ref{bc.2}) can be evaluated using a generalisation of the Selberg
integral due to Aomoto \cite{Ao-87} to give (\ref{bc.1}).
Alternatively, the $O(1)$ term in (\ref{bc.1}) can be substituted in (\ref{4})
to deduce the $O(t)$ term.
\hfill $\square$

Let us now reconcile Proposition \ref{pr6} with the result 
that (\ref{xx1}) satisfies (\ref{4}) with parameters (\ref{xx1a}).
For this purpose we note from (\ref{Et}) that with $\mu = 0$, $b_2 =
b_3$. Then the exponents in (\ref{Et1}) simplify, and we see that
\begin{equation}\label{3.12}
  U_N^{\rm J}(t;a,b,0) = - t b_2^2 + {1 \over 2}(b_1 b_4 + b_2^2) 
  + t(t-1){d \over dt} \log\tilde{E}_N^{\rm J}(t;a,b,0).
\end{equation}
Interchanging $a$ and $b$ as required by (\ref{xx1}) changes the sign of
$b_4$, so with $\mathbf{b}$ given as in (\ref{Et}) with
$\mu = 0$, we see from (\ref{3.12}) that
\begin{equation*}
  -U_N^{\rm J}(1-t;b,a,0) = - t b_2^2 - {1 \over 2}(b_1 b_4 - b_2^2)
   + t(t-1){d \over dt} \log\tilde{E}_N^{\rm J}(1-t;a,b,0).
\end{equation*}
But in general, if $h(t)$ satisfies (\ref{4}) with parameters $\mathbf{b}$, 
then $-h(1-t)$ also satisfies (\ref{4}) but with an odd number of
$b_1,\dots,b_4$ reversed in sign. Choosing to reverse the sign of $b_4$,
it follows that
\begin{equation*}
  - t b_2^2 + {1 \over 2} (b_1 b_4 + b_2^2) 
  + t(t-1){d \over dt} \log\tilde{E}_N^{\rm J}(1-t;a,b,0)
\end{equation*}
satisfies (\ref{4}) with $\mathbf{b}$ given by (a suitable permutation of)
(\ref{xx1a}), which is in agreement with (\ref{xx1}).

A corollary of Proposition \ref{pr6}, which follows from (\ref{ju}) and
the second equality in (\ref{E1}), is the formula
\begin{align}\label{1.41g}
  U_N^{\rm J}(t;a,\mu,b)
  & = t(t-1) {d \over dt} \log \Big(
             (t-1)^{e_2'[\mathbf{b}^*] - {1 \over 2} e_2[\mathbf{b}^*]}
              t^{{1 \over 2} e_2[\mathbf{b}^*]}
             \Big\langle 
                 \prod_{l=1}^N z_l^{N+a+b/2}|1+z_l|^b (1+t z_l)^\mu 
             \Big\rangle_{{\rm CUE}_N} \Big), \nonumber \\
  \mathbf{b}^*
  & = \Big( - (a+\mu)/2, b+N+(a+\mu)/2, N+(a+\mu)/2, (a-\mu)/2 \Big).
\end{align} 

Comparison of the $ \tau$-function solution not analytic at the origin (\ref{cc}) 
with (\ref{1.17a}) shows, for $ N = 1 $
\begin{align}\label{ccA}
  \tilde{E}_1^{\rm J}(t;a,b,\mu;\xi)
  & \propto f^{\rm J}(-\mu,-\mu-a-1-b,-\mu+a;t)
  \nonumber \\
  & = \tau_3\Big(t; 1+{1 \over 2}(a+b), {1 \over 2}(b-a), -{1 \over 2}(a+b),
                     -\mu -1-{1 \over 2} (a+b) \Big).
\end{align}
Furthermore comparing the $n$-dimensional integral (\ref{3.20a}) with 
(\ref{1.17a}) allows us to characterise the latter in terms of a solution of 
(\ref{4}). 

\begin{prop}\label{p9}
Let $\tau_3[n](t) = \tau_3(t;b_1,b_2,b_3+n,b_4) \Big |_{b_1+b_3=0}$ refer 
to the $\tau$-function sequence (\ref{2.35a}) with $\tau_3[0] = 1$ and 
$\tau_3[1](t)$ given by (\ref{3.03a}). Then we have
\begin{equation}\label{3.22a}
  \tilde{E}_N^{\rm J}(t;a,b,\mu;\xi) = C \tau_3(t;\hat{\mathbf{b}}), \quad
  \hat{\mathbf{b}} = \Big( {1 \over 2}(a+b)+N, {1 \over 2}(b-a),
                          -{1 \over 2}(a+b), -{1 \over 2}(a+b)-N-\mu \Big)
\end{equation}
and consequently
\begin{equation}\label{uE}
  {e}_2'[\hat{\mathbf{b}}] t - {1 \over 2} e_2[\hat{\mathbf{b}}] 
   + t(t-1){d \over dt} \log\tilde{E}_N^{\rm J}(t;a,b,\mu;\xi)
   = \hat{U}_N^{\rm J}(t;a,b,\mu;\xi)
\end{equation}
where $\hat{U}_N^{\rm J}(t;a,b,\mu;\xi)$ satisfies the Jimbo-Miwa--Okamoto 
$\sigma$-form of the \PVI equation (\ref{4}) with $\mathbf{b}$ specified as 
in (\ref{3.22a}), and $e_2'[\hat{\mathbf{b}}]$, $e_2[\hat{\mathbf{b}}]$ 
defined as in Proposition \ref{p1}. In the case $\xi = 1$ we have the 
boundary condition
\begin{equation}\label{bc.3}
\begin{split}
  \hat{U}^{\rm J}_N(t;a,b,\mu;1)
  & \mathop{\sim}\limits_{t \to 0} 
    - {1 \over 2} e_2[\hat{\mathbf{b}}] - N(a+\mu+N) 
    + \Big( e_2'[\hat{\mathbf{b}}] + (N+\mu+a)N + bN{a+N \over a+\mu+2N} \Big)t
  \cr
  & \mathop{\sim}\limits_{t \to 0}
    -{1 \over 2}N^2-{1 \over 2}(\mu+a-b)N+{1 \over 4}b(a+b)-{1 \over 4}\mu(a-b)
    +\left( -bN-{1 \over 4}(a+b)^2+bN {a+N \over a+\mu+2N} \right)t
\end{split}
\end{equation}
with $\hat{U}^{\rm J}_N$ a power series in $t$ about $t=0$,
while in the case $\xi = 0$
\begin{equation}\label{bc.4}
\begin{split}
  \hat{U}^{\rm J}_N(t;a,b,\mu;0)
  & \mathop{\sim}\limits_{|t| \to \infty} 
    \Big( e_2'[\hat{\mathbf{b}}] + N\mu\Big) t + O(1)
  \cr
  & \mathop{\sim}\limits_{|t| \to \infty} 
    -\left( N+{1 \over 2}(a+b) \right)^2 t + O(1)
\end{split}
\end{equation}
with $\hat{U}^{\rm J}_N$, apart from the leading term, a power series in
$1/t$.
\end{prop}

\noindent
Proof. \quad The only remaining task is the specification of the boundary
conditions (\ref{bc.3}) and (\ref{bc.4}). The first of these follows from the
fact that $\tilde{E}_N^{\rm J}(t;a,b,\mu;1) = \tilde{E}_N^{\rm J}(t;a,b,\mu)$ 
and then substituting (\ref{bc.2}) in (\ref{uE}), while the second follows 
from the fact that $\tilde{E}_N(t;a,b,\mu;0) = F_N^{\rm J}(t;a,b,\mu)$ and 
noting from (\ref{F1}) that $ F_N^{\rm J}(t;a,b,\mu) \sim t^{N \mu}$ as
$t \to \infty$. \hfill $\square$

Comparing the definition (\ref{1.17a}) with (\ref{E1}) shows
\begin{equation}\label{1.21c}
  \tilde{E}_N^{\rm J}(t;a,b,\mu;1) =\tilde{E}_N^{\rm J}(t;a,b,\mu)
\end{equation}
so the $\sigma$-function $\hat{U}$ in (\ref{uE}) must be related to the $\sigma$-function 
$U$ in (\ref{Et1}). To explore this point, writing $\mathbf{b}$ in (\ref{Et}) 
as $\mathbf{b} = (b_1,b_2,b_3,b_4)$ we see from (\ref{3.22a}) that
\begin{equation}\label{1.21d}
  \hat{\mathbf{b}}= (b_3, -b_4, b_1, -b_2).
\end{equation}
The differential equation (\ref{4}) is unchanged by the replacement of
$\mathbf{b}$ by (\ref{1.21d}), so $\hat{U}$ and $U$ in fact satisfy the 
same equation. In fact $\hat{U}$ and $U$ are in this case the same function, 
as the identity (\ref{1.21c}) together with the readily verified formula
\begin{equation*}
  e_2'[\mathbf{b}] t - {1 \over 2} e_2[\mathbf{b}] 
  - (t-1)(b_1 + b_3)(b_2 + b_4) 
  = e_2'[\hat{\mathbf{b}}]t - {1 \over 2} e_2[\hat{\mathbf{b}}]
\end{equation*}
show that the left hand sides of (\ref{Et1}) and (\ref{uE}) agree.

Another special case of (\ref{1.17a}) of interest is the coefficient of 
$\xi^N$,
\begin{equation*}
  [\xi^N] \tilde{E}_N^{\rm J}(s;a,b,\mu;\xi) 
  = {(-1)^N \over C} \int_s^1 dx_1 \, x_1^a(1-x_1)^b(s-x_1)^\mu \cdots
                     \int_s^1 dx_N \, x_N^a(1-x_N)^b(s-x_N)^\mu 
                     \prod_{1 \le j < k \le N} (x_k - x_j)^2.
\end{equation*}
This exhibits the functional property
\begin{equation*}
  [\xi^N] \tilde{E}_N^{\rm J}({1 \over s};a,b,\mu;\xi) 
  = (-1)^N s^{-N(a+b+\mu+N)}[\xi^N] \tilde{E}_N^{\rm J}(s;a,b,\mu;\xi).
\end{equation*}
The $\tau$-function evaluation (\ref{3.22a}) then implies 
\begin{equation}\label{2.12e}
  \tau_3({1 \over t};\hat{\mathbf{b}}) 
  = (-1)^N t^{(\hat{b}_3 + \hat{b}_4)(\hat{b}_1 + \hat{b}_3)}
           \tau_3(t;\bar{\mathbf{b}})
\end{equation}
where
\begin{equation}\label{2.12f}
  \bar{\mathbf{b}} = \Big( 
      {1 \over 2}(\hat{b}_1 - \hat{b}_2 + \hat{b}_3 - \hat{b}_4), 
      {1 \over 2}(-\hat{b}_1 + \hat{b}_2 + \hat{b}_3 - \hat{b}_4), 
      {1 \over 2}(\hat{b}_1 + \hat{b}_2 + \hat{b}_3 + \hat{b}_4),
      {1 \over 2}(-\hat{b}_1 - \hat{b}_2 + \hat{b}_3 + \hat{b}_4) \Big)
\end{equation}
($\bar{\mathbf{b}}$ is obtained from $\hat{\mathbf{b}}$ by simply
interchanging $b$ and $\mu$ in the latter; recall (\ref{1.21d}) and
(\ref{Et})). This result can be understood within the context of the
general theory of the \PVI equation. Thus the mapping
\begin{equation*}
  (q,p,H,t;\hat{\mathbf{b}})
  \mapsto
  ({1 \over q},(b_1+b_3)q -q^2p, -{H \over t^2}+\Phi(t),{1 \over t};
   \bar{\mathbf{b}}), \: \:
   \Phi(t) = - {1 \over t}(\hat{b}_3 + \hat{b}_1)(\hat{b}_3 + \hat{b}_4)
\end{equation*}
has been identified in \cite{Ok_1987a} as a canonical transformation of
the \PVI system
(the value of $\Phi(t)$ was not given explicitly in \cite{Ok_1987a}).
Recalling (\ref{2.12a}), up to a proportionality constant this
immediately implies (\ref{2.12e}).

A consequence of the difference equation (\ref{dPV_Ham})
is that a difference equation for $ U_N^{\rm J}(t;a,b,\mu;\xi) $ with respect to
$ \mu $ can be found.

\begin{prop}
The Jimbo-Miwa-Okamoto $\sigma$-function $ U_N^{\rm J}(t;a,b,\mu;\xi) $ satisfies a
third order difference equation in $ \mu $, for all $ N, a, b, t $
\begin{multline}
  \left[ (N+a+\mu+1)U-(N+a+\mu)\uU + \quarter(a-b)^2 t
         -\quarter(a^2+b^2) - \quarter(a+b)N \right]
  \\ \times
  \left[ (N+b+\mu+1)U-(N+b+\mu)\uU + \quarter(a-b)^2 t
         +\half ab + \quarter(a+b)N \right]
  \\ \times
  \left[ \uU-\dU+(2N+2\mu+a+b)(t-\half) \right]
  \left[ \uuU-U+(2N+2\mu+2+a+b)(t-\half) \right]
  \\ = t(t-1) \Big[
    (N+a+\mu)(N+b+\mu)(\uU-\dU)-(2N+2\mu+a+b)U
  \\
    -\quarter(a-b)^2(2N+2\mu+a+b)t
    +\quarter (b-a)(2Nb +(b-a)\mu +b(a+b)) \Big]
  \\ \times \Big[
    (N+a+\mu+1)(N+b+\mu+1)(\uuU-U)-(2N+2\mu+2+a+b)\uU
  \\
    -\quarter(a-b)^2(2N+2\mu+2+a+b)t
    +\quarter (b-a)(2Nb +(b-a)(\mu+1) +b(a+b)) \Big] ,
\label{UJ_recur}
\end{multline}
where
$ U := U_N^{\rm J}(t;a,b,\mu;\xi), \dU := U_{N}^{\rm J}(t;a,b,\mu-1;\xi),
  \uU := U_{N}^{\rm J}(t;a,b,\mu+1;\xi) $ etc.
\end{prop}

\noindent
Proof. \quad This follows from (\ref{dPV_Ham}) using the parameters in
(\ref{Et}) with $ b_{2} \leftrightarrow b_{3} $ and the invariance of $ h $ or
$ U $ under this interchange, and the relation
\begin{equation}
    U_N^{\rm J}(t;a,b,\mu;\xi) = K[\mu]
  - \left[ b(N+\mu)+\quarter (a+b)^2 \right]t
  - \half N^2 -\half N(\mu+a-b)-\quarter\mu(a-b)+\quarter b(a+b)  ,
\end{equation}
and the shifted variants. \hfill $\square$

One can easily verify that (\ref{UJ_recur}) is invariant under the transformations
$ a \leftrightarrow b $, $ t \mapsto 1-t $ and 
$ U_N^{\rm J} \mapsto -U_N^{\rm J} $.

\subsection{The CyUE and cJUE}
Comparison of the $ \tau$-function solution (\ref{5m}) with the average 
(\ref{1.15a}) shows, for $ N = 1 $
\begin{align}
  \tilde{E}_1^{\rm Cy}(t;(\eta_1,\eta_2),\mu;\xi)
  & \propto
    f^{\rm Cy}(-\mu, 2\eta_1 -\mu -1, \eta_1 +i\eta_2 -\mu;
                {1+it \over 2})
  \nonumber \\
  & = \tau_3 \Big(
      {1+it \over 2}; - \eta_1+1, i\eta_2, \eta_1, \eta_1 -\mu -1 \Big),
\end{align}
where in obtaining the second line use has been made of (\ref{abc}) with
$b_3 = - b_1$. 
For $ n \geq 1 $ comparison of (\ref{tca1}) with (\ref{1.15a}) allows us to 
deduce the analogue of Proposition \ref{pr6} for $\tilde{E}_N^{\rm Cy}$.

\begin{prop}\label{p12}
Let $\tau_3[n](t) = \tau_3(t;b_1,b_2,b_3+n,b_4) \Big |_{b_1+b_3=0}$
refer to the $\tau$-function sequence (\ref{2.35a}) with $\tau_3[0] = 1$ 
and $\tau_3[1](t)$ given by (\ref{3.03}). Then we have
\begin{equation}\label{Etc}
  \tilde{E}_N^{\rm Cy}(t;(\eta_1,\eta_2),\mu;\xi) 
  = C \tau_3({it +1 \over 2};\mathbf{b}), \quad
    \mathbf{b} = ( N-\eta_1, i\eta_2, \eta_1, -\mu+\eta_1-N )
\end{equation}
and consequently
\begin{equation}\label{Et1c}
  (t^2+1){d \over dt} \log\Big(
         (it-1)^{e_2'[\mathbf{b}] - {1 \over 2} e_2[\mathbf{b}]}
         (it+1)^{{1 \over 2} e_2[\mathbf{b}]}
         \tilde{E}_N^{\rm Cy}(t;(\eta_1,\eta_2),\mu;\xi) \Big)
  = U_N^{\rm Cy}(t;(\eta_1,\eta_2),\mu;\xi)
\end{equation}
where $U_N^{\rm Cy}(t;(\eta_1,\eta_2),\mu;\xi)$ satisfies the transformed
Jimbo-Miwa--Okamoto $\sigma$-form of the \PVI equation (\ref{iv}) with 
$\mathbf{b}$ specified as in (\ref{Etc}), and $e_2'[\mathbf{b}]$,
$e_2[\mathbf{b}]$ defined as in Proposition \ref{p1}. In the case $\xi=1$
we have the boundary condition
\begin{equation}\label{bcC}
\begin{split}
  U_N^{\rm Cy}(t;(\eta_1,\eta_2),\mu;1) 
  & \: \mathop{\sim}\limits_{t \to -\infty} \:
    \Big( e_2'[\mathbf{b}] + N(N+\mu-2\eta_1) \Big)t
  \cr
  & \: \mathop{\sim}\limits_{t \to -\infty} \:
    -\eta_1^2 t - {\eta_2(N-\eta_1)(N+\mu-\eta_1) \over \eta_1} + O(1/t)
\end{split}
\end{equation}
while in the case $\xi=0$
\begin{equation}\label{bcC1}
\begin{split}
  U_N^{\rm Cy}(t;(\eta_1,\eta_2),\mu;0)
  & \: \mathop{\sim}\limits_{t \to - \infty} \:
    \Big( e_2'[\mathbf{b}] + N \mu \Big)t + O(1)
  \cr
  & \: \mathop{\sim}\limits_{t \to - \infty} \:
    -(N-\eta_1)^2 t + O(1)
\end{split}
\end{equation}
\end{prop}

We see from (\ref{Et1c}) that with $\eta_2=0$ we have $b_2=0$. This in
turn implies that $e_2[\mathbf{b}] = e_2'[\mathbf{b}]$ and so
(\ref{Et1c}) reduces to
\begin{equation}\label{Et1e}
  (t^2+1){d \over dt} \log\Big( (t^2+1)^{e_2'[\mathbf{b}]/2}
         \tilde{E}_N^{\rm Cy}(t;(\eta_1,0),\mu;\xi) \Big)
  = U_N^{\rm Cy}(t;(\eta_1,0),\mu;\xi).
\end{equation}
That this with $\mu=0$, $\eta_1=N+a$, $\mathbf{b} = (-a,0,N+a,a)$
satisfies (\ref{iv}) is in precise agreement with the fact noted earlier
that (\ref{iv}) with the substitution (\ref{iva}) gives the
equation (\ref{yy2}) satisfied by (\ref{yy1}).

According to (\ref{1.28a}), we can write down from Proposition \ref{p12} an
analogous result for 
$\tilde{E}_N^{\rm cJ}(\phi;(\omega_1,\omega_2),\mu;\xi)$. Using the fact that 
for $\phi \to 0$ this quantity tends to a constant, and evaluating the latter 
in terms of $M_N(a,b)$ as specified by (\ref{ju1}), we have
\begin{multline}\label{3.35d}
  \tilde{E}_N^{\rm cJ}(\phi;(\omega_1,\omega_2),\mu;\xi^*) 
  = {M_N(\omega_1-i\omega_2+\mu/2,\omega_1+i\omega_2+\mu/2) \over 
     M_N(\omega_1,\omega_1)}
  \\ \times 
     \exp \Big\{ -{1 \over 2} \int_0^\phi \Big( 
      U_N^{\rm Cy}(\cot {\theta \over 2}; (N+\omega_1+\mu/2,\omega_2),\mu;\xi) 
      + \omega_2(N+\omega_1-\half\mu)
      + (\omega_1+\half\mu)^2\cot{\theta \over 2} \Big) d\theta \Big\}.
\end{multline}
In the case $\omega_2=0$ the average $\tilde{E}^{\rm cJ}_N$ satisfies a 
functional relation which according to (\ref{3.35d}) must be related to a 
corresponding functional relation of $U_N^{\rm Cy}$. Thus, in the case 
$\omega_2=0$ the integrand of the multi-dimensional integral specifying
$\tilde{E}_N^{\rm cJ}$ is periodic, which in turn allows the change of
variable $\theta_l \mapsto \theta_l - \phi$. From the latter it follows
\begin{equation}\label{11.s}
  \tilde{E}_N^{\rm cJ}(\phi;(\omega_1,0),2\mu;\xi) =
  \tilde{E}_N^{\rm cJ}(\phi;(\mu,0),2\omega_1;\xi).
\end{equation}
To exhibit this symmetry in (\ref{3.35d}), we note from (\ref{Etc})
that in the special case $\omega_2=0$ we must substitute in (\ref{3.35d})
\begin{equation}\label{11.s1}
  \mathbf{b} = (-(\omega_1+\mu/2),0,N+\omega_1+\mu/2,\omega_1-\mu/2), \quad
  e_2'[\mathbf{b}] = e_2[\mathbf{b}], \quad
  e_2'[\mathbf{b}] + N\mu = - (\omega_1+\mu/2)^2.
\end{equation}
Replacing $\mu$ by $2 \mu$ as required by the left hand side of (\ref{11.s})
we see that all quantities in (\ref{11.s1}) are then symmetric in  
$\mu$ and $\omega_1$ as required by the right hand side of (\ref{11.s}),
except for the component $b_4 = \omega_1 - \mu/2$, which is antisymmetric
under these operations. However we see the equation (\ref{4}) in the case
$b_2=0$ is unchanged by the mapping $b_4 \mapsto - b_4$ and so indeed
(\ref{3.35d}) is consistent with (\ref{11.s}).

With $\xi^*$ specified as in (\ref{1.20a}), the average defining 
$ \tilde{E}_N^{\rm cJ}(\phi;(\omega_1,\omega_2),2\mu;\xi^*)$ is proportional 
to the CUE average
\begin{equation}\label{12a}
  \Big\langle \prod_{l=1}^N(1-\xi^*\chi^{(l)}_{(\pi-\phi,\pi)})
  e^{\omega_2 \theta_l}|1+z_l|^{2\omega_1} \Big( {1 \over t z_l} \Big)^\mu
  (1+t z_l)^{2\mu} \Big\rangle_{\rm CUE} \Big|_{t = e^{i \phi}}
  =: A_N(t;\omega_1,\omega_2,\mu;\xi^*)\Big|_{t = e^{i \phi}}.
\end{equation}
It then follows from (\ref{3.35d}) that
\begin{multline}\label{12a'}
  - it{d \over dt} \log A_N(t;\omega_1,\omega_2,\mu;\xi^*) \\
  \quad = {1 \over 2} \Big( 
    U_N^{\rm Cy} \big( i{t+1 \over t-1}; (N+\omega_1+\mu,\omega_2), 2\mu;\xi \big)
           + i (e_2'[\mathbf{b}] - e_2[\mathbf{b}])
           - i {t+1 \over t-1}(e_2'[\mathbf{b}] + 2N\mu) \Big)
\end{multline}
where $U_N^{\rm Cy}$ satisfies (\ref{iv}) with
\begin{equation}\label{12a1}
  \mathbf{b} = (-(\omega_1+\mu),i\omega_2, N+\omega_1+\mu,\omega_1 -\mu)
   =: \mathbf{b}^{\rm Cy}.
\end{equation}
But, in the case $\xi^* = 0$, the same average (\ref{12a}) results from
(\ref{1.41g}) upon introducing a factor of $t^{-\mu N/2}$ into the average and 
making the replacements
\begin{equation*}
  b \mapsto 2\omega_1, \qquad \mu \mapsto 2\mu, \qquad 
  a \mapsto - (N+\omega_1+i\omega_2+\mu).
\end{equation*}
The quantity $U_N^{\rm J}$ in (\ref{1.41g}) then satisfies (\ref{4}) with
\begin{equation}\label{12a2}
  \mathbf{b} = \Big( {1 \over 2}(N+\omega-\mu), 
                     \bar{\omega} + {1 \over 2}(N+\omega+\mu),
                     {1 \over 2}(N-\omega+\mu),
                      -\mu-{1 \over 2}(N+\omega+\mu) \Big), \quad
  \omega = \omega_1 +i\omega_2.
\end{equation}
It follows that
\begin{equation}\label{12a3}
  t(t-1){d \over dt} \log A_N(t;\omega_1,\omega_2,\mu;0) = 
  U_N^{\rm J}(t; -(N+\omega_1+i\omega_2+\mu),2\mu,2\omega_1) - C_1t + C_2
\end{equation}
where
\begin{equation}\label{ccb}
  C_1 = e_2'[\mathbf{b}] + \mu N, \qquad
  C_2 = {1 \over 2} e_2[\mathbf{b}] + \mu N.
\end{equation}
Comparing (\ref{12a'}) and (\ref{12a3}) shows
\begin{multline}\label{12a4}
  i{(t-1) \over 2} \Big( 
     U_N^{\rm Cy} \big( i{t+1 \over t-1};(N+\omega_1+\mu,\omega_2),2\mu;0 \big)
     + i(e_2'[\mathbf{b}] - e_2[\mathbf{b}])
     - i{t+1 \over t-1}(e_2'[\mathbf{b}] + 2N\mu) \Big) \\ 
   \qquad = U_N^{\rm J}(t; -(N+\omega_1+i\omega_2+\mu),2\mu,2\omega_1) - C_1t + C_2.
\end{multline}
In fact (\ref{12a4}) is a special case of the following transformation
property of (\ref{4}).

\begin{prop}\label{pct}
Let $h$ satisfy (\ref{4}), and put
\begin{equation}\label{12a5}
h(t) =  i (t-1) {1 \over 2} f \Big ( i {t + 1 \over t - 1} \Big )
- {1 \over 4} (b_1^2  + b_2^2 + b_3^2 + b_4^2) t.
\end{equation}
Then $f(s)$ satisfies (\ref{iv}) with $\mathbf{b} \mapsto \bar{\mathbf{b}}$,
where
\begin{equation}\label{12a6}
\begin{array}{cc}
  \bar{b}_1 = {1 \over 2}(b_1 - b_2 + b_3 + b_4), \quad &
  \bar{b}_2 = {1 \over 2}(b_1 - b_2 - b_3 - b_4)  \\
  \bar{b}_3 = {1 \over 2}(b_1 + b_2 + b_3 - b_4), \quad &
  \bar{b}_4 = {1 \over 2}(b_1 + b_2 - b_3 + b_4)
\end{array}
\end{equation}
(or any permutation of these values).
\end{prop}

\noindent
Proof. \quad Substituting (\ref{12a5}) in (\ref{4}), changing variables to
\begin{equation*}
s =  i {t+1 \over t - 1}
\end{equation*}
and equating like terms with (\ref{iv}) modified so that
$h \mapsto f$ and $b \mapsto \bar{b}$, shows the statement of the proposition
is correct provided
\begin{align}\label{12a7}
  \bar{b}_1 \bar{b}_2 \bar{b}_3 \bar{b}_4
  & = {1 \over 16}( e_1[\mathbf{b}^2])^2 - {1 \over 2} b_1 b_2 b_3 b_4
       - {1 \over 4} e_2[\mathbf{b}^2] \nonumber \\
  -e_2[\bar{\mathbf{b}}^2]
  & = -{3 \over 8}( e_1[\mathbf{b}^2])^2 - 3 b_1 b_2 b_3 b_4
       + {1 \over 2} e_2[\mathbf{b}^2] \nonumber \\
  -e_3[\bar{\mathbf{b}}^2]
  & = - {1 \over 2}e_1[\mathbf{b}^2] b_1 b_2 b_3 b_4
      - {1 \over 16}(e_1[\mathbf{b}^2])^2
      + {1 \over 4} e_1[\mathbf{b}^2]e_2[\mathbf{b}^2] - e_3[\mathbf{b}^2].
\end{align}
Direct substitution of (\ref{12a6}) into these equations verifies their
validity. \hfill $\square$

We can check immediately that $\mathbf{b}^{\rm Cy}$ in (\ref{12a1}) is
related to $\mathbf{b}$ as in (\ref{12a2}) according to 
(\ref{12a6}), and furthermore that
\begin{equation*}
{(t-1) \over 2}(e_2'[\mathbf{b}^{\rm Cy}] - 
e_2[\mathbf{b}^{\rm Cy}]) + {1 \over 2} (1 + t) (
e_2'[\mathbf{b}^{\rm Cy}] + 2N\mu) + C_1 t -C_2 + {1 \over 4}(
b_1^2 + b_2^2 + b_3^2 + b_4^2) t = 0,
\end{equation*}
which together show (\ref{12a4}) follows from Proposition \ref{pct}.

An analogous result of the difference equation in $ \mu $ for the Jacobi case
(\ref{UJ_recur}) holds also for the Cauchy Jimbo-Miwa-Okamoto $\sigma$-function
$ U_N^{\rm Cy}(t;(\eta_1,\eta_2),\mu;\xi) $.

\begin{prop}
The Jimbo-Miwa-Okamoto $\sigma$-function $ U_N^{\rm Cy}(t;(\eta_1,\eta_2),\mu;\xi) $ 
satisfies a third order difference equation in $ \mu $, for all 
$ N, \eta_1, \eta_2, t $
\begin{multline}
  \left[ (N+\mu+1-2\eta_1)U-(N+\mu-2\eta_1)\uU + \eta^2_1 t
         + \eta_2(N-\eta_1) \right]
  \\ \times
  \left[ (N+\mu+1)U-(N+\mu)\uU + \eta^2_1 t
         - \eta_2(N-\eta_1) \right]
  \\ \times
  \left[ \uU-\dU+2(N+\mu-\eta_1)t \right]
  \left[ \uuU-U+2(N+\mu+1-\eta_1)t \right]
  \\ = (1+t^2) \Big[
    (N+\mu)(N+\mu-2\eta_1)(\uU-\dU) - 2(N+\mu-\eta_1)U
    - 2\eta^2_1(N+\mu-\eta_1)t - 2\eta_1\eta_2(N-\eta_1) \Big]
  \\ \times \Big[
    (N+\mu+1)(N+\mu+1-2\eta_1)(\uuU-U) - 2(N+\mu+1-\eta_1)\uU
    - 2\eta^2_1(N+\mu+1-\eta_1)t - 2\eta_1\eta_2(N-\eta_1) \Big] ,
\label{UCy_recur}
\end{multline}
where
$   U := U_{N}^{\rm Cy}(t;(\eta_1,\eta_2),\mu;\xi), 
  \dU := U_{N}^{\rm Cy}(t;(\eta_1,\eta_2),\mu-1;\xi),
  \uU := U_{N}^{\rm Cy}(t;(\eta_1,\eta_2),\mu+1;\xi) $ etc.
\end{prop}

\noindent
Proof. \quad This follows from (\ref{dPV_Ham}) using the parameters in
(\ref{Etc}) with $ b_{3} \leftrightarrow b_{4} $ and the invariance of $ h $ 
or $ U $ under this interchange, and the relation
\begin{equation}
    U_N^{\rm Cy}(t;(\eta_1,\eta_2),\mu;\xi) = K[\mu]
  - \left[ (N-\eta_1)^2+\mu N \right]t
  + \half\left[ (N-\eta_1)^2+\mu N+i\eta_2(\mu-\eta_1) \right] ,
\end{equation}
and the shifted variants. Finally one has to transform the variables
$ t \mapsto (1+it)/2 $ and $ U_N^{\rm Cy} \mapsto iU_N^{\rm Cy}/2 $.
\hfill $\square$

\subsection{Duality relations}
In our previous studies of the average (\ref{a4}) in the Gaussian and
Laguerre cases we have exhibited a duality in $\mu$ and $N$ which
in fact extends to random matrix ensembles in which the exponent
$2$ in the product of differences (\ref{a1}) is replaced by a continuous
parameter $\beta$. The same holds true of the average (\ref{ju}).
Let us define by C$\beta$E${}_N$ the eigenvalue PDF proportional to
\begin{equation}\label{cb}
\prod_{1 \le j < k \le N} | z_k - z_j |^\beta, \qquad
z_j = e^{i \theta_j}, \quad - \pi < \theta_j \le \pi.
\end{equation}
When $\beta = 2$ this is the CUE, while the cases $\beta = 1$ and
$\beta = 4$ are known in the random matrix literature as the COE
(circular orthogonal ensemble) and CSE
(circular symplectic ensemble) respectively.
Similarly, let us define by J$\beta$E${}_n$ the eigenvalue PDF
proportional to
\begin{equation*}
\prod_{l=1}^n x_l^{\lambda_1}(1 - x_l)^{\lambda_2}
\prod_{1 \le j < k \le n} |x_k - x_j|^\beta, \qquad
0 \le x_k \le 1.
\end{equation*}
In a previous study
the two particle distribution functions for the ensembles
C$\beta$E${}_N$ in the cases
$\beta$ even have been expressed as a $\beta$-dimensional integral
\cite{Fo-94}, with the latter having the form of an average in the
ensemble J$(4/\beta)$E${}_\beta$.
Generalising the derivation of this identity (see section
\ref{sgh}) gives
\begin{equation}\label{1.31}
\Big \langle \prod_{l=1}^N z_l^{(\eta_1-\eta_2)/2}
|1+z_l|^{\eta_1+\eta_2}
(1 + t z_l)^m \Big \rangle_{{\rm C}\beta{\rm E}_N}
\propto
\Big \langle \prod_{l=1}^m (1 - (1 - t)x_l)^N \Big \rangle_{{\rm J}
(4/\beta) {\rm E}_m}
\Big |_{\lambda_1 = 2(\eta_2-m+1)/\beta - 1 \atop
\lambda_2 = 2(\eta_1 +1)/\beta - 1}.
\end{equation}
In the case $\beta = 2$ it then follows from (\ref{ju}) that
for $\mu \in \zz_{\ge 0}$
\begin{equation}\label{1.31b}
\Big \langle \prod_{l=1}^N z_l^{(\eta_1-\eta_2)/2}
|1+z_l|^{\eta_1+\eta_2}
(1 + t z_l)^\mu \Big \rangle_{{\rm CUE}_N}
\propto
\Big \langle \prod_{l=1}^\mu z_l^{(\eta_1+2\eta_2)/2}
|1+z_l|^{\eta_1}
(1 + (1-t) z_l)^N \Big \rangle_{{\rm CUE}_\mu}.
\end{equation}
As we have characterised the averages in (\ref{1.31b}) in terms of
solutions of the $\sigma$ form of the \PVI equation (\ref{4}), it must
be that both sides of (\ref{1.31b}) satisfy (\ref{4}) with the same
parameters $\mathbf{b}$.

To check this, we note that according to (\ref{1.41g})
\begin{multline}\label{wa}
  U_N^{\rm J}(t;-N-\eta_2,\mu,\eta_1+\eta_2) \\
  = t(t-1) {d \over dt}
  \log \Big( (t-1)^{e_2'[\mathbf{b}] - {1 \over 2} e_2[\mathbf{b}]}
  t^{{1 \over 2} e_2[\mathbf{b}]} \Big \langle
  \prod_{l=1}^N z_l^{(\eta_1 - \eta_2)/2} |1 + z_l|^{\eta_1 + \eta_2}
  (1 + t z_l)^\mu \Big \rangle_{\rm CUE_{N}} \Big)
\end{multline}
satisfies (\ref{4}) with
\begin{equation*}
\mathbf{b}= \Big ( {N+\eta_2 - \mu \over 2}, \eta_1 + {N+\eta_2+\mu
\over 2}, {N - \eta_2 + \mu \over 2}, - {N + \eta_2 + \mu \over 2}
\Big ) =: (b_1, b_2, b_3, b_4).
\end{equation*}
Now the right hand side of (\ref{1.31b}) is obtained from the left hand side of
(\ref{1.31b}) by the mappings
\begin{equation*}
t \mapsto 1-t, \quad \mu \leftrightarrow N, \quad
\eta_1 \mapsto \eta_1 + \eta_2, \quad \eta_2 \mapsto - \eta_2
\end{equation*}
which when applied to (\ref{wa}) tells us that
\begin{multline}\label{wa1}
  U_N^{\rm J}(1-t;-\mu+\eta_2,N,\eta_1) \\
  = - t(1-t) {d \over dt} \log \Big( 
       (t-1)^{{1 \over 2}e_2[\mathbf{b}]}
        t^{e_2'[\mathbf{b}] - {1 \over 2} e_2[\mathbf{b}]}
      \Big \langle
      \prod_{l=1}^N z_l^{(\eta_1 - \eta_2)/2} |1+z_l|^{\eta_1 +\eta_2}
                   (1+t z_l)^\mu \Big \rangle_{\rm CUE_{N}} \Big)
\end{multline}
satisfies (\ref{4}) with $\mathbf{b}$ replaced by
\begin{equation*}
\tilde{\mathbf{b}}:= (-b_1, b_2, -b_4, - b_3).
\end{equation*}
But we have already commented (recall the paragraph including (\ref{3.12}))
that $-h(1-t)$ satisfies (\ref{4}) with the sign of an odd number of the
$b$'s reversed. Furthermore (\ref{4}) is symmetric in the $b$'s so we
deduce $t(t-1)$ times the logarithmic derivative of the averages in
(\ref{wa}) and (\ref{wa1}) satisfy the same equation provided
\begin{equation*}
t e_2'[\mathbf{b}] - {1 \over 2} e_2[\mathbf{b}] =
t e_2'[\tilde{\mathbf{b}}] -  e_2'[\tilde{\mathbf{b}}]
+{1 \over 2} e_2[\tilde{\mathbf{b}}],
\end{equation*}
which is readily verified.

Another manifestation of the duality relation is that for $\xi = 0$ and
$\mu$ a positive integer, $\tilde{E}_N^{\rm J}(t;a,b,\mu;\xi)$ as specified
by (\ref{1.17a}) can be written as a $\mu \times \mu$ determinant.

\begin{prop}\label{p14}
Let $\tau_3[\mu](t)$, $\mu \in \zz_{\ge 0}$, denote the $\tau$-function
sequence (\ref{ob}) with
\begin{equation*}
\tau_3[1](t) \propto {}_2 F_1(-N, N+a+b+1;1+a;t).
\end{equation*}
Then
\begin{equation*}
\tilde{E}_N^{\rm J}(t;a,b,\mu;0) \propto \tau_3[\mu](t)
\propto t^{-\mu(\mu - 1)/2}
\det \Big [ {}_2 F_1(-N-j,N+a+b+1+k;1+a;t) \Big ]_{j,k=0,\dots,\mu-1}.
\end{equation*}
\end{prop}

\noindent
Proof. \quad When $\mu = 1$ we have
\begin{equation}\label{3.55a}
\tilde{E}_N^{\rm J}(t;a,b,1;0) = \Big \langle \prod_{l=1}^N(t - x_l)
\Big \rangle_{\rm JUE} \propto P_N^{(a,b)}(1-2t)
\propto {}_2 F_1(-N, N+a+b+1;1+a;t)
\end{equation}
where the first proportionality follows from a standard result in
\cite{ops_Sz} ($P_N^{(a,b)}$ denotes the Jacobi polynomial of degree $N$).
Recalling Proposition \ref{pr3}, we therefore have
\begin{equation*}
\tilde{E}_N^{\rm J}(t;a,b,1;0) \propto \tau_3[1](t;\mathbf{b}), \quad
\mathbf{b} = \Big ( - N - {1 \over 2}(a+b), -{1 \over 2} (b-a),
N + {1 \over 2} (a+b)+1, {1 \over 2} (a+b) \Big ).
\end{equation*}
Thus
\begin{equation}\label{d.4}
 \tau_3[\mu](t) = \tau_3(t;\mathbf{b}), \quad
\mathbf{b} = (-\hat{b}_1, - \hat{b}_2, - \hat{b}_4, - \hat{b}_3)
\end{equation}
where $\hat{\mathbf{b}} = (\hat{b}_1,  \hat{b}_2,  \hat{b}_3,  \hat{b}_3)$
is specified by (\ref{3.22a}). The operation of reversing the signs of
all the $\hat{b}$'s and interchanging $\hat{b}_3$ and $\hat{b}_4$ leaves
$e_2'[\hat{\mathbf{b}}]$, $e_2[\hat{\mathbf{b}}]$ and the differential
equation (\ref{4}) unchanged. It follows from (\ref{3}) that
$\tau_3(t;\mathbf{b})$ itself is unchanged, so in fact
\begin{equation*}
\tau_3[\mu](t) = \tau_3(t;\hat{\mathbf{b}}),
\end{equation*}
which is just the equation (\ref{3.22a}). That the case $\xi = 0$ of
(\ref{3.22a}) is singled out by (\ref{d.4}) follows from the latter
being a polynomial in $t$. \hfill $\square$

\subsection{Relationship to generalised hypergeometric functions}\label{sgh}
In this section we will show that $\tilde{E}_N^{\rm J}(s;a,b,\mu)$ can be
identified as an integral representation of a generalised hypergeometric
function \cite{Ka_1993}
based on Jack polynomials evaluated at a special point. To present
this theory requires some notation. Let $\kappa := (\kappa_1,\dots,
\kappa_N)$ denote a partition so that $\kappa_i \ge \kappa_j$
$(i < j)$ and $\kappa_i \in \zz_{\ge 0}$. Let $m_\kappa$ denote the
monomial symmetric function corresponding to the partition $\kappa$
(e.g.~if $\kappa = (2,1)$ then $m_\kappa = z_1^2z_2 + z_1 z_2^2$), and for
partitions $|\kappa| = |\mu|$ define the dominance partial ordering by the
statement that $\kappa > \mu$ if $\kappa \ne \mu$ and
$\sum_{j=1}^p \kappa_j \ge \sum_{j=1}^p \mu_j$ for each $p=1,\dots,N$.
Introduce the Jack polynomial $P_\kappa^{(1/\alpha)}(z_1,\dots,z_N) =:
P_\kappa^{(1/\alpha)}(z)$ as the unique homogeneous polynomial of degree
$|\kappa|$ with the structure
\begin{equation*}
P_\kappa^{(1/\alpha)}(z) = m_\kappa + \sum_{\mu < \kappa} a_{\kappa \mu}
m_\mu
\end{equation*}
(the $a_{\kappa \mu}$ are some coefficients in $\qq(\alpha)$) and which
satisfy the orthogonality
\begin{equation*}
\langle P_\kappa^{(1/\alpha)}, P_\rho^{(1/\alpha)} \rangle^{(\alpha)}
\propto \delta_{\kappa, \rho}
\end{equation*}
where
\begin{equation*}
\langle f, g \rangle^{(\alpha)} :=
\int_{-1/2}^{1/2} dx_1 \cdots \int_{-1/2}^{1/2} dx_N \,
\overline{f(z_1,\dots,z_N)} g(z_1,\dots,z_N)
\prod_{1 \le j < k \le N} | z_k - z_j |^{2 \alpha}, \quad
z_j := e^{2 \pi i x_j}.
\end{equation*}
We remark that when $\alpha = 1$ the Jack polynomial coincides with the
Schur polynomial. Introduce the generalised factorial function
\begin{equation*}
[u]_{\kappa}^{(\alpha)} =
\prod_{j=1}^N {\Gamma(u - (j-1)/\alpha + \kappa_j) \over
\Gamma(u - (j-1)/\alpha)}.
\end{equation*}
Let
\begin{equation*}
d_\kappa' = \prod_{(i,j) \in \kappa}
\Big ( \alpha (a(i,j) + 1) + l(i,j)  \Big ),
\end{equation*}
where the notation $(i,j) \in \kappa$ refers to the diagram of $\kappa$, in
which each part $\kappa_i$ becomes the nodes $(i,j)$,
$1 \le j \le \kappa_i$ on a square lattice labelled as is conventional
for a matrix. The quantity $a(i,j)$ is the so called arm length (the number of
nodes in row $i$ to the right of column $j$), while $l(i,j)$ is the leg
length (number of nodes in column $j$ below row $i$). Define the
renormalised Jack polynomial
\begin{equation}\label{cpc}
C_\kappa^{(\alpha)}(z) := {\alpha^{|\kappa|} | \kappa|! \over d_\kappa'}
P_\kappa^{(\alpha)}(z).
\end{equation}
Then the generalised hypergeometric function ${}_p^{} F_q^{(\alpha)}$
based on the Jack polynomial (\ref{cpc}) is specified by the series
\begin{equation}\label{cpc1}
  {}_p^{} F_q^{(\alpha)}(a_1,\dots,a_p;b_1,\dots,b_q;z)
  := \sum_{\kappa} {1 \over |\kappa|!}
    {[a_1]_\kappa^{(\alpha)} \cdots [a_p]_\kappa^{(\alpha)} \over
     [b_1]_\kappa^{(\alpha)} \cdots [b_q]_\kappa^{(\alpha)} }
     C_\kappa^{(\alpha)}(z)
\end{equation}
(when $N=1$ this reduces to the classical definition of ${}_p F_q$).

The relevance of the generalised hypergeometric functions to the present
study is that it was shown in \cite{Fo-93} that
\begin{align}\label{iti}
  & {1 \over C} 
     \int_0^1 dx_1 \, x_1^{\lambda_1} (1-x_1)^{\lambda_2} (1-t x_1)^{-r} \cdots 
     \int_0^1 dx_N \, x_N^{\lambda_1} (1-x_N)^{\lambda_2} (1-t x_N)^{-r}
     \prod_{1 \leq j<k \leq N} | x_j - x_k |^{2/\alpha}
  \nonumber \\ \qquad
  & := \Big\langle \prod_{l=1}^N (1-t x_l)^{-r}
       \Big\rangle_{{\rm J}(2/\alpha){\rm E}_N}
  \nonumber \\
  &  = {}_2^{} F_1^{(\alpha)}(r, {1 \over \alpha}(N-1)+\lambda_1 +1;
                    {2 \over \alpha}(N-1) +\lambda_1 +\lambda_2 +2;
                    t_1,\dots, t_N) \Big|_{t_1 = \cdots = t_N = t}
\end{align}
In the case $\alpha = 1$, $\lambda_1 = a$, $\lambda_2 = \mu$, $r= - b$, the
integrand in (\ref{iti}) coincides with the integral in the second equality
of (\ref{E1}). Hence, by Proposition \ref{pr6} we have that
\begin{equation}\label{r3d}
\tau_3(t;\mathbf{b}) =  {}_2^{} F_1^{(1)}(-b, N+a; 2N + a + \mu;
t_1,\dots,t_N) \Big |_{t_1 = \cdots = t_N = t}
\end{equation}
where $\mathbf{b}$ is specified by (\ref{Et})
and $\tau_3$ is normalised so that $\tau_3(0;\mathbf{b}) = 1$. 
This, after inserting the known value of $P_\kappa^{(1)}$
evaluated with all arguments equal in (\ref{cpc1}), can seen to be
in agreement with a conjecture of Noumi et al.~\cite{NOOU_1998} subsequently 
proved by Taneda \cite{Ta_2001} using a different argument to that here. 

For future reference we note that for $-r =: m$, $m \le N$,
a non-negative integer, a result of Kaneko \cite{Ka_1993} gives that the
average in (\ref{iti}) has the alternative generalised hypergeometric
function evaluation
\begin{multline}\label{itj}
  \Big\langle \prod_{l=1}^N (1-t x_l)^m \Big\rangle_{{\rm J}(2/\alpha){\rm E}_N}
  \\
  = {}_2^{} F_1^{(1/\alpha)}(
              -N,-(N-1)-\alpha(\lambda_1+1);
              -2(N-1)-\alpha(\lambda_1 + \lambda_2 + 2); t_1, \dots, t_m)
      \Big|_{t_1 = \cdots = t_m = t}
  \\
  \propto {}_2 F_1^{(1/\alpha)}(-N,-(N-1)-\alpha(\lambda_1+1);
                 \alpha(\lambda_2 + m); 1 - t_1, \dots, 1 - t_m)
            \Big|_{t_1 = \cdots = t_m = t}
\end{multline}
where the final formula follows from an identity in
\cite{Fo-93}. Note that the role of $m$ and $N$ is interchanged relative
to (\ref{iti}). In fact it is the equality of (\ref{iti}) and
(\ref{itj}) which implies the duality relation (\ref{1.31}).
Also from \cite{Ka_1993} we have that
\begin{equation}\label{r3aa}
  \Big\langle \prod_{l=1}^N (t-x_l)^m \Big\rangle_{{\rm J}
  (2/\alpha){\rm E}_N} \propto
  {}_2^{} F_1^{(1/\alpha)}(-N, \alpha(\lambda_1 +\lambda_2 +m+1)+N-1;
  \alpha(\lambda_1 + m);t_1,\dots,t_m) \Big|_{t_1= \cdots = t_m = t}.
\end{equation}
In the case $\alpha = 1$, and with $\lambda_1=a$, $\lambda_2=b$,
$m=\mu$ this coincides with the definition (\ref{1.17a}) of
$\tilde{E}_N^{\rm J}(t;a,b,\mu;0)$. Hence from Proposition \ref{p9} we have a
formula similar to (\ref{r3d}) valid for $ \mu, N \in \zz_{\ge 0}$
\begin{equation}\label{r3ab}
  \tau_3(t;\hat{\mathbf{b}}) \propto
  {}_2^{} F_1^{(1)}(-N, N+a+b+\mu; a+\mu;t_1,\dots,t_{\mu})
          \Big|_{t_1= \cdots = t_{\mu} = t}.
\end{equation}

Results relating
generalised hypergeometric functions to $\tau$-functions can also be found in 
\cite{AvM_2001a}.

\section{\PV scaling limit}
\setcounter{equation}{0}
\subsection{Scaling limit of the cJUE at the spectrum singularity}
We now turn our attention to the scaled limit (\ref{vpu}). 
To anticipate that it is well defined, one notes that
the integrand defining the average has an interpretation in classical
statistical mechanics as a log-gas. 
Thus it can be written in the form of a Boltzmann
factor $e^{-\beta U}$ with inverse temperature $\beta = 2$ and potential
energy
\begin{equation}\label{Bf}
  U = - \sum_{j=1}^N \Big( {\omega_2 \over 2}\theta_l +\omega_1 \log|1+z_l|
      + \mu\log| e^{i(\pi - X/N)}-z_l| \Big)
      - \sum_{1 \le j < k \le N} \log|z_k-z_j|.
\end{equation}
The classical particle system corresponding to (\ref{Bf}) has $N$ identical
mobile charges interacting via a repulsive logarithmic potential, and subject
to the discontinuous external potential $-(\omega_2/2) \theta_l$
$(-\pi < \theta < \pi)$. The mobile charges also interact with impurity charges
at $\theta = \pi$ (of strength $\omega_1$) and at $\theta = \pi - X/N$ (of 
strength $\mu$). In random matrix language these impurities correspond to 
spectrum singularities of degenerate eigenvalues.
We anticipate (\ref{vpu}) to be well defined because changing
variables $\theta_l \mapsto \theta_l/N$ gives a system which has spacing
between particles of order unity, and $X$ is measured in units of this
spacing.

In the notation of (\ref{12a}), the average in (\ref{vpu}) is denoted
$A_N(t;\omega_1,\omega_2,\mu;\xi^*)$. It follows from (\ref{12a'}), (\ref{12a4})
and Proposition \ref{pct} that
\begin{equation*}
  t(t-1){d \over dt} \log A_N(t;\omega_1,\omega_2,\mu;\xi^*) 
  = \sigma(t) - C_1 t + C_2
\end{equation*}
where $\sigma(t)$ satisfies (\ref{4}) with $\mathbf{b}$ given by (\ref{12a2})
and $C_1$, $C_2$ given by (\ref{ccb}). Since with $t = e^{iX/N}$,
\begin{equation*}
  t(t-1){d \over dt} \: \sim \: X {d \over dX}
\end{equation*}
it follows from the definition of $u$ in (\ref{vpu}) that
\begin{equation*}
  \lim_{N \to \infty} \Big( (\sigma(t) - C_1 t + C_2) \Big|_{t = e^{iX/N}}
                      \Big ) = u(X;(\omega_1,\omega_2),\mu;\xi).
\end{equation*}
This limit can be taken directly in the differential equation
(\ref{4}).

\begin{prop}\label{pho}
Let $C_1, C_2$ be defined as in (\ref{ccb}), let $\mathbf{b}$ be given
by (\ref{12a2}), and replace $h$ in (\ref{4}) by $u + C_1 t - C_2$. Then
with the change of variable $t=e^{iX/N}$, as $N \to \infty$ the leading
terms of (\ref{4}) are of $O(N^2)$, and give that
\begin{equation}\label{vp3a}
  h(t) = u(it;(\omega_1,\omega_2),\mu;\xi)
         + {i \omega_2 \over 2}t + 2\omega_1 \mu + \omega_2^2/2
\end{equation}
satisfies (\ref{vp}) with
\begin{equation}\label{vp3}
  \tilde{v}_1 = \mu, \: \: \tilde{v}_2 = - \mu, \: \:
  \tilde{v}_3 = {\omega}, \: \: \tilde{v}_4 = - \bar{\omega}, \: \:
  \tilde{v}_j := v_j + {i\omega_2 \over 2}, \: \: \omega:= \omega_1 + i\omega_2.
\end{equation}
\end{prop}

\noindent
Proof. \quad Direct substitution with the change of variables $t = e^{iX/N}$
and expanding in $N$ for $N$ large shows that the leading order term is
proportional to $N^2$. Equating the coefficient of $N^2$ to zero gives
\begin{multline*}
  - \Big( X {d^2 u \over dX^2} \Big)^2
  - 4X \Big( {d u \over dX} \Big)^3 + 4u \Big( {d u \over dX} \Big)^2
  + [4(\mu + \omega_1)^2 - 4\omega_2 X - X^2] \Big( {d u \over dX} \Big)^2 \\
  + 4(\omega_2 + {X \over 2}) u {d u \over dX}
  + 8 \mu(\omega_1 \omega_2 + \mu \omega_2 + {\omega_1 \over 2}X) {d u \over d X}
  - u^2 - 4\omega_1 \mu u + (2 \mu \omega_2)^2 = 0.
\end{multline*}
Changing variables $X \mapsto i t$ and comparing the resulting equation with
(\ref{vp}) after the replacement $h_V \mapsto h + a_1 t + a_2$ shows that
the equations for $u(it)$ and $h$ coincide provided
\begin{equation*}
  a_1 =  {1 \over 2} i \omega_2, \qquad 
  a_2 = {1 \over 2} \omega_2^2 + 2 \omega_1 \mu
\end{equation*}
and the equations
\begin{equation}\label{4.56'}
  e_1[\tilde{\mathbf{v}}] = 2i\omega_2,  \quad  
  e_2[\tilde{\mathbf{v}}] = - (\mu^2 + \omega_1^2 + \omega_2^2) \quad
  e_3[\tilde{\mathbf{v}}] = -2i\mu^2 \omega_2,  \quad 
  e_4[\tilde{\mathbf{v}}] = \mu^2(\omega_1^2 + \omega_2^2)
\end{equation}
hold. Direct substitution verifies (\ref{4.56'}) is satisfied by
(\ref{vp3}).
\hfill $\square$

The boundary condition which (\ref{vp3a}) is to satisfy can be predicted
in the cases $\xi = 0$ and $\xi = 1$ from the log-gas interpretation of the
average in (\ref{vpu}). Consider first the case $\xi = 0$. Then the potential
energy function (\ref{Bf}) is being averaged over the whole circle. For
large $X$ (after the scaled limit has been taken), to leading order the
impurity charge originally at $z = e^{i(\pi - X/N)}$ on the circle is
expected to decouple from the impurity charge at $z=e^{i \pi}$, meaning
that the average in (\ref{vpu}) will be independent of $X$. However, for
this to happen the potential energy must be modified to include the term
$-2\omega_1 \mu \log | 1+ e^{-i X/N}|$ corresponding to the interaction between
the two impurity charges (see \cite{rmt_Fo} for similar applications of this
argument). This leads to the prediction
\begin{equation}
  u(X;(\omega_1,\omega_2),\mu;0) \: \mathop{\sim}\limits_{X \to \infty} \:
  - 2\omega_1 \mu
  \quad \text{and thus} \quad
  h(t) \: \mathop{\sim}\limits_{t \to -i\infty} \:
  {i \omega_2 \over 2}t + {\omega_2^2 \over 2}.
\end{equation}
In the case $\xi = 1$, the average (\ref{vpu}) corresponds to excluding
the particles from the interval $(\pi - X/N, \pi)$. To leading order one 
would expect the large $X$ behaviour to be independent of the parameters 
$\omega_1,\omega_2,\mu$ and thus to be given by the $\omega_1=\omega_2=\mu=0$ 
results \cite{Wi-1994,DIZ-1997} which implies
\begin{equation}
  u(X;(\omega_1,\omega_2),\mu;1) \: \mathop{\sim}\limits_{X \to \infty} \:
  - {X^2 \over 16} + O(1) \quad
  \text{and thus} \quad
  h(t) \: \mathop{\sim}\limits_{t \to -i\infty} \:
  {t^2 \over 16} + {i\omega_2 \over 2}t + O(1).
\end{equation}

\subsection{Multi-dimensional integral solutions}
As commented in the Introduction, the differential equation (\ref{vp}) is
the analogue of (\ref{4}) for the PV system. Specifically, one considers 
 \cite{Ok_1987b} a
Hamiltonian $H_{V}$ associated with the Painlev\'e V equation,
\begin{equation*}
  t H_{V} = q(q-1)^2 p^2 - \Big \{ (v_2 - v_1)(q-1)^2 - 2(v_1 + v_2) q (q-1)
  + tq \Big \}p + (v_3 - v_1)(v_4 - v_1)(q-1)
\end{equation*}
where the parameters $v_1,\dots, v_4$ are constrained by
\begin{equation*}
v_1 + v_2 + v_3 + v_4 = 0.
\end{equation*}
It is well known  that eliminating $p$ in the Hamilton equations
(\ref{1.1}) gives the general \PV equation with $\delta = -1/2$ in the
usual notation. Our interest is in the further fact that the auxiliary
Hamiltonian
\begin{equation*}
h_{V} = t H_V + (v_3 - v_1)(v_4 - v_1) - v_1 t - 2v_1^2
\end{equation*}
satisfies (\ref{vp}), which is the analogue of the fact that (\ref{3})
satisfies (\ref{4}). 
Using the Hamiltonian formalism, we have previously
constructed determinant/multi-dimensional integral solutions of (\ref{vp})
\cite{FW_2002a}. To make use of these results, we first note that with 
\begin{equation}\label{vpy}
h_V = \sigma - v_1 t - 2v_1^2
\end{equation}
it follows from (\ref{vp}) that the quantity $\sigma$ satisfies
\begin{equation}\label{vpx}
(t\sigma'')^2 - [\sigma - t \sigma' + 2 (\sigma')^2 +
(\nu_0 + \nu_1 + \nu_2 + \nu_3 ) \sigma']^2 +
4(\nu_0 + \sigma')(\nu_1 + \sigma')(\nu_2 + \sigma')
(\nu_3 + \sigma') = 0
\end{equation}
with
\begin{equation}\label{nuv}
\nu_0 = 0, \quad \nu_1 = v_2 - v_1, \quad \nu_2 = v_3 - v_1, \quad
\nu_3 = v_4 - v_1.
\end{equation}
Now, a family of multi-dimensional integral solutions of (\ref{vpx})
found in \cite[Prop.~3.1 together with (1.45)]{FW_2002a} states that
\begin{equation}\label{rl}
t {d \over dt} \log \Big( t^{an + n^2} \big\langle e^{-t \sum_{j=1}^n x_j}
\big\rangle_{{\rm JUE}_n} \Big) = U_n^{\rm L}(t;a,b)
\end{equation}
where $U_n^{\rm L}(t;a,b)$ satisfies (\ref{vpx}) with
\begin{equation}\label{rla}
\nu_0=0, \quad \nu_1=-b, \quad \nu_3 = n+a, \quad \nu_3=n.
\end{equation}

The relevance of (\ref{rl}) for the present purposes is that in the special
case $m \in \zz^+$ it follows from (\ref{1.31b}) that
\begin{equation}\label{vpu2}
\lim_{N \to \infty}
X {d \over dX} \log \Big \langle \prod_{l=1}^N z_l^{(\eta_1 - \eta_2)/2}
|1 + z_l |^{\eta_1 + \eta_2}(1 + e^{-X/N} z_l)^m
\Big \rangle_{{\rm CUE}_N} =
X {d \over dX} \log \big\langle e^{-X \sum_{j=1}^m x_j} \big\rangle_{{\rm JUE}_m}
\Big |_{\lambda_1 = \eta_2 - m \atop \lambda_2 = \eta_1}.
\end{equation}
We recognise the left hand side of (\ref{vpu2}) as a rewrite of the left hand
side of (\ref{vpu}) in the case $\xi=\xi^*$, where
$\xi^*=0$ for $m$ even and $\xi^*=2$ for $m$ odd.  
Consequently
\begin{equation}\label{4.12a}
X {d \over dX} \log \langle e^{-X \sum_{j=1}^m x_j}
\rangle_{{\rm JUE}_m} \Big |_{\lambda_1 =a \atop \lambda_2 = b} 
= u\Big (iX;({a+b+m \over 2}, i {b-a \over 2}), {m \over 2};\xi^* \Big ) 
- {m \over 2} X.
\end{equation}
It thus follows from (\ref{vpu2}) and Proposition \ref{pho} that
\begin{equation}\label{htt}
t {d \over dt} \log
\langle e^{-t \sum_{j=1}^n x_j}
\rangle_{{\rm JUE}_n} = h(t) + {1 \over 4} (b-a-2n) t -
{n \over 2}(a+b+n) + {1 \over 8}(b-a)^2
\end{equation}
where $h(t)$ satisfies (\ref{vpx}) with $v_1 = n/2 + (b-a)/4$ and
$\nu_0, \dots, \nu_3$ as in (\ref{rla}). Use of (\ref{vpy}) to
substitute for $h(t)$ in (\ref{htt}) reclaims the fact that
(\ref{rl}) satisfies (\ref{vpx}) with parameters (\ref{rla}).

\subsection{Generalised hypergeometric function expressions}
The generalised hypergeometric function ${}_2^{}F_1^{(\alpha)}$ has
the confluence property \cite{Ya_1992}
\begin{equation}\label{su}
\lim_{L \to \infty} {}_2^{}F_1^{(\alpha)}(L,b;c;x/L) =
{}_1 ^{}F_1^{(\alpha)}(b;c;x),
\end{equation}
which follows easily from the series definition (\ref{cpc1}). Applying 
(\ref{su}) to (\ref{iti}) with $t \mapsto -t/r$, $r \to \infty$ gives
\cite{Fo-93a}
\begin{equation}\label{su1}
\Big \langle \prod_{l=1}^n e^{-t x_l} \Big \rangle_{{\rm J}
(2/\alpha){\rm E}_n} =
{}_1^{} F_1^{(\alpha)} \Big ( {1 \over \alpha}(n-1) + \lambda_1 + 1;
{2 \over \alpha}(n-1) + \lambda_1 + \lambda_2 + 2; - t_1, \dots,
- t_n \Big ) \Big |_{t_1 = \cdots = t_n = t}.
\end{equation}
In the case $\alpha = 1$, $\lambda_1 = a$, $\lambda_2 = b$ we can
substitute (\ref{su1}) in (\ref{rl}) to deduce 
\begin{equation}\label{su2}
t {d \over dt} \log t^{an+ n^2} {}_1^{} F_1^{(1)}(n+a, 2n + a + b;
-t_1, \dots, - t_n) \Big |_{t_1 = \cdots = t_n = t} =
U_n^{\rm L}(t;a,b)
\end{equation}
where $U_n^{\rm L}(t;a,b)$ satisfies (\ref{vpx}) with
\begin{equation*}
\nu_0 = 0, \quad \nu_1 = -b, \quad \nu_2 = n+a, \quad \nu_3 = n.
\end{equation*}
Furthermore, applying (\ref{1.19a}) to the left hand side of (\ref{su1})
we deduce \cite{Fo-93a}
\begin{equation}\label{su2a}
\Big \langle \prod_{l=1}^n z_l^{(a'-b')/2} |1 + z_l |^{a'+b'}
e^{-t z_l}  \Big \rangle_{{\rm C}(2/\alpha){\rm E}_n}
\propto
{}_1^{} F_1^{(\alpha)} \Big ( -b';
{1 \over \alpha}(n-1) + a' + 1;  t_1, \dots,
 t_n \Big ) \Big |_{t_1 = \cdots = t_n = t}
\end{equation}

\section{Applications}
\setcounter{equation}{0}
\subsection{Eigenvalue distributions in ensembles with 
unitary symmetry}\label{s5.1}
Let a sequence of $N$ eigenvalues $x_1,\dots,x_N$ have joint distribution
$p_N(x_1,\dots,x_N)$ and let the support of these eigenvalues be an
interval $I$. Let $I_0 \subset I$, and consider the multi-dimensional
integral
\begin{equation}\label{7.1}
E_N(I_0;\xi) := \Big ( \int_I - \xi \int_{I_0} \Big ) dx_1 \cdots
\Big ( \int_I - \xi \int_{I_0} \Big ) dx_N \, p_N(x_1,\dots,x_N).
\end{equation}
We see immediately that the probability $E_{N,n}(I_0)$ of there being
exactly $n$ eigenvalues in the interval $I_0$ is related to
(\ref{7.1}) by
\begin{equation*}
E_{N,n}(I_0) = {(-1)^n \over n!} {\partial^n \over \partial \xi^n}
E_N(I_0;\xi) \Big |_{\xi = 1}.
\end{equation*}
Equivalently
\begin{equation}\label{5.1b}
\sum_{n=0}^\infty (1 - \xi)^n E_{N,n}(I_0) = E_N(I_0;\xi).
\end{equation}
In particular, in an obvious notation it follows from (\ref{1.17a})
and (\ref{pw4}) that
\begin{align}
  E_{N,n}^{\rm J}((s,1);a,b)
  & = {(-1)^n \over n!} {\partial^n \over \partial \xi^n}
       \tilde{E}_N^{\rm J}(s;a,b,0;\xi) \Big|_{\xi = 1}
  \label{7.2} \\
  E_{N,n}^{\rm cJ}((\pi - \phi, \pi);(\omega_1,\omega_2))
  & = {(-1)^n \over n!} {\partial^n \over \partial \xi^n}
       \tilde{E}_N^{\rm cJ}(\phi;(\omega_1,\omega_2),0;\xi) \Big|_{\xi = 1}.
  \label{7.3}
\end{align}
According to Proposition \ref{p9}, and the fact that with $s=1$,
$\tilde{E}_N^{\rm J}$ in (\ref{7.2}) equals unity, we have
\begin{equation}\label{5.4}
E_{N,n}^{\rm J}((s,1);a,b) = {(-1)^n \over n!}
{\partial^n \over \partial \xi^n}
\exp \Big \{ - \int_s^1\Big (\hat{U}_N^{\rm J}(t;a,b,0;\xi)
- e_2'[\hat{\mathbf{b}}] t + {1 \over 2} e_2[\hat{\mathbf{b}}]
\Big ) {dt \over t(t-1)} \Big \} \Big |_{\xi = 1}
\end{equation}
where $\hat{U}_N^{\rm J}$ is specified as in Proposition \ref{p9}
with boundary condition (\ref{1.34r}). The result (\ref{3.35d'}) 
with $\mu = 0$ gives an analogous evaluation of (\ref{7.3}).

Suppose now that $I_0 = (s,t)$. Then the quantity
\begin{equation}\label{7.4}
E_N(I_0;s;\xi) := (N+1)
\Big ( \int_I - \xi \int_{I_0} \Big ) dx_1 \cdots
\Big ( \int_I - \xi \int_{I_0} \Big ) dx_N \,
p_{N+1}(s,x_1,\dots,x_N)
\end{equation}
is related to the probability density $p_{N,n}(I_0,s)$ of there being an
eigenvalue at $s$ as well as $n$ eigenvalues in the interval
$I_0 = (s,t)$. Thus we have
\begin{equation*}
p_{N,n}(I_0,s) = {(-1)^n \over n!} {\partial^n \over \partial \xi^n}
E_N(I_0;s;\xi) \Big |_{\xi = 1}.
\end{equation*}
Now for the JUE with $I_0 = (s,1)$ the quantity (\ref{7.4}) is
proportional to $s^a(1-s)^b$ times $\tilde{E}_N^{\rm J}(s;a,b,2;\xi)$ as
specified by (\ref{1.17a}). 
Writing the proportionality in terms of the integral (\ref{nj}), it follows
from Proposition \ref{p9} that
\begin{multline}\label{5.6}
  p_{N,n}^{\rm J}((s,1);a,b)
  = (N+1) {J_N(a,b+2) \over J_{N+1}(a,b)} s^a (1-s)^b {(-1)^n \over n!} \\
  \times {\partial^n \over \partial \xi^n}
     \exp \Big\{ - \int_s^1\Big(
                   \hat{U}_N^{\rm J}(t;a,b,2;\xi)
                   - e_2'[\hat{\mathbf{b}}]t + {1 \over 2} e_2[\hat{\mathbf{b}}]
                           \Big) {dt \over t(t-1)}
          \Big\} \Big|_{\xi = 1}
\end{multline}
where $\hat{U}_N^{\rm J}$ satisfies the differential equation
specified as in Proposition \ref{p9}. The boundary condition which
$\hat{U}_N^{\rm J}$ must satisfy is given by the following result.

\begin{prop}\label{p18}
We have
\begin{equation}\label{fr0}
\hat{U}_N^{\rm J}(t;a,b,2;\xi) 
\: \mathop{\sim}\limits_{t \to 1^-} \:
e_2'[\hat{\mathbf{b}}] - {1 \over 2} e_2[\hat{\mathbf{b}}] +
O(1-t) + d_0 (1 - t)^{b+3}\left[ 1 + O(1-t) \right] + \cdots
\end{equation}
where
\begin{equation}\label{fr0a}
d_0 = - 2 \xi {1 \over (b+1)(b+2) \Gamma(b+3) \Gamma(b+4)}
{\Gamma(a+b+N+3) \over \Gamma(a+N)} {\Gamma(b+N+3) \over
\Gamma(N)}
\end{equation}
and the terms $ O(1-t)$ are analytic in $1-t$.
\end{prop}

\noindent
Proof. \quad The constant term follows immediately from the requirement that
the integrand in (\ref{5.6}) be integrable at $t=1$. To deduce the 
explicit form of the leading non analytic term we note that the analogue
of (\ref{5.1b}) applied to (\ref{5.6}) gives
\begin{multline}\label{fr1}
  p_{N-1,0}^{\rm J}((s,1);a,b) + (1-\xi) p_{N-1,1}^{\rm J}((s,1);a,b)
  \\
  \mathop{\sim}\limits_{s \to 1^-}
   N {J_{N-1}(a,b+2) \over J_{N}(a,b)} s^a (1-s)^b
     \exp \Big\{ - \int_s^1\Big(
                   \hat{U}_{N}^{\rm J}(t;a,b,2;\xi)
                   - e_2'[\hat{\mathbf{b}}]t + {1 \over 2} e_2[\hat{\mathbf{b}}]
                           \Big) \Big|_{N \mapsto N-1} {dt \over t(t-1)}
          \Big \} .
\end{multline}
But it follows from the definitions that
\begin{align}
  p_{N-1,0}^{\rm J}((s,1);a,b) 
  & \mathop{\sim}\limits_{s \to 1^-}
    \rho_{N,1}^{\rm J}(s) - \int_s^1  \rho_{N,2}^{\rm J}(s,t) \, dt + \cdots
  \label{fr3a} \\
  p_{N-1,1}^{\rm J}((s,1);a,b) 
  & \mathop{\sim}\limits_{s \to 1^-}
    \int_s^1  \rho_{N,2}^{\rm J}(s,t) \, dt + \cdots
  \label{fr3b} 
\end{align}
where $\rho_{N,n}^{\rm J}$ denotes the $n$-point distribution function in the
JUE with $N$ eigenvalues. Hence, making use also of the ansatz (\ref{fr0}),
(\ref{fr1}) reduces to
\begin{multline}\label{fr4}
  \rho_{N,1}^{\rm J}(s) - \xi \int_s^1 
  \rho_{N,2}^{\rm J}(s,t) \, dt + \cdots \\
  \mathop{\sim}\limits_{s \to 1^-}
    N {J_{N-1}(a,b+2) \over J_{N}(a,b)} s^a (1-s)^b
    \Big\{ 1 + O(1-s) + \cdots
  + {d_0 |_{N \mapsto N-1} \over b+3} (1-s)^{b+3}\left[ 1+O(1-s) \right] + \cdots
    \Big\}.
\end{multline}
For the first term on the left hand side of (\ref{fr4}) we note
from the definition of $ \rho_{N,1}^{\rm J}(s)$ as a multiple integral that
\begin{equation*}
\rho_{N,1}^{\rm J}(s) \: \mathop{\sim}\limits_{s \to 1^-} \:
N {J_{N-1}(a,b+2) \over J_N(a,b)} s^a (1 - s)^b,
\end{equation*}
which is in agreement with the first term on the right hand side of
(\ref{fr4}). For the term  proportional to $\xi$ we note from the
definition of $ \rho_{N,2}^{\rm J}(s,t)$ as a multiple integral that
\begin{equation*}
\rho_{N,2}^{\rm J}(s,t)  \: \mathop{\sim}\limits_{s,t \to 1^-} \:
N(N-1) {J_{N-2}(a,b+4) \over J_N(a,b)} (1-s)^b (1 - t)^b (s-t)^2,
\end{equation*}
which gives
\begin{equation*}
\int_s^1 \rho_{N,2}^{\rm J}(s,t) \, dt \,
 \mathop{\sim}\limits_{s \to 1^-} \,
N(N-1)  {J_{N-2}(a,b+4) \over J_N(a,b)} (1-s)^{2b+3}
{2 \over (b+1)(b+2)(b+3)}.
\end{equation*}
Substituting in (\ref{fr4}) and
making use of (\ref{jna}) shows this is consistent with the value of $d_0$
in (\ref{fr0a}). \hfill $\square$

The probability $p_{N,n}$ for the
JUE with $I_0 = (0,s)$ can be similarly characterised. In fact by changing
variables $x_l \mapsto 1 - x_l$ in (\ref{1.17a}) we see that
\begin{equation}\label{5.6a}
p^{\rm J}_{N,n}((0,s);a,b) = p^{\rm J}_{N,n}((s,1);b,a).
\end{equation} 

Again with $I_0 = (s,t)$ we introduce the quantity
\begin{equation}\label{7.5}
E_N(I_0;(s,t);\xi) := (N+2)(N+1)\Big ( \int_I - \xi \int_{I_0} \Big )
dx_1 \cdots \Big ( \int_I - \xi \int_{I_0} \Big )dx_N\,
p_{N+2}(s,t,x_1,\dots,x_N).
\end{equation}
We have
\begin{equation*}
p_{N,n}(I_0,(s,t)) = 
{1 \over \rho_{N+2}(t)}
{(-1)^n \over n!} {\partial^n \over \partial \xi^n}
E_N(I_0;(s,t);\xi) \Big |_{\xi = 1}
\end{equation*}
where $p_{N,n}(I_0,(s,t))$ denotes the probability density of there being
an eigenvalue at $s$ and $n$ eigenvalues in $I_0=(s,t)$, given there is
an eigenvalue at $t$ ($\rho_{N+2}(t)$ denotes the eigenvalue density at
$t$). For the CUE the eigenvalue density is a constant. Choosing
$(s,t) = (\pi - \phi, \pi)$, (\ref{7.5}) is proportional to
$\sin^2\phi /2$ times $\tilde{E}_N^{\rm cJ}(\phi;(1,0),2;\xi)$ as
specified by (\ref{pw4}). Using (\ref{3.35d'}) and (\ref{mna}) 
we deduce
\begin{multline}\label{ro}
  \Big( {2 \pi \over N} \Big)
  p_{N-2,n}^{\rm CUE}\Big( {2 \pi X \over N} \Big) \\
  = {1 \over 3} (N^2 - 1) \sin^2 {\pi X \over N}
  {(-1)^n \over n!} {\partial^n \over \partial \xi^n}
  \exp \Big\{ -\frac{1}{2} \int_0^{2\pi X /N}
        \left( U_{N-2}^{\rm Cy}(\cot{\theta \over 2};(N,0),2;\xi) 
               +4\cot{\theta \over 2} \right) d\theta \Big\} \Big|_{\xi=1}.
\end{multline}
The boundary condition can be deduced by adapting a strategy analogous to
that used in the proof of Proposition \ref{p18}. We find
\begin{multline}
  \frac{1}{2} \left( U_{N-2}^{\rm Cy}(\cot{\theta \over 2};(N,0),2;\xi) 
                     +4\cot{\theta \over 2} \right) \\
  \: \sim \:
  {2 \over 15}(N^2-4)\tan{\theta \over 2} + O(\tan^3 {\theta \over 2})
  + {(\xi-1) \over 540 \pi} {(N+2)! \over (N-3)!} \left(
             \tan^4 {\theta \over 2} + O(\tan^6 {\theta \over 2}) \right)
\end{multline}
where are term $ O(\tan^3 {\theta \over 2})$ are odd in $\theta$ while all
terms $O(\tan^6 {\theta \over 2})$ are even in $\theta$.

The eigenvalue density is another statistical property of the eigenvalues
accessible from the average (\ref{1.17a}) (for the JUE) and (\ref{pw4})
(for the cJUE). We must choose $\mu=2$ and $\xi = 0$, and multiply
(\ref{1.17a}) by $x^a (1 - x)^b$ and (\ref{pw4}) by 
$e^{\omega_2 \phi}|1 + e^{i \phi}|^{2\omega_1}$. We then obtain expressions 
proportional to the definition of the density in an ensemble of $N+1$ 
eigenvalues, $\rho_{N+1}$ say. Making use of Proposition \ref{p9} it therefore 
follows that
\begin{equation}\label{8.1}
\rho_{N+1}^{\rm J}(s) =
(N+1) {J_N(a+2,b) \over J_{N+1}(a,b)} s^a (1 - s)^b
\exp \Big \{ \int_0^s \Big (\hat{U}_N^{\rm J}(t;a,b,2;0)
- e_2'[\hat{\mathbf{b}}] t + {1 \over 2} e_2[\hat{\mathbf{b}}]
\Big ) {dt \over t(t-1)} \Big \}
\end{equation}
where $\hat{U}_N^{\rm J}$ is specified as in Proposition \ref{p9}
with boundary condition (\ref{r.1}). An analogous formula can be written
down for $\rho_{N+1}^{\rm cJ}(\phi)$ using (\ref{3.35d'}).

The term in (\ref{8.1}) involving the exponential function is proportional
to $\tilde{E}_N^{\rm J}(s;a,b,2;0)$. According to Proposition \ref{p14}
and (\ref{3.55a})
\begin{equation}\label{jpi}
\tilde{E}_N^{\rm J}(s;a,b,2;0) \propto s^{-1}
\left | \begin{array}{cc} P_N^{(a,b)}(1-2s) & P_N^{(a,b+1)}(1-2s) \\
P_{N+1}^{(a,b-1)}(1-2s) & P_{N+1}^{(a,b)}(1-2s) \end{array} \right |.
\end{equation}
On the other hand, from the fact that $\{P_n^{(a,b)}(1-2s) \}_{n=0,1,\dots}$
are orthogonal with respect to the weight function $s^a(1-s)^b$,
$(0 < s < 1)$, well known direct integration methods give that
\begin{equation}\label{jpi1}
\tilde{E}_N^{\rm J}(s;a,b,2;0) \propto
\left | \begin{array}{cc} P_{N+1}^{(a,b)}(1-2s) & {d \over
ds} P_{N+1}^{(a,b)}(1-2s) \\
P_{N}^{(a,b)}(1-2s) & {d \over ds} P_{N}^{(a,b)}(1-2s) \end{array} \right |.
\end{equation}
Using Jacobi polynomial identities, (\ref{jpi1}) can be reduced to
(\ref{jpi}).

The above results apply to the finite $N$ ensembles. Also of interest are
scaled $N \to \infty$ limits of these results. As mentioned in the
introduction, we are restricting attention of such limits to the
spectrum singularity. One example of the latter type is
\begin{equation*}
p_n^{\rm bulk}(X) := \lim_{N \to \infty} {2 \pi \over N}
p_{N-2,n}^{\rm CUE}
\Big ( {2 \pi X \over N} \Big ).
\end{equation*}
For its evaluation we read off from (\ref{ro}) that
\begin{equation*}
  p_0^{\rm bulk}(X) \mathop{\sim}\limits_{X \to 0} {\pi^2 \over 3} X^2,
\end{equation*}
and note that the scaled limit of (\ref{7.5}) for the CUE is proportional
to the average in (\ref{vpu}) with
\begin{equation}\label{4.26a}
  \omega_2=0, \quad \omega_1=\mu=1.
\end{equation}
Thus
\begin{equation}\label{ntn}
p_n^{\rm bulk}(X) = {\pi^2 \over 3} X^2 {(-1)^n \over n!}
{\partial^n \over \partial \xi^n}
\exp \Big \{\int_0^{2 \pi X} u(t;(1,0),1;\xi)
{dt \over t} \Big \} \Big |_{\xi = 1}
\end{equation}
where according to Proposition \ref{pho} the quantity
$h(t) = u(it;(1,0),1;\xi) + 2$ satisfies 
(\ref{vp}).  
Substituting the parameter values (\ref{4.26a}) in
(\ref{vp3}) it follows from (\ref{vp3a}) and (\ref{vp}) that $u$ itself
satisfies
\begin{equation}\label{ntn1}
  (s u'')^2 + (u - su') \{ u - su' + 4 - 4 (u')^2 \} - 16 (u')^2 = 0
\end{equation}
with boundary condition
\begin{equation}\label{ntn2}
  u(t;(1,0),1;\xi) \mathop{\sim}\limits_{t \to 0} 
  - {1 \over 15} t^2 + O(t^4) 
  - {(\xi - 1) \over 8640\pi}\left( t^5 + O(t^7) \right)
\end{equation}
where all terms $O(t^4)$ are even, while all terms $O(t^7)$ are odd.
In the case $n=0$, $p_n^{\rm bulk}(X)$ is the distribution of consecutive
neighbour spacings in the infinite CUE, scaled so that the mean eigenvalue
spacing is unity. The formula (\ref{ntn}) gives
\begin{equation}\label{4.29}
p_0^{\rm bulk}(X) = {\pi^2 \over 3} X^2
\exp \Big (  \int_0^{2 \pi X}
{h(-it;\mathbf{v}) - 2 \over t} \, dt \Big ), \quad
\mathbf{v} = (1,-1,1,-1).
\end{equation}
Previous studies have given two alternative Painlev\'e
$\sigma$-function evaluations of this same quantity. The first, due to Jimbo
et al.~\cite{JMMS_1980}, can be written
\begin{equation}\label{4.30}
p_0^{\rm bulk}(X) = {d^2 \over dX^2} \exp \Big ( \int_0^{2 \pi X}
{h(-it;\mathbf{v}) \over t} \, dt \Big ), \quad 
\mathbf{v} = (0,0,0,0)
\end{equation}
while the second, due to the present authors \cite{FW-00}, reads
\begin{equation}\label{4.31}
p_0^{\rm bulk}(X) = - {d \over dX} \exp \Big ( \int_0^{2 \pi X}
{h(-it;\mathbf{v}) \over t} \, dt \Big ), \quad
\mathbf{v} = (0,0,1,-1) \quad\text{or}\quad (1,-1,0,0)
\end{equation}
(the boundary conditions in (\ref{4.30}) and (\ref{4.31}) are not required
for the present purpose). In each of (\ref{4.29})--(\ref{4.31}) the
function $h(t;\mathbf{v})$ satisfies (\ref{vp}) with
$\mathbf{v}$ as specified.

The equality between (\ref{4.30}) and (\ref{4.31}) implies the identity
\begin{equation}\label{ei}
h(s;(1,-1,0,0)) = -1 + h(s;(0,0,0,0)) +
s {h'(s;(0,0,0,0)) \over h(s;(0,0,0,0))}.
\end{equation}
A direct proof of this identity was given in \cite{WF_2000} using the formulas
of \cite{CS-93} for $h_V$ and $h_V'$ in (\ref{vp}) in terms of a particular
\PV transcendent and its derivative. Analogous to (\ref{ei}), the equality
between (\ref{4.31}) and (\ref{4.29}) implies the identity
\begin{equation}\label{ei1}
h(s;(1,-1,1,-1)) = -1 + h(s;(0,0,1,-1)) +
s {h'(s;(0,0,1,-1)) \over h(s;(0,0,1,-1))}.
\end{equation}
In fact both (\ref{ei}) and (\ref{ei1}) are particular examples of an
identity between functions satisfying the $\sigma$-form of the
\PV equation proved in our paper \cite{FW-00}. With
$\sigma = \sigma(t;\mathbf{v})$ specified as a solution of
(\ref{vpx}) with $\mathbf{v}$ given by 
\begin{equation}\label{ei5}
\mathbf{v} = \Big ( - {1 \over 4} (2N + a - \mu), - {1 \over 4} (2N + a
+ 3 \mu), {1 \over 4}(2N+3a+\mu), {1 \over 4}(2N-a+\mu) \Big )
\end{equation}
the identity states
\begin{equation}\label{ei2}
\sigma(t;\mathbf{v}) \Big |_{\mu = 2} = - (a+1+2N) + t +
\Big ( \sigma(t;\mathbf{v}) + t 
{ \sigma'(t;\mathbf{v}) \over \sigma(t;\mathbf{v}) } \Big )
\Big |_{\mu = 0 \atop N \mapsto N + 1}.
\end{equation}
Now, from (\ref{vpy}),
\begin{equation}\label{vpq}
\begin{array}{cc}
   h(t;(1,-1,1,-1)) = \sigma(t;(1,-1,1,-1)) - t -2, \quad&
   h(t;(1,-1,0,0)) = \sigma(t;(1,-1,0,0)) - t -2, \\ 
   h(t;(0,0,1,-1)) = \sigma(t;(0,0,1,-1)), \quad&
   h(t;(0,0,0,0)) = \sigma(t;(0,0,0,0))
\end{array}
\end{equation}
To deduce (\ref{ei}) from (\ref{ei2}) we set $\mu=2$, $N=-1$, $a=0$ in
(\ref{ei5}),  giving $\mathbf{v} = (1,-1,0,0)$ on the left hand side,
then set $\mu=0$, $N=0$, $a=0$ giving
$\mathbf{v} = (0,0,0,0)$ on the right hand side of (\ref{ei2}). The identity
(\ref{ei}) then follows by making use of 
the second and fourth formulas in (\ref{vpq}). Similarly,
(\ref{ei1}) follows from (\ref{ei2}) by choosing $\mu = 2$, $N=-2$,
$a=2$, then $\mu=0$, $N=-1$, $a=2$,
and making use of the first and third identities in (\ref{vpq}).

There is an analogue of (\ref{ei2}) for functions satisfying the
$\sigma$-form (\ref{4}) of the \PVI equation. This can be obtained by
substituting in the formula
\begin{equation}
p_{N,n}^{\rm J}((s,1);a,b) = {d \over ds} E^{\rm J}_{N+1,n}((s,1);a,b)
\end{equation}
the exact evaluations (\ref{5.6}) and (\ref{5.4}), and taking the logarithmic 
derivative. We find
\begin{multline}\label{wf}
  \hat{U}_N^{\rm J}(s;a,b,2;\xi) =
   - \Big(a+b+2 + e_2'[\hat{\mathbf{b}}] \Big|_{\mu=0 \atop N \mapsto N+1}
          - e_2'[\hat{\mathbf{b}}] \Big |_{\mu=2} \Big)s
   + \Big(a+1 + {1 \over 2} e_2[\hat{\mathbf{b}}] \Big|_{\mu=0 \atop N \mapsto N+1}
   - {1 \over 2} e_2[\hat{\mathbf{b}}] \Big|_{\mu=2} \Big)
   \\
   + \hat{U}_{N+1}^{\rm J}(s;a,b,0;\xi)
   + s(s-1) {d \over ds} \log \Big(
   \hat{U}_{N+1}^{\rm J}(s;a,b,0;\xi)
        - e_2'[\hat{\mathbf{b}}] \Big|_{\mu=0 \atop N \mapsto N+1} s 
        + {1 \over 2} e_2[\hat{\mathbf{b}}] \Big|_{\mu=0 \atop N \mapsto N+1} \Big)
   \\
   = \half-s + \hat{U}_{N+1}^{\rm J}(s;a,b,0;\xi)
   + s(s-1) {d \over ds} \log \Big(                            
   \hat{U}_{N+1}^{\rm J}(s;a,b,0;\xi)
        + \big( N+1+{1 \over 2}(a+b) \big)^2(s-\half)+{1\over 8}(a^2-b^2) \Big).
\end{multline}

An alternative derivation of (\ref{wf}) can be given.
Following the method first introduced in \cite{WF_2000}, and also used in
\cite{FW_2002a} to derive (\ref{ei2}), we use (\ref{2.28}) and the action
\begin{equation*}
s_1 \mathbf{b} = (b_1,b_2,b_4,b_3)
\end{equation*}
to note that
\begin{equation*}
s_1 T_3^{-1} s_1  T_3^{-1}(b_1,b_2,b_3+1,b_4-1) =
(b_1,b_2,b_3,b_4-2).
\end{equation*}
This means that if we temporarily reorder the components in (\ref{1.34e}),
\begin{equation*}
\mathbf{b} \mapsto
\Big ( {1 \over 2}(b-a), - {1 \over 2}(a+b), {1 \over 2}(a+b) + N,
- {1 \over 2}(a+b) - N - \mu \Big )
\end{equation*}
(as (\ref{4}) is symmetric in the $\mathbf{b}$'s, this does not affect
the $\hat{U}^{\rm J}$, which are defined as solutions of (\ref{4})),
then
\begin{equation}\label{wf1}
\hat{U}_N^{\rm J}(s;a,b,2;\xi) = s_1 T_3^{-1} s_1 T_3^{-1}
\hat{U}_{N+1}^{\rm J}(s;a,b,0;\xi)
\end{equation}
Using (\ref{2}), (\ref{3}), (\ref{4.1}), Table \ref{t1}, (\ref{rb})
and (\ref{T1}), we can use computer algebra to verify that the right
hand side of (\ref{wf1}) agrees with the right hand side of (\ref{wf}),
thus providing an independent derivation of this result.

\subsection{Relationship to the hard edge gap probability in the scaled
LOE and LSE}
The Laguerre orthogonal ensemble (LOE) refers to the PDF
\begin{equation}\label{5.24}
{1 \over C} \prod_{l=1}^N x_l^a e^{-x_l/2} \prod_{1 \le j < k \le N}
|x_k - x_j|, \quad x_l \ge 0.
\end{equation}
For $a= n - N - 1$ ($n \ge N$), this is realised as the joint 
eigenvalue distribution of random matrices $A^TA$, where $A$ is an
$n \times N$ matrix with independent, identically distributed standard
Gaussian random variables. Scaling by $x_l \mapsto X_l/4N$, and taking
the limit $N \to \infty$ gives well defined distributions in the vicinity
of $x=0$ (the hard edge) \cite{Fo-93}. One particular quantity of interest
in this scaled limit is $E_1^{\rm hard}(s;a)$, the probability that the
interval $(0,s)$ is free of eigenvalues. This was evaluated in terms of
a Painlev\'e V transcendent in \cite{Fo-00} and expressed as a $\tau$-function
for a \PV system in \cite{FW_2002c}. The latter result can be written
\begin{equation}\label{5.24a}
{d \over dx} \log E_1^{\rm hard}(x^2;(a-1)/2) = \tilde{h}_V(x)
\end{equation}
where
\begin{equation*}
\sigma_V(x) := x \tilde{h}_V(x) + {1\over 4} x^2 - {(a-1) \over 2} x +
{a(a-1) \over 4}
\end{equation*}
satisfies (\ref{vpx}) with $t \mapsto 2x$ and
\begin{equation}\label{vav}
v_1 = - v_2 = {1 \over 4} (a-1), \qquad
v_3 = - v_4 = {1 \over 4} (a+1).
\end{equation}
On the other hand, if follows from Proposition \ref{pho}
that
\begin{equation*}
  u(2ix;({1 \over 4}(a+1),0), {1 \over 4}(a-1); \xi )
  - {1 \over 2}(a-1) x + {1 \over 4} a (a-1)
\end{equation*}
also satisfies (\ref{vpx}) with $t \mapsto 2x$ and parameters (\ref{vav}).
Hence
\begin{equation}\label{vav1}
x {d \over dx} \log E_1^{\rm hard}(x^2;(a-1)/2) + {1\over 4} x^2
\doteq u(2ix;({1 \over 4}(a+1),0), {1 \over 4}(a-1); \xi ),
\end{equation}
where we use the symbol $\doteq$ to mean that both sides
satisfy the same differential
equation.

In the special case $(a-1)/2 = m$, $m \in \zz_{\ge 0}$, if follows from
the definition of $E_1^{\rm hard}(s;(a-1)/2)$ that it is analytic in $s$.
With this value of $(a-1)/2$, the right hand side of (\ref{vav1}) reads
$u(2ix;((m+1)/2,0),m/2;\xi)$. But we know from (\ref{4.12a}) that $u$ with
these parameters, analytic in $x$, is given by an $m$-dimensional integral,
thus leading to the following result.

\begin{prop}\label{p16}
For $m \in \zz_{\ge 0}$,
\begin{equation}\label{5.27}
E_1^{\rm hard}(s^2;m) = e^{-s^2/8 + ms}
\Big \langle e^{-2s \sum_{j=1}^m x_j} \Big \rangle_{{\rm JUE}_m}
\Big |_{\lambda_1 = \lambda_2 = 1/2}.
\end{equation}
\end{prop}

\noindent
Proof. The only point which remains to be checked is that for $u$ in 
(\ref{vav1}) given by (\ref{4.12a}), both sides have the same small $x$
behaviour, thus determining the boundary condition in their characterisation
as the solution of the same differential equation. This is the same as showing
that both sides of (\ref{5.27}) have the same small $s$ behaviour. Now,
with the scaled density of the LOE ensemble (\ref{5.24}) denoted
$\rho_a(X)$, it's easy to see from the definitions that
\begin{equation*}
{d \over ds} E_1^{\rm hard}(s^2,m) \: \sim \: -2s \rho_{2m}(s^2).
\end{equation*}
But we know that \cite{FNH-98}
\begin{align*}
  \rho_a(s^2) & = K^{\rm hard}(s^2,s^2) + {J_{a+1}(s) \over 4s}
                  \Big( 1 -\int_0^s J_{a-1}(v) \, dv \Big),
  \nonumber \\
  K^{\rm hard}(s^2,s^2) & := {1 \over 4}
               \Big( (J_a(s))^2 - J_{a+1}(s) J_{a-1}(s) \Big).
\end{align*}
For small $s$ the term $ J_{a+1}(s)/4s $ dominates, implying the result
\begin{equation}\label{cxw}
{d \over ds} E_1^{\rm hard}(s^2,m) \: \sim \: - {s^{2m+1} \over 2^{2m+2}
\Gamma(2m+2)}.
\end{equation} 

To determine the small $s$ behaviour of the right hand side of (\ref{5.27}),
we note that the JUE${}_m$ average with $\lambda_1 = \lambda_2 = 1/2$
coincides with an average over the unitary symplectic group USp$(m)$ upon
the change of variables $\lambda_j = {1 \over 2} (\cos \theta_j + 1)$.
Explicitly (see e.g.~\cite[eq.~(1.26)]{FW_2002a})
\begin{equation}\label{5.27b}
e^{ms} \Big \langle e^{-2s \sum_{j=1}^m x_j} \Big \rangle_{{\rm JUE}_m}
\Big |_{\lambda_1 = \lambda_2 = 1/2} =
\Big \langle e^{(s/2) {\rm Tr} \, U} \Big \rangle_{U \in {\rm USp}(m)}.
\end{equation}
Furthermore, it is known from the work of Rains \cite{Ra_1998} (see also
\cite{BF-2001}) that
\begin{equation}\label{5.27e}
\Big \langle e^{(s/2) {\rm Tr} \, U} \Big \rangle_{U \in {\rm USp}(m)} =
\sum_{n=0}^\infty {(s/2)^{2n} \over 2^{2n} n!}
{f_{nm}^{\rm inv} \over (2n-1)!!}
\end{equation}
where $f_{nm}^{\rm inv}$ denotes the number of fixed point free involutions
of $\{1,2,\dots,2n\}$ constrained so that the length of the maximum
decreasing subsequence is less than or equal to $2m$. Following
\cite{AvM_2001a}, the small $s$ behaviour of (\ref{5.27e}) follows by noting
that from the definition, for $m \le n$ we have
$f_{nm}^{\rm inv} = 2^n (2n-1)!!$ (the number of fixed point free
involutions without constraint) while
$f_{m+1 \, m}^{\rm inv} = 2^{m+1} (2m+1)!! -1$. Hence
\begin{equation*}
\Big \langle e^{(s/2) {\rm Tr} \, U} \Big \rangle_{U \in {\rm USp}(m)} =
\exp \Big \{ s^2/8 -
{(s/2)^{2(m+1)} \over 2^{m+1} (m+1)! (2m+1)!!} +
O(s^{2(m+2)}) \Big \}.
\end{equation*}
The derivative of this expression multiplied by $e^{-s^2/8}$ has leading
order behaviour (\ref{cxw}), as required. \hfill $\square$

Note that substituting (\ref{5.27b}) in (\ref{5.27}) gives the identity
\begin{equation}\label{5.27c}
E_1^{\rm hard}((2s)^2;m) = e^{-s^2/2}  \langle e^{s {\rm Tr} \, U}
 \rangle_{U \in {\rm USp}(m)}.
\end{equation}

Consider next the Laguerre symplectic ensemble (LSE) which refers to the
PDF
\begin{equation*}
{1 \over C} \prod_{l=1}^N x_l^a e^{-x_l} \prod_{1 \le j < k \le N}
(x_k - x_j)^4, \qquad x_l \ge 0.
\end{equation*}
The corresponding scaled hard edge gap probability $E_4^{\rm hard}(s;a)$
has been shown in \cite{FW_2002c} to have the $\tau$-function evaluation
\begin{equation}\label{tft}
E_4^{\rm hard}(x^2;a+1) = {1 \over 2} \Big (
\tilde{\tau}_{V}(x) + \tilde{\tau}_{V}^*(x) \Big )
\end{equation}
where
\begin{equation*}
{d \over dx} \log \tilde{\tau}_{V}(x) = 
\tilde{h}_V(x) \qquad {\rm and} \qquad
{d \over dx} \log \tilde{\tau}_{V}^*(x) \doteq 
\tilde{h}_V(x)
\end{equation*}
(recall the meaning of $\doteq$ from (\ref{vav1})). Thus
\begin{equation}\label{tft1}
\tilde{\tau}_{V}(x) = E_1^{\rm hard}(x^2;(a-1)/2)
\end{equation}
and the logarithmic derivative of both $\tau$-functions in (\ref{tft})
satisfy the same differential equation, differing only in the boundary
condition. From \cite{FW_2002c} we require
\begin{equation}\label{5.27p1}
x {d \over dx} \log  \tilde{\tau}_{V}^*(x) \: \mathop{\sim}\limits_{x \to 0^+}
\:  {x \over 2} J_a(x) \: \sim \:
{x^{a+1} \over 2^{a+1} \Gamma(a+1)}.
\end{equation}

Analogous to the results (\ref{5.27}) and (\ref{5.27c}), 
$\tilde{\tau}_{V}^*(x)$ with $(a-1)/2 = m$, $m \in \zz_{\ge 0}$, can be
written as an average over the JUE, or equivalently as an average over a
classical group. To see this, we first note from Proposition \ref{pho}
that
\begin{equation*}
  u(it;(\omega_1,0),\mu;\xi) \doteq u(it;(\mu,0),\omega_1;\xi),
\end{equation*}
and use this together with (\ref{vav1}) and (\ref{tft1}) to deduce
\begin{equation*}
x {d \over dx} \log  \tilde{\tau}_{V}^*(x) + {1 \over 4} x^2
\doteq u(2ix;({1 \over 4}(a-1),0),
{1 \over 4}(a+1);\xi ).
\end{equation*}
But for $(a-1)/2 = m$ and thus $(a+1)/4 = (m+1)/2$, we see from (\ref{4.12a})
that
\begin{equation}\label{sm.1}
u(2ix;({1 \over 2}(m-1),0), {1 \over 2}(m+1);\xi^*) \doteq
x {d \over dx} \log \Big (
e^{(m+1)x} \Big \langle e^{-2x \sum_{j=1}^{m+1} x_j} 
\Big \rangle_{{\rm JUE}_{m+1}}
\Big |_{\lambda_1 = \lambda_2 = -1/2} \Big ).
\end{equation}
Because, as will be checked below,
the small $x$ behaviour of (\ref{sm.1}) coincides with
(\ref{5.27p1}), we can read off the expression for 
$ \tilde{\tau}_{V}^*(x)$ as an average over the JUE.

\begin{prop}
For $(a-1)/2 =m$, $m \in \zz_{\ge 0}$,
\begin{equation}\label{sm.2}
\tilde{\tau}_{V}^*(x) = e^{-x^2/8 + (m+1)x}
\Big \langle e^{-2x \sum_{j=1}^{m+1} x_j} \Big \rangle_{{\rm JUE}_{m+1}}
\Big |_{\lambda_1 = \lambda_2 = -1/2}.
\end{equation}
\end{prop}

\noindent
Proof. \quad We have to verify the assertion that (\ref{sm.2}) is consistent
with (\ref{5.27p1}). Again following \cite{AvM_2001a}, we proceed in an analogous
fashion to the proof of Proposition \ref{p16}. Now, in the case
$\lambda_1 = \lambda_2 = -1/2$, the JUE${}_{m+1}$ average coincides with
an average over matrices in $O^+(2m+2)$ --- $(2m+2) \times (2m+2)$
orthogonal matrices with determinant $+1$. Analogous to (\ref{5.27b}) we
have
\begin{equation}\label{sm.3}
e^{(m+1)x} 
\Big \langle e^{-2x \sum_{j=1}^{m+1} x_j} \Big \rangle_{{\rm JUE}_{m+1}}
\Big |_{\lambda_1 = \lambda_2 = -1/2} =
\Big \langle e^{(x/2) {\rm Tr}U} \Big \rangle_{U \in O^+(2m+2)}.
\end{equation}
But we know from \cite{AvM_2001a} (see also \cite{VM-2001}) that
\begin{equation*}
\Big \langle e^{ x {\rm Tr} U} \Big \rangle_{U \in O^{\pm}(l)} =
\exp \Big ( {x^2 \over 2} \pm {x^l \over l!} + O(x^{l+1}) \Big ).
\end{equation*}
Thus
\begin{equation*}
  \tilde{\tau}_{V}^*(x) 
   = \exp \Big( {(\half x)^{2m+2} \over (2m+2)!} + O( x^{2m+3}) \Big),
\end{equation*}
which, recalling $(a-1)/2 = m$, is consistent with (\ref{5.27p1}).
\hfill $\square$

We can express $E_4^{\rm hard}(x^2;a+1)$ in a form analogous to
(\ref{5.27c}). For this purpose we recall that the joint PDF of the
eigenvalues from the group USp$(l)$, coincides with the joint PDF of
the eigenvalues from the group O${}^-(2l+2)$ (not including the two
fixed eigenvalues $\lambda = \pm 1$). Using this fact to rewrite the average
in (\ref{5.27c}), substituting in (\ref{tft1}), substituting (\ref{sm.3})
in (\ref{sm.2}), and substituting the results in (\ref{tft}) shows
\begin{align}
  E_4^{\rm hard}(x^2;2m+2) 
  & = {1 \over 2}e^{-x^2/8} \Big(
       \Big\langle e^{(x/2) {\rm Tr}U} \Big\rangle_{U \in O^+(2m+2)} +
       \Big\langle e^{(x/2) {\rm Tr}U} \Big\rangle_{U \in O^-(2m+2)} \Big)
  \cr
  & = e^{-x^2/8}\Big\langle e^{(x/2) {\rm Tr}U} \Big\rangle_{U \in O(2m+2)}.
\end{align}

\subsection{Last passage percolation}
Johansson \cite{Jo-2000} has introduced the following probabilistic model.
Define on each site $(i,j)$ of the lattice $\zz_{\ge 0}^2$ a non-negative
integer variable $w(i,j)$, chosen independently with the geometric distribution
$g(q^2)$ (thus Pr$(g(q^2) = k) = (1 - q^2) q^{2k}$, $k=0,1,\dots,$). Let
$(0,0)$ wu/wr $(M-1,N-1)$ denote the set of all weakly up / weakly right
paths (meaning a sequence $\{(i_r, j_r) \}_{r=1}^l$ with
$i_1 \le i_2 \le \cdots \le i_l$ and
$j_1 \le j_2 \le \cdots \le j_l$) on the lattice starting at
$(0,0)$ and finishing at
$(M-1,N-1)$, moving only upwards or to the right. A quantity of
interest is the random variable
\begin{equation}\label{pg}
  G^{\rm wu/wr}((M-1,N-1);q^2) :=
    \mathop{\rm max}\limits_{\pi \in (0,0)\, {\rm wu/wr} \,(M-1,N-1)}
  \sum_{(i,j) \in \pi} w(i,j)
\end{equation}
which with the integer variables $w(i,j)$ regarded as waiting times
represents the last passage time of directed percolation paths
$(0,0) \, {\rm wu/wr} \, (M-1,N-1)$. The relevance to the present study
is that Baik and Rains \cite{BR-2001,Ba-2001} have shown 
\begin{equation}\label{pgw}
{\rm Pr}\Big ( G^{\rm wu/wr}((M-1,N-1);q^2) \le l \Big ) =
(1 - q^2)^{MN}
\Big \langle \prod_{j=1}^l (1 + q^2 z_j)^N (1 + 1/z_j)^M
\Big \rangle_{{\rm CUE}_l}.
\end{equation}
Using the results of this paper, (\ref{pgw}) can be expressed both as a
${}_2^{} F_1^{(1)}$ hypergeometric function and as a $\tau$-function for
a \PVI system.

\begin{prop}
We have
\begin{align}\label{itk}
  & {\rm Pr}\Big( G^{\rm wu/wr}((M-1,N-1);q^2) \le l \Big) 
  \\
  & = (1 - q^2)^{MN} \, {}_2^{} F_1^{(1)}( - N, -M;l;t_1,\dots, t_l)
                        \Big|_{t_1 = \cdots = t_l = q^2}
  \nonumber \\
  & = (1 - q^2)^{MN} {J_N(M-N,0) \over J_N(M-N,l)}
      {}_2^{} F_1^{(1)}( -l, M;M+N;1-t_1,\dots, 1-t_N)
                        \Big|_{t_1 = \cdots = t_N = q^2}
  \nonumber \\
  & = (1 - q^2)^{MN} \tau_3(q^2;\hat{\mathbf{b}})
                        \Big|_{\mu=l, \, a=0 \atop b = -M-N-l}
\end{align}
where the second equality is valid for $l \ge N$.
\end{prop}

\noindent
Proof. \quad The average in (\ref{pgw}) coincides with the CUE average
(\ref{ju}) provided 
\begin{equation*}
(N,a',b',\mu,t) = (l,0,M,N,q^2).
\end{equation*}
With $a,b$ specified in terms of $a',b'$ in the line below (\ref{ju}),
these values imply $(a,b) = (-l-M,M)$. The ${}_2^{} F_1^{(1)}$ evaluations
now follow from the specification of the right hand side of (\ref{pgw})
with the above parameters, and the results (\ref{iti}) and (\ref{itj}).
For the normalisation in the second equality we have used the general
formula \cite{Ya_1992}
\begin{equation*}
 {}_2^{}
F_1^{(1)}(r,N+\lambda_1;2N+\lambda_1+\lambda_2;t_1,\dots,t_N)
\Big |_{t_1 = \cdots = t_N = 1} =
{J_N(\lambda_1, \lambda_2 -r) \over J_N(\lambda_1,\lambda_2)}.
\end{equation*}
The final equality follows from (\ref{r3ab}).
\hfill $\square$

The second equality in (\ref{itk}) is well suited to taking the exponential limit
\begin{equation}\label{caz}
q = e^{-1/2L}, \quad l = Ls, \quad L \to \infty
\end{equation}
in which the discrete 
geometric random variables $w(i,j)$ in (\ref{pg}) tend to
continuous exponential variables, with unit mean, and each random variable
again confined to the lattice sites.

\begin{prop}
We have
\begin{align}\label{ti}
  g^{\rm wu/wr}((M-1,N-1);s)
  & := \lim_{L \to \infty} {\rm Pr} \Big(
        G^{{\rm wu/wr}}((M-1,N-1);e^{-1/L}) \le Ls \Big)
  \nonumber \\
  & = \Big( \prod_{j=0}^{N-1} {\Gamma(1 + j) \over \Gamma(M+1+j)} \Big)
       s^{MN} {}_1^{} F_1^{(1)}(M;M+N;s_1,\dots,s_N) \Big|_{s_1 = \cdots = s_N = -s}
  \nonumber \\
  & = {1 \over \prod_{j=0}^{N-1}(M-N+j)! j!}
       \int_0^sdx_1 \, x_1^{M-N} e^{-x_1} \cdots
       \int_0^sdx_N \, x_N^{M-N} e^{-x_N} \,
       \prod_{j<k}(x_k - x_j)^2
\end{align}
where in the final equality it is assumed $M > N$. Furthermore
\begin{equation}\label{gU}
s {d \over ds} \log g^{\rm wu/wr}((M-1,N-1);s) =
U_N^{\rm L}(t;M-N,0)
\end{equation}
where $U_N^{\rm L}(t;a,0)$ satisfies (\ref{vpx}) with
\begin{equation*}
\nu_0 =0, \quad \nu_1=0, \quad \nu_2=N+a, \quad \nu_3 = N.
\end{equation*}
\end{prop}

\noindent
Proof. \quad We see from (\ref{jna}) that in the limit (\ref{caz})
\begin{equation*}
(1- q^2)^{MN} {J_N(M-N,0) \over J_N(M-N,l)} \, \sim \,
s^{MN} \prod_{j=0}^{N-1} {\Gamma(1+j) \over
\Gamma(M+1+j)},
\end{equation*}
while it follows from (\ref{su}) that
\begin{equation*}
  {}_2^{}F_1^{(1)}(-l,M;M+N;1-t_1,\dots, 1-t_N)\Big|_{t_1 =\cdots =t_N =q^2}
  \, \sim \,
  {}_1^{}F_1^{(1)}(M;M+N;s_1,\dots,s_N)\Big|_{s_1 =\cdots =s_N = -s}. 
\end{equation*}
To obtain the final equality in (\ref{ti}) we make use of
(\ref{su1}) and (\ref{jna}) in the equality before. The formula
(\ref{gU}) follows from the first equality in (\ref{ti}) and (\ref{su2}).
\hfill $\square$

The final equality in (\ref{ti}) was first derived by Johansson \cite{Jo-2000}
using a different method. 
In words this equality says that the limiting last passage
time cumulative distribution is equal to the probability of there being
no eigenvalues in the interval $(s,\infty)$ of the Laguerre unitary
ensemble with $a=M-N$ (the general Laguerre weight being $x^ae^{-x}$).
With this identity established, (\ref{gU}) is equivalent to a result first
derived in \cite{TW_1994b}.

Formulas analogous to (\ref{pgw}) are known for certain variations of the
original Johansson model. One such variation involves $0,1$ Bernoulli random
variables with distribution $b(q^2)$ at each site
(thus ${\rm Pr}(b(q^2) = 0) = q^2/(1+q^2)$, ${\rm Pr}(b(q^2) = 1) = 1/(1+q^2)
$). Furthermore, the set of weakly up / weakly right paths is replaced by
the set of weakly up / strictly right paths (to be denoted wu/sr),
meaning a sequence $\{i_r,j_r) \}_{r=1}^l$ with
$i_1 < i_2 < \cdots < i_l$ and $j_1 \le j_2 \le \cdots \le j_l$, or the
set of strictly up / strictly right paths (to be denoted su/sr), meaning
a sequence $\{i_r,j_r) \}_{r=1}^l$ with
$i_1 < i_2 < \cdots < i_l$ and $j_1 < j_2 < \cdots < j_l$. Thus with
$B^{\rm wu/sr}$ and $B^{\rm su/sr}$ defined analogously to (\ref{pg}), one
has \cite{BR-2001,GTW-2000a,Ba-2001}
\begin{align}
  {\rm Pr}\Big( B^{\rm wu/sr}((M-1,N-1);q^2) \le l \Big)
  & = {1 \over (1 + q^2)^{MN}}
       \Big\langle \prod_{j=1}^l (1+1/z_l)^M (1-q^2z_l)^{-N}
       \Big\rangle_{{\rm CUE}_l}
  \label{pgw1} \\
  {\rm Pr}\Big( B^{\rm su/sr}((M-1,N-1);q^2) \le l \Big)
  & = (1 - q^2)^{MN}
       \Big\langle \prod_{j=1}^l (1+1/z_l)^{-M} (1+q^2z_l)^{-N}
       \Big\rangle_{{\rm CUE}_l}
  \label{pgw2}
\end{align}
where in (\ref{pgw2}) the contours of integration in the CUE${}_l$ average
must be deformed so that the point $z=-1$ lies inside  the contour while
$z = -1/q^2$ lies outside.

\begin{prop}
We have
\begin{align}
  {\rm Pr}\Big( B^{\rm wu/sr}((M-1,N-1);q^2) \le l \Big) 
  & = {1 \over (1 + q^2)^{MN}} \,
      {}_2^{} F_1^{(1)}(N,-M;l;t_1,\dots,t_l) \Big|_{t_1 = \cdots = t_l = -q^2}
  \nonumber \\
  & = {1 \over (1 + q^2)^{MN}} \, \tau_3(-q^2;\hat{\mathbf{b}})
                               \Big|_{\mu =l, \, N \mapsto -N \atop a=0, \, b=N-M-l}
\end{align}
\begin{align}
  {\rm Pr}\Big( B^{\rm su/sr}((M-1,N-1);q^2) \le l \Big)
  & = (1 - q^2)^{MN} 
      {}_2^{} F_1^{(1)}(N,M;l;t_1,\dots,t_l) \Big|_{t_1 = \cdots = t_l = q^2}
  \nonumber \\
  & = (1 - q^2)^{MN} \tau_3(q^2;\hat{\mathbf{b}})
                     \Big|_{\mu =l, \, N \mapsto -N \atop a=0, \, b=N+M-l}.
\end{align}
\end{prop}

\noindent
Proof. \quad The averages (\ref{pgw1}) and (\ref{pgw2}) result from the CUE 
average (\ref{ju}) by setting
\begin{equation}
(N,a',b',\mu,t) = \left \{ \begin{array}{l}
(l,0,M,-N,-q^2) \quad {\rm wu/sr} \\
(l,0,-M,-N,q^2) \quad {\rm su/sr} \end{array} \right.
\end{equation}
With $a',b'$ so specified, it follows from the equations in the line below
(\ref{ju}) that $(a,b)$ equals $(-l-M,M)$ and $(-l+M,-M)$ respectively.
The ${}_2^{}F_1^{(1)}$ evaluations
now follow from the specification of the left hand side of (\ref{pgw})
with the above parameters, and the results (\ref{iti}) and (\ref{itj}),
and the $\tau$-function evaluations in turn 
follow from (\ref{r3ab}). \hfill $\square$

We remark that for $l \ge N$ it follows from the definitions that
\begin{equation*}
{\rm Pr}\Big ( B^{\rm wu/sr}((M-1,N-1);q^2) \le l \Big ) =
{\rm Pr}\Big ( B^{\rm su/sr}((M-1,N-1);q^2) \le l \Big ) = 1.
\end{equation*}
These equalities for $l=N$ can be read off from the
above ${}_2^{}F_1^{(1)}$ evaluations by noting the general formula
\cite{Ya_1992}
\begin{equation*}
{}_2^{}F_1^{(1)}(a,b;a;t_1,\dots, t_m) =
{}_1^{}F_0^{(1)}(b;t_1,\dots, t_m) = \prod_{j=1}^m(1 - t_j)^{-b}.
\end{equation*}

Underlying the Johansson probabilistic model and its generalisations is
the Knuth correspondence between integer matrices, generalised permutations
and semi-standard tableaux (see \cite[\S 2.1]{Jo-2000}). In fact the
random variable (\ref{pg}) in the Johansson model is equivalent to the
random variable specifying the length of the longest increasing
subsequence of random generalised permutations with a certain probability
measure, or alternatively the length of the first row of a certain
class of random Young diagrams. The notion of measures on Young diagrams
also occurs in the work of Borodin and Olshanski on representations of
the infinite symmetric group (see e.g.~\cite{BO-2000b,BO-2001b}). In 
particular, for
random Young diagrams distributed according to a so called $z$-measure,
the probability that the first row length $Y(z;\xi)$ is less than or
equal to $l$ is shown to be given by
\begin{equation}\label{bob}
{\rm Pr}\Big ( Y(z;\xi) \le l \Big ) = (1 - \xi)^{|z|^2}
\Big \langle \prod_{j=1}^l(1 + \sqrt{\xi} z_j)^z
(1 + \sqrt{\xi}/z_j)^{\bar{z}} \Big \rangle_{{\rm CUE}_l}.
\end{equation}
Note that with $z=N$, $\xi = q^2$, this coincides with the case $M=N$ of
(\ref{pgw}). Comparing (\ref{bob}) with (\ref{pw4}), we see that in the
case $z$ real
\begin{equation}\label{bob1}
{\rm Pr}\Big ( Y(z;e^{-t})  \le l \Big ) = (1 - e^{-t})^{z^2}
\tilde{E}^{\rm cJ}_{l}(-it;(0,0),z;0).
\end{equation}
Hence we have the \PVI $\sigma$-function evaluation (\ref{3.35d}).
Complimentary to this result, we draw attention to the recent work of
Borodin \cite{Bo_2001}
(see \cite{BB_2002} for still more recent developments), 
who through a newly developed theory of discrete integral
operators and discrete
Riemann-Hilbert problems, has derived a difference equation for
${\rm Pr}\Big ( Y(z;e^{-t})  \le l \Big )$ in the discrete variable $l$
involving two auxiliary quantities which satisfy a particular case of
the discrete \PV system (\ref{dPv_1}), (\ref{dPv_2}). 

\subsection{Symmetrised last passage percolation and distribution of the
largest eigenvalue in the finite LOE and LSE}
Consider the original Johansson probabilistic model introduced in the
previous section. Choose the weights $w(i,j)$ for $i < j$ as independent
geometric random variables with distribution $g(q^2)$, and impose the
symmetry that $w(i,j) = w(j,i)$ for $i > j$. Choose the weights $w(i,i)$
on the diagonal independently with distribution $g(q)$, and consider
the random variable (\ref{pg}) with $M=N$. By the symmetry constraint the
set of paths in (\ref{pg}) can be restricted to the triangular region
$i \le j$, so we denote (\ref{pg}) in this case by 
$G^{\rm wu/wr}_{\rm triangle}(N;q^2)$.

Baik and Rains \cite{BR-2001} have shown that
\begin{equation}\label{ac0}
  {\rm Pr} \Big ( G^{\rm wu/wr}_{\rm triangle}(N;q^2) \le l \Big ) 
  = {1 \over 2} (1-q)^N (1-q^2)^{N(N-1)/2}
     \Big\langle \det \Big[ (1+U)(1+qU)^N \Big] \Big\rangle_{O^+(l)}.
\end{equation}
The significance of (\ref{ac0}) from the viewpoint
of the present study is that it can be written in the form
(\ref{F1}). To see this we recall that for $l$ even all eigenvalues of
$O^+(l)$ come in complex conjugate pairs $e^{\pm i \theta}$, while for
$l$ odd one eigenvalue of $O^+(l)$ is equal to $+1$ and the
rest come in complex conjugate pairs. Hence
\begin{equation*}
\det \Big [ (1 + U)(1 + qU)^N \Big ] = \chi_{l,N}
\prod_{j=1}^{[l/2]}(2 + 2\cos \theta_j)
(1 + q^2 + 2q \cos \theta_j)^N
\end{equation*}
where $\chi_{l,N} = 1$, $l$ even, $\chi_{l,N} = 2(1 + q)^N$,
$l$ odd. Introducing the variables
\begin{equation}\label{irn}
\lambda_j = {1 \over 2}(\cos \theta_j + 1)
\end{equation}
this reads
\begin{equation*}
\det \Big [ (1 + U)(1 + qU)^N \Big ] = 2^{2(N+1)[l/2]} q^{N[l/2]}
\chi_{l,N}
\prod_{j=1}^{[l/2]} \lambda_j \Big ( {1 \over 4 q} (1 - q)^2 + \lambda_j
\Big )^N.
\end{equation*}
Furthermore, in terms of the variables (\ref{irn}), the PDF of the
eigenvalues $\{ e^{\pm i \theta_j} \}_{j=1,\dots,[l/2]}$ for matrices
$U \in  O^+(l)$ is of the form (\ref{a1}) with $N$ replaced by $N^*$,
$w(x) = x^a(1-x)^b$ (Jacobi weight) and parameters
\begin{equation*}
(N^*,a,b) = ([l/2],-1/2, (-1)^{l-1}/2).
\end{equation*}
Thus we can rewrite (\ref{ac0}) to read
\begin{equation}\label{cc3}
{\rm Pr} 
\Big ( G^{\rm wu/wr}_{\rm triangle}(N;q^2) \le l \Big ) =
C_{l,N}(q)
\Big \langle \prod_{j=1}^{[l/2]} \lambda_j
\Big ( {1 \over 4 q} (1 - q)^2 + \lambda_j \Big )^N
\Big \rangle_{{\rm JUE}_{[l/2]}}
\Big |_{a = -1/2 \atop b = (-1)^{l-1}/2}
\end{equation}
where
\begin{equation}\label{cc3b}
C_{l,N}(q) :=
2^{2(N+1)[l/2]} q^{N[l/2]}
\chi_{l,N} {1 \over 2} (1 - q)^N (1 - q^2)^{N(N-1)/2}
\end{equation}
The result (\ref{r3aa}) can be used to express (\ref{cc3}) as a generalised
hypergeometric function.

\begin{prop}\label{p20}
With $C_{l,N}(q)$ specified by (\ref{cc3b}) and $J_n(a,b)$ specified by 
(\ref{jna})
\begin{multline}\label{5.45}
  {\rm Pr} \Big( G^{\rm wu/wr}_{\rm triangle}(N;q^2) \le l \Big) =
   C_{l,N}(q) {J_{[l/2]}(1/2+N, (-1)^{l-1}/2) \over
               J_{[l/2]}(-1/2, (-1)^{l-1}/2) } \\
   \times {}_2^{}F_1^{(1)}\Big( 
                 -[l/2],N+[l/2]+(1+(-1)^{l-1})/2;N+1/2;t_1,\dots,t_N 
                          \Big) \Big|_{t_1= \cdots = t_N = -(1-q)^2/4q}
\end{multline}
valid for $[l/2] \ge N$.
\end{prop}
The expression (\ref{5.45}) is well suited to taking the exponential limit
(\ref{caz}).
\begin{prop}\label{p21}
We have
\begin{align}\label{pp2}
  g_{\rm triangle}^{\rm wu/wr}(N;s)
  & := \lim_{L \to \infty} 
       {\rm Pr} \Big( G_{\rm triangle}^{\rm wu/wr}(N; e^{-1/L}) \le Ls \Big) 
  \nonumber \\
  & = \prod_{j=0}^N {\Gamma(j+1) \over \Gamma(2j+1)} s^{N(N+1)/2} e^{-Ns/4} \,
      {}_0^{} F_1^{(1)}(\underline{\:\:} ;N+1/2;t_1,\dots,t_N) \Big|_{t_1=\cdots=t_N=s^2/64} 
\nonumber \\
\end{align}
\end{prop}

\noindent
Proof. \quad Suppose for definiteness that $l$ is even.
In the limit (\ref{caz}) we can check from the definition (\ref{cc3b})
and the evaluation (\ref{jna}), making use of Stirling's formula,
that
\begin{equation*}
C_{l,N}(q)
{J_{[l/2]}(1/2+N, (-1)^{l-1}/2) \over
J_{[l/2]}(-1/2, (-1)^{l-1}/2) } \: \sim \: 
\prod_{j=0}^N {\Gamma(j+1) \over \Gamma(2j+1)}
s^{N(N+1)/2} e^{-Ns/4}.
\end{equation*}
Also, analogous to the result (\ref{su}), we can see from the series definition 
(\ref{cpc1}) that in the limit (\ref{caz})
\begin{multline}
  {}_2^{} F_1^{(1)}\Big( -[l/2],N + [l/2] + (1 + (-1)^{l-1})/2;
  N + 1/2; t_1,\dots,t_N \Big ) \Big|_{t_1= \cdots = t_N = - (1 - q)^2/4q} \\
  \quad \sim \, 
   {}_0^{} F_1^{(1)}(\underline{\: \:} ;N+1/2;t_1,\dots,t_N) 
                     \Big|_{t_1=\cdots=t_N=s^2/64}.
\end{multline}
\quad \hfill $\square$

In \cite{Fo_1994a} the generalised hypergeometric function 
\begin{equation*}
  {}_0^{}F_1^{(1)}(\underline{\: \:} ;N+\mu;t_1,\dots,t_N)\Big|_{t_1=\cdots=t_N=t},
\end{equation*}
for $\mu =0$ and $\mu=2$ has been evaluated as an $N$-dimensional determinant.
Generalising the working therein gives that for general $\mu$
\begin{equation}\label{sbs}
{}_0^{} F^{(1)}_1(\underline{\: \:}
;N+\mu;t_1,\dots,t_N) \Big |_{t_1=\cdots=t_N=s^2/4}
= \prod_{j=0}^{N-1} {\Gamma(\mu+j+1) \over \Gamma(j+1)} \,
\Big ( {2 \over s} \Big )^{N \mu}
\det \Big [ I_{j-k+\mu}(s) \Big ]_{j,k=0,\dots,N-1}.
\end{equation}
Furthermore, we know from \cite{FW_2002a} that
\begin{equation}
\prod_{j=0}^{N-1} {\Gamma(\mu+j+1) \over \Gamma(j+1)} \,
\Big ( {2 \over \sqrt{t}} \Big )^{N \mu}
\det \Big [ I_{j-k+\mu}(\sqrt{t}) \Big ]_{j,k=0,\dots,N-1} =
\exp \int_0^t {v(s;N,\mu) + s/4 \over s} \, ds
\end{equation}
where
\begin{equation}\label{5.62}
  v(t;N,\mu) = - \Big ( \sigma_{\III}(t) + \mu(\mu+N)/2 \Big ).
\end{equation}
The function $\sigma_{\III}(t)$ satisfies the differential equation
\begin{equation}\label{sbs2}
  (t\sigma_{\III}'')^2 - v_1 v_2 (\sigma_{\III}')^2
  + \sigma_{\III}'(4 \sigma_{\III}' - 1)(\sigma_{\III} - t \sigma_{\III}')
  - {1 \over 4^3} (v_1 - v_2)^2 = 0,
\end{equation}
which is the Jimbo-Miwa-Okamoto $\sigma$-form of the \PIIIa equation
(a simple change of variables relates \PIII to \PIIIa \cite{Ok_1987c}) with
parameters
\begin{equation}\label{sbs2a}
  (v_1, v_2) = (\mu+N, - \mu + N)
\end{equation}
and subject to the boundary conditions
\begin{equation}\label{5.64a}
  \sigma_{\III}(t) \, \mathop{\sim}\limits_{t \to \infty} \,
  {t \over 4} - {Nt^{1/2} \over 2} 
   + \Big( {N^2 \over 4} - {\mu^2 \over 2} \Big).
\end{equation} 
The boundary condition (\ref{5.64a}) does not distinguish solutions with
$\mu \mapsto - \mu$ (note from (\ref{sbs2a}) that under this mapping
$v_1 \leftrightarrow v_2$, and this latter transformation leaves
(\ref{sbs2}) unchanged). However, it follows from (\ref{sbs})--(\ref{5.62})
that
\begin{equation}\label{dbc}
  \sigma_{\III}(t) \, \mathop{\sim}\limits_{t \to 0^+} \,
  - {\mu(\mu+N) \over 2} + {\mu \over 2 (\mu+N)} t + O(t^2).
\end{equation}
This boundary condition does distinguish solutions with $\mu \mapsto - \mu$.

Collecting together the above results allows the evaluation (\ref{pp2}) of
$g_{\rm triangle}^{\rm wu/wr}(N;s)$ to be further developed.

\begin{prop}\label{p22}
We have
\begin{align}
  g_{\rm triangle}^{\rm wu/wr}(N;s)
  & = 2^{3N/2} {\Gamma(N+1) \over \Gamma(2N+1)}
         \Big( \prod_{j=0}^{N-1} {\Gamma(j+3/2) \over \Gamma(2j+1)} \Big)
      s^{N^2/2} e^{-Ns/4} \det \Big[ I_{j-k+1/2}(s/4) \Big]_{j,k=0,\dots,N-1}
  \nonumber \\
  & = \prod_{j=0}^N {\Gamma(j+1) \over \Gamma(2j+1)}
       s^{N(N+1)/2} \exp \int_0^{(s/4)^2}
       (v(t;N,1/2) + t/4 - Nt^{1/2}/2) {dt \over t}
  \nonumber \\
  & = \exp\left( -\int_{(s/4)^2}^\infty
      \Big(-\sigma_{\III}(t) \Big|_{v_1 = N+1/2 \atop v_2 = N-1/2}
       + t/4 - Nt^{1/2}/2 + (N^2-1/2)/4 \Big) {dt \over t} \right)
\end{align}
\end{prop}

In the Okamoto $\tau$-function theory of the \PIIIa system \cite{Ok_1987c}, the
transcendent $-\sigma_{\III}(t)/t$ is an auxiliary Hamiltonian, being
equal to the original Hamiltonian plus a function of $t$. Changing
the function of $t$ doesn't alter this property, so we have that
\begin{equation}\label{pla}
  g_{\rm triangle}^{\rm wu/wr}(N;s)
  = \tau_{\III}((s/4)^2) \Big|_{v_1 = N+1/2 \atop v_2 = N-1/2}
\end{equation}
where
\begin{equation}\label{pla1}
   t{d \over dt} \log\tau_{\III}(t) \Big|_{v_1 = N+1/2 \atop v_2 = N-1/2}
  \doteq 
    - \sigma_{\III}(t) \Big|_{v_1 = N+1/2 \atop v_2 = N-1/2}
    + t/4 - Nt^{1/2}/2 + (N^2-1/2)/4.
\end{equation}

The symmetrised Johansson model has been generalised by Baik and Rains
\cite{BR-2001,BR-2001a} to include a parameter $\alpha$ in the geometric
distribution of the waiting times $w(i,i)$ of the diagonal sites. Thus
these waiting times are chosen to have distribution $g(\alpha q)$, with the
case $\alpha = 1$ corresponding to the original symmetrised Johansson model.
With the random variable (\ref{pg}) now denoted 
$ G_{\rm triangle}^{\rm wu/wr}(N;\alpha,q^2) $, one then has \cite{Ba-2001}
\begin{multline}\label{tf}
  {\rm Pr}\Big( G_{\rm triangle}^{\rm wu/wr}(N;\alpha,q^2) \le l \Big)
  = {1 \over 2} (1 - \alpha q)^N (1 - q^2)^{N(N-1)/2} \\
    \times \Big(  \Big\langle \det \Big[(1 + \alpha U)(1 + q U)^N \Big]
                  \Big\rangle_{U \in O^{+}(l)}
                + \Big\langle \det \Big[(1 + \alpha U)(1 + q U)^N \Big]
                  \Big\rangle_{U \in O^{-}(l)} \Big)
\end{multline}
The analogue of Propositions \ref{p20}, \ref{p21} and \ref{p22}
can be obtained
for the case $\alpha = 0$.

\begin{prop}\label{p20a}
Let $\chi^+_{l,N} = 1$, $l$ even and $\chi^+_{l,N} = (1+q)^N$, $l$ odd, and
put
\begin{equation*}
C_{l,N}^+(q) = 2^{2N[l/2]} q^{N[l/2]} \chi_{l,N}^+
{1 \over 2} (1 - q^2)^{N(N-1)/2}.
\end{equation*}
Let $\chi^-_{l,N} = (1 - q^2)^N$, $l$ even and $\chi^-_{l,N} = (1 - q)^N$,
$l$ odd, and put
\begin{equation*}
C_{l,N}^-(q) =  2^{2N[(l-1)/2]} q^{N[(l-1)/2]}  \chi_{l,N}^-
{1 \over 2} (1 - q^2)^{N(N-1)/2}.
\end{equation*}
We have
\begin{multline}\label{pqp}
  {\rm Pr} \Big( G^{\rm wu/wr}_{\rm triangle}(N;0,q^2) \le l \Big)
  = C_{l,N}^+(q) {J_{[l/2]}(-1/2+N, (-1)^{l-1}/2) \over
                  J_{[l/2]}(-1/2, (-1)^{l-1}/2) } \\
    \times {}_2^{} F_1^{(1)}\Big( -[l/2],N + [l/2] + ((-1)^{l-1}-1)/2;
            N - 1/2; t_1,\dots,t_N \Big) \Big|_{t_1= \cdots = t_N = - (1 - q)^2/4q} \\
    + C_{l,N}^-(q) {J_{[(l-1)/2]}(1/2+N, (-1)^{l}/2) \over
                    J_{[l/2]}(1/2, (-1)^{l}/2) } \\
    \times {}_2^{} F_1^{(1)}\Big ( -[(l-1)/2],N + [(l-1)/2] + (1+(-1)^{l})/2;
            N + 1/2; t_1,\dots,t_N \Big) \Big|_{t_1= \cdots = t_N = - (1 - q)^2/4q}.
\end{multline}
valid for $[l/2] \ge N$.
\end{prop}

\noindent
Proof. \quad The rewrite of the average over $O^+(l)$ in (\ref{tf}) is done
in the same way as in going from (\ref{ac0}) to (\ref{5.45}). To rewrite
the average over $O^-(l)$ in (\ref{tf}) we must first recall that for
$l$ even there are eigenvalues $\pm 1$, while the remaining eigenvalues
come in complex conjugate pairs $e^{\pm i \theta_j}$, and for $l$ odd there is
an eigenvalue $-1$, with the remaining eigenvalues also coming in
complex conjugate pairs $e^{\pm i \theta_j}$. Furthermore, 
in terms of the variables (\ref{irn}), the PDF of the
eigenvalues $\{ e^{\pm i \theta_j} \}_{j=1,\dots,[(l-1)/2]}$ 
is of the form (\ref{a1}) with $N$ replaced by $N^*$,
$w(x) = x^a(1-x)^b$  and parameters
\begin{equation*}
(N^*,a,b) = ([(l-1)/2],1/2, (-1)^{l}/2).
\end{equation*}
Using these facts, we proceed as in the rewrite of the $O^+(l)$ average.
\hfill $\square$

\begin{prop}
We have
\begin{align}\label{pp2a}
  g_{\rm triangle}^{\rm wu/wr}(N;0,s)
  & := \lim_{L \to \infty} {\rm Pr} \Big(
             G_{\rm triangle}^{\rm wu/wr}(N; 0, e^{-1/L}) \le Ls \Big)
  \nonumber \\
  & =  {1 \over 2} \left\{ \prod_{j=0}^N {\Gamma(j+1) \over \Gamma(2j+1)}
        s^{N(N+1)/2} e^{-Ns/4} {}_0^{} F_1^{(1)}(\underline{\: \:};
                         N+1/2;t_1,\dots,t_N) \Big |_{t_1 = \cdots = t_N = s^2/64}
                   \right.
  \nonumber \\
  & \quad +        \left.
      \prod_{j=1}^N {\Gamma(j) \over \Gamma(2j-1)}
        s^{N(N-1)/2} e^{-Ns/4} {}_0^{} F_1^{(1)}(\underline{\: \:};
                         N-1/2;t_1,\dots,t_N) \Big|_{t_1 = \cdots = t_N = s^2/64}
                   \right\}
  \nonumber \\
  & = {1 \over 2}  \left\{ \exp\left( -\int_{(s/4)^2}^\infty
         \Big(- \sigma_{\III}(t) \Big|_{v_1 = N + 1/2 \atop v_2 = N - 1/2}
               + t/4 - Nt^{1/2}/2 + (N^2-1/2)/4 \Big) {dt \over t} \right)
                   \right.
  \nonumber \\
  & \quad +        \left.
         \exp\left( -\int_{(s/4)^2}^\infty
         \Big(- \sigma_{\III}(t) \Big|_{v_1 = N - 1/2 \atop v_2 = N + 1/2}
               + t/4 - Nt^{1/2}/2 + (N^2-1/2)/4 \Big) {dt \over t} \right)
                   \right\} 
\end{align}
\end{prop}

\noindent
Proof. \quad
These formulas follow from Proposition \ref{p20a} by following the working
which led from Proposition \ref{p20} to Propositions \ref{p21} and
\ref{p22}. The two terms in (\ref{pp2a}) correspond to the two terms in
(\ref{pqp}) but in the reverse order. \hfill $\square$

Baik \cite{Ba-2002} has recently evaluated the scaled limit of (\ref{tf})
for general $\alpha$ (the variable $\alpha$ is also involved in the
scaling). The evaluation is given in terms of the solution of a 
Riemann-Hilbert problem. It is pointed out in \cite{Ba-2002} that this has
consequence with regard to the distribution of the largest eigenvalue
in an interpolating Laguerre ensemble, which passes continuously from
the LOE${}_{N}$ to the LSE${}_{N/2}$. This follows from the identity
\cite{BR-2001} (see also \cite{FR-2002})
\begin{multline}\label{nc}
  g_{\rm triangle}^{\rm wu/wr}(N;A,s) :=
     \lim_{L \to \infty} {\rm Pr} \Big(
           G_{\rm triangle}^{\rm wu/wr}(N; e^{A/2L}, e^{-1/L}) \le Ls \Big) \\
  = {1 \over C} \int_0^s dx_1 \cdots \int_0^s dx_N \,
                \chi_{x_1 > x_2 > \cdots > x_N \ge 0}
                 e^{-(1/2) \sum_{j=1}^N x_j} e^{(A/2) \sum_{j=1}^N (-1)^{j-1} x_j}
                \prod_{j<k}(x_j - x_k)
\end{multline}
where $\chi_T = 1$ if $T$ is true and $\chi_T=0$ otherwise. The case
$A=0$ corresponds to $\alpha=1$. Now, with $A=0$ the integral in (\ref{nc})
is precisely the gap probability 
$E_1((s,\infty);N;0)$
(no eigenvalues in the interval $(s,
\infty)$) for the LOE${}_N$ with parameter $a=0$. Equating (\ref{nc}) in this
case with (\ref{pla}) then gives the $\tau$-function evaluation
\begin{equation}\label{bai1}
  E_1^{\rm LOE}((s,\infty),N;0) 
  = \tau_{\III}((s/4)^2) \Big|_{v_1 = N+1/2 \atop v_2 = N-1/2}.
\end{equation}

Taking the $A \to - \infty$ limit in (\ref{nc})
corresponds to $\alpha = 0$. We see from
(\ref{nc}) that for $N$ even 
\begin{equation*}
\lim_{A \to - \infty} g_{\rm triangle}^{\rm wu/wr}(N;A,s) =
E_4^{\rm LSE}((s,\infty);N/2;0),
\end{equation*}
where $E_4^{\rm LSE}((s,\infty);n;a)$ denotes the probability that there
are no eigenvalues in the LSE${}_n$ with weight $x^a e^{-x}$. Equating
(\ref{nc}) in this case with (\ref{pla}) then gives
\begin{equation}\label{nce}
  E_4^{\rm LSE}((s,\infty);N/2;0)
  = {1 \over 2} \Big( 
     \tau_{\III}((s/4)^2) \Big|_{v_1 = N+1/2 \atop v_2 = N-1/2}
   + \tau_{\III}((s/4)^2) \Big|_{v_1 = N-1/2 \atop v_2 = N+1/2} \Big).
\end{equation}
As the differential equation (\ref{sbs2}) is unchanged by the interchange 
$ v_1 \leftrightarrow v_2 $ the Hamiltonians for both $\tau$-functions in
(\ref{nce}) satisfy the same differential equation.
It is interesting to compare (\ref{nce}) with the structural formula
\cite{FR-2001}
\begin{equation}\label{nce1}
E_4^{\rm LSE}((s,\infty);N/2;0) =
{1 \over 2} \Big (
E_1^{\rm LOE}((s,\infty);N;0) +
{E_2^{\rm LUE}((s,\infty);N;0) \over E_1^{\rm LOE}((s,\infty);N;0) }
\Big ).
\end{equation}
Doing this tells us that
\begin{equation}\label{nce2}
  E_2^{\rm LUE}((s,\infty);N;0)
  = \tau_{\III}((s/4)^2) \Big|_{v_1 = N + 1/2 \atop v_2 = N - 1/2}
    \tau_{\III}((s/4)^2) \Big|_{v_1 = N - 1/2 \atop v_2 = N + 1/2}.
\end{equation}
On the other hand we know from \cite{TW_1994} that
\begin{equation}\label{nce3}
  E_2^{\rm LUE}((s,\infty);N;0) = \exp \Big ( - \int_s^\infty
  \sigma(t) \Big |_{\nu_0=\nu_1=0 \atop \nu_2=\nu_3=N} \,
  {dt \over t} \Big ) =
  \tau_V(s) \Big |_{\nu_0=\nu_1=0 \atop \nu_2=\nu_3=N},
\end{equation}
where $\sigma(t)$ satisfies (\ref{vpx}), and so
\begin{equation}\label{nce4}
  \tau_V(s) \Big|_{\nu_0=\nu_1=0 \atop \nu_2=\nu_3=N}
   = \tau_{\III}((s/4)^2) \Big|_{v_1 = N + 1/2 \atop v_2 = N - 1/2}
     \tau_{\III}((s/4)^2) \Big|_{v_1 = N - 1/2 \atop v_2 = N + 1/2}.
\end{equation}
Similar $\tau$-function identities to (\ref{nce4}) have occured in the works 
\cite{Fo-99,Wi_2001a}. The Painlev\'e transcendent evaluations of
$E_1^{\rm LOE}((s,\infty);N;0)$ and $E_4^{\rm LSE}((s,\infty);N/2;0)$
given in \cite{Ba-2002} differ from (\ref{bai1}) and (\ref{nce}),
involving instead of $\tau$-functions for the PIII' system, the
transformed PV transcendent found in \cite{TW_1994} to be simply related
to the derivative of $\sigma(t) \Big |_{\nu_0=\nu_1=0 \atop \nu_2=\nu_3=N}$. 

\subsection{Diagonal-diagonal correlation in the 2d Ising model}
The square lattice Ising model has one of two possible states,
$\sigma_{ij} = \pm 1$, on each site $(i,j)$ of the two-dimensional
square lattice (see e.g.~\cite{Ba-82}). 
The infinite square lattice of states is achieved by
considering a sequence of finite lattices, of dimension
$(2N+1) \times (2N+1)$ say, centred about the origin. The joint
probability density function for a particular configuration of states
in the finite system is given by
\begin{equation*}
{1 \over Z_{2N+1}}
\exp \Big [ K_1 \sum_{j=-N}^N \sum_{i=-N}^{N-1}
\sigma_{ij} \sigma_{i+1 \, j} +
K_2 \sum_{i=-N}^N \sum_{j=-N}^{N-1} \sigma_{ij}
\sigma_{i \, j+1} \Big ],
\end{equation*}
where $ Z_{2N+1}$ is the normalisation. Thus there is a coupling between
nearest neighbours in the horizontal and vertical direction. In the
limit $N \to \infty$, an unpublished result of Onsager
(see \cite{McCW_1973}) gives that the diagonal spin-spin correlation has the
Toeplitz form
\begin{equation}\label{5.84'}
\langle \sigma_{00} \sigma_{nn} \rangle = \det [ a_{i-j} ]_{i,j=1,\dots,n}
\end{equation}
where
\begin{equation}\label{rfc}
a_p = {1 \over 2 \pi} \int_0^{2 \pi} d \theta \,
e^{-i p \theta}
\Big [ {1 - (1/k) e^{-i \theta} \over
1 - (1/k) e^{i \theta} } \Big ]^{1/2}, \quad
k := \sinh 2K_1 \sinh 2K_2 .
\end{equation}
Regarding the Fourier integral (\ref{rfc}) as a contour integral, 
deforming the contour of integration and then writing the 
Toeplitz determinant as an average of the CUE we see that
\begin{equation}
\langle \sigma_{00} \sigma_{nn} \rangle = \left \{
\begin{array}{l}
\langle \prod_{l=1}^n z_l^{1/4} |1 + z_l |^{-1/2}
(1 + (1/k^2) z_l )^{1/2} \rangle_{{\rm CUE}_n}, \quad
1/k^2 \le 1 \\
k^{-n}
\langle \prod_{l=1}^n z_l^{-3/4} |1 + z_l |^{-1/2}
(1 + k^2 z_l )^{1/2} \rangle_{{\rm CUE}_n}, \quad
k^2 \le 1 . \end{array}\right.
\end{equation}
We can now read off from (\ref{1.41g}) the following results.

\begin{prop}
Let $s=1/k^2$. We have
\begin{equation}
- {1 \over 4}(1+n^2) s + {1 \over 8} + s(s-1) {d \over ds} \log
\langle \sigma_{00} \sigma_{nn} \rangle =
U^{\rm J}_n(s;1/2-n,1/2,-1/2)
\end{equation}
where $U^{\rm J}_n$ satisfies the \PVI equation in $\sigma$-form
(\ref{4}) with parameters
\begin{equation*}
\mathbf{b} = 
\Big ({1 \over 2}(n-1), {n \over 2}, {1 \over 2} (n+1), - {n \over 2}  
\Big ).
\end{equation*}
Let $t = k^2$. We have
\begin{equation}\label{sfu}
- {n^2 \over 4} t - {1 \over 8} + t(t-1) {d \over dt} \log
\langle \sigma_{00} \sigma_{nn} \rangle =
U^{\rm J}_n(t;-1/2-n,1/2,-1/2)
\end{equation}
where $U^{\rm J}_n$ satisfies the \PVI equation in $\sigma$-form
(\ref{4}) with parameters
\begin{equation*}
\mathbf{b} = 
\Big ({n \over 2}, {n - 1 \over 2}, {n \over 2}, - {n +1 \over 2}  
\Big ).
\end{equation*}
\end{prop}

It is straightforward to check from (\ref{sfu}) and (\ref{4}) that
\begin{equation*}
\sigma(t) :=  t(t-1) {d \over dt} \log
\langle \sigma_{00} \sigma_{nn} \rangle - {1 \over 4}
\end{equation*}
satisfies the differential equation
\begin{equation*}
  \Big( t(t-1) {d^2 \sigma \over d t^2} \Big)^2 =
  n^2 \Big( (t-1) {d \sigma \over dt} - \sigma \Big)^2
  - 4{d \sigma \over dt} \Big( (t-1){d \sigma \over dt}
                               -\sigma -{1 \over 4} \Big)
     \Big( t{d \sigma \over dt} - \sigma \Big).
\end{equation*}
This is a result due to Jimbo and Miwa \cite{JM_1980}. Our point therefore
is not a new characterisation of $\langle \sigma_{00} \sigma_{nn} \rangle$,
but rather the fact that the known characterisation fits 
our development of the Okamoto $\tau$ function theory of \PVI.
We note too that a recent result of Borodin \cite{Bo_2001} can be used to
give a recurrence for $\langle \sigma_{00} \sigma_{nn} \rangle$ as a
function of $n$. This recurrence applies to all Toeplitz determinants
\begin{align}
  q_n^{(z,z',\xi)} 
  & := (1 - \xi)^{zz'} \det [ g_{i-j} ]_{i,j=1,\dots,n}
  \nonumber\\
  g_p 
  & = {1\over 2 \pi} \int_0^{2 \pi} d\theta \, 
       e^{-i p\theta} (1 - \sqrt{\xi} e^{i \theta})^z
       (1 - \sqrt{\xi} e^{-i \theta})^{z'}.
\end{align}\label{Ising_toeplitz}
Comparison with (\ref{5.84'}) and (\ref{rfc}) shows
\begin{equation*}
\langle \sigma_{00} \sigma_{nn} \rangle =
{1 \over 1 - (1/k)^2} q_n^{(-1/2,1/2,1/k^2)}.
\end{equation*}
Recent work of Adler and van Moerbeke \cite{AvM_2002} gives a different recurrence
relation for (\ref{Ising_toeplitz}) to that of Borodin, and thus a further
recurrence for $ \langle \sigma_{00} \sigma_{nn} \rangle $.
Explicit computation of $\langle \sigma_{00} \sigma_{nn} \rangle$ as a
power series in $k^2$, for which both the differential and difference
equation characterisations are well suited, is a crucial step in recent 
very high precision numerical studies of the zero field susceptibility 
\cite{Ni-1999,Ni-2000,ONGP-2001a,ONGP-2001b}. 

\subsection*{Acknowledgements}
This research has been supported by the Australian Research Council.
We thank W. Orrick for pointing out the paper \cite{JM_1980}, and J. Baik for
sketching his results on the evaluation of 
$ E^{\rm LOE}_{1}((s,\infty);N;0) $ and $ E^{\rm LSE}_{4}((s,\infty);N/2;0) $.

\bibliographystyle{plain}
\bibliography{moment,random_matrices,nonlinear}

\end{document}